\documentclass[sigconf]{acmart}

\usepackage{ellipsis}

\AtBeginDocument{%
  \providecommand\BibTeX{{%
    \normalfont B\kern-0.5em{\scshape i\kern-0.25em b}\kern-0.8em\TeX}}}

\copyrightyear{2021} 
\acmYear{2021} 
\setcopyright{licensedusgovmixed}\acmConference[FAccT '21]{ACM Conference on Fairness, Accountability, and Transparency}{March 1--10, 2021}{Virtual Event, Canada}
\acmBooktitle{Conference on Fairness, Accountability, and Transparency (FAccT '21), March 3--10, 2021, Virtual Event, Canada}
\acmPrice{15.00}
\acmDOI{10.1145/3442188.3445937}
\acmISBN{978-1-4503-8309-7/21/03}

\acmConference[FAccT'2021]{FAccT '21: The ACM Conference on Fairness, Accountability, and Transparency}{March, 2021}{Toronto, CA (Virtual)}
\acmBooktitle{FAccT '21: ACM Conference on Fairness, Accountability, and Transparency, March 2021, Toronto, CA (Virtual)}

\begin{document}

\title{Outlining Traceability: A Principle for Operationalizing Accountability in Computing Systems}

\author{Joshua A. Kroll}
\email{jkroll@nps.edu}
\affiliation{
  \institution{Naval Postgraduate School}
  \city{Monterey, CA}
}
\begin{abstract}
Accountability is widely understood as a goal for well governed
computer systems, and is a sought-after value in many governance
contexts. But how can it be achieved? Recent work on standards for
governable artificial intelligence systems offers a related principle:
traceability. Traceability requires establishing not only how a system
worked but how it was created and for what purpose, in a way that
explains why a system has particular dynamics or behaviors.  It
connects records of how the system was constructed and what the system
did mechanically to the broader goals of governance, in a way that
highlights human understanding of that mechanical operation and the
decision processes underlying it. We examine the various ways in which
the principle of traceability has been articulated in AI principles
and other policy documents from around the world, distill from these a
set of requirements on software systems driven by the principle, and
systematize the technologies available to meet those
requirements. From our map of requirements to supporting tools,
techniques, and procedures, we identify gaps and needs separating what
traceability requires from the toolbox available for
practitioners. This map reframes existing discussions around
accountability and transparency, using the principle of traceability
to show how, when, and why transparency can be deployed to serve
accountability goals and thereby improve the normative fidelity of
systems and their development processes.
\end{abstract}




\maketitle

\section{Introduction}
\label{sec:intro}
Accountability is a long sought-after value in decision-making
systems, especially when those systems are
computerized~\cite{jabbra1989public, nissenbaum1996accountability,
  friedman1999trust, mulgan2003holding, breaux2006enforceability,
  weitzner2007information, feigenbaum2012systematizing,
  nature-editorial, kroll2017penn, raji2020closing,
  kroll2020accountability}. It is a multifaceted concept, presenting
challenges for operationalization~\cite{wieringa2020account}. Work to
date on operationalizing accountability in computing systems focuses on
keeping records along the dimensions of time, information, and
action~\cite{feigenbaum2012systematizing}, on ensuring that
misbehavior in a protocol can be attributed to a specific
participant~\cite{argyraki2007accountability, haeberlen2007peerreview,
  kuesters2010accountability, kunnemann2019automated}, and on
demonstrating partial information about the contents of those records
to affected people or to oversight entities~\cite{kroll2017penn}. But,
despite work on requirements
engineering~\cite{breaux2006enforceability, breaux2009distributed},
there remains no clear approach to defining what requirements exist
for generating and maintaining records, neither regarding what
records must contain nor regarding how to store them. Further, there
is minimal work describing how records lead to accountability in the
sense of responsibility or in the sense of fidelity to values and
norms, such as fairness, nondiscrimination, or
safety~\cite{kroll2020accountability}. However, the related principle
of \emph{traceability} has recently been espoused by many
organizations as an approach to making software systems robustly
transparent in order to facilitate accountability. This article
explores the notion of traceability, asking how it might be
operationalized in practical systems and how it serves the broader
value of accountability.

Traceability refers broadly to the idea that the outputs of a computer
system can be understood through the process by which that system was
designed and developed~\cite{kroll2018fallacy}. Specifically,
traceability requires that transparency about the development process
and its goals be tied to outcomes through the auditability of the
methods used in both the creation and the operation of the system
itself. This includes ensuring the existence and legibility of records
related to technology choice, design procedure, development process,
operation, data sources, and system documentation. In this way,
traceability unifies many desirable goals in the governance of
computing systems, relating transparency about system design,
construction (e.g., components and data), and operation to the
\emph{auditability} of the system to investigate its properties and to
establish responsibility for system outcomes. Tracing how a system
functions must go beyond providing a simple, mechanical description of
how operations led from input to output or what the system's dynamics
and behaviors are. Instead, traceability demands an answer to the
question of why the system works the way it does and what decisions
led to that design or operation. In other words, traceability relates
the objects of transparency (disclosures about a system or records
created within that system) to the goals of accountability (holding
the designers, developers, and operators of a computer system
responsible for that system's behaviors and ultimately assessing that
the system reflects and upholds desired norms).

In this article, we unpack the concept of traceability analytically,
examining several ways the principle of traceability has been
articulated in official policy guidance and consider how it might be
operationalized in real systems to achieve the lofty goal of
connecting the operation of a computer system to normative concerns
such as fidelity to policy goals or embodiment of desired values. In
addition, we systematize existing technological tools which can help
achieve traceability and suggest a set of system requirements implied
by the principle and its goals. We find that traceability ties
together a number of existing threads of research around the
responsible and ethical use of software technology for socially
important applications. For example, traceability relates work on
accounting for computing systems
concretely~\cite{argyraki2007accountability, csar,
  haeberlen2007peerreview}, calls for better governance of computer
systems~\cite{citron2007technological, citron2008open,
  mulligan2018saving, citron2020automated}, demands for
reproducibility of software design and
construction~\cite{stodden2014best, stodden2014implementing,
  reproducible-builds,nikitin2017chainiac}, work on the security of
software supply chains~\cite{ellison2010evaluating}, and the
burgeoning field of trustworthy artificial
intelligence~\cite{brundage2020toward,floridi2019establishing}.

In another sense, traceability is a human factors problem. For a
system to be traceable, key information about the development and
operation of the system must be understandable to relevant
stakeholders. Thus, traceability relates to questions of whether
systems can be adequately understood for their given purpose,
including an understanding of their provenance and development and of
their operation~\cite{doshi2017accountability}. We refrain, however,
from saying that traceability requires explainability (at least as
that concept is understood within computer science as a mechanical
description of the mapping between inputs and outputs) --- it does
not. As noted above, a mechanical tracing of inputs to outputs, even
if causal, does not describe a system's origins or answer questions
about why it has particular dynamics or exhibits particular
behavior. Establishing appropriate human factors evaluation will
depend on an understanding of stakeholders and their
needs~\cite{young2019toward, wolf2018, lee2019procedural,
  lee2019webuildai, lee2020human}, along with robust explanations for
operational behavior that are appropriately causal, contrastive, and
selective~\cite{miller2017explanation}.

Ultimately, we find that traceability provides a grounding for other
ethical principles and normative desiderata within computing
systems. Traceability is made more easily amenable to measurement and
assessment than other principles because it is clear when efforts
towards traceability have been undertaken and when those efforts have
been successful. This is not true for other commonly espoused but more
abstract principles (such as fairness and equitability or even
transparency and accountability). Partly, this is because
traceability, like transparency, is well understood as an
\emph{instrumental} value that serves as a path to achieving other
goals, rather than an end to achieve in itself. That is not to say
traceability is not worthy of pursuit, but rather to say that its
pursuit serves other ends.

Achieving traceability may be more concrete than achieving other
ethical goals such as fairness, but it is by no means an easy
task. Although the path to reaching traceable systems is
straightforward, few if any systems can claim to have successfully
navigated it. We identify gaps in achieving real-world traceability in
systems at the level understood within policy documents as a
requirement or obligation on the entities bound by those
policies. These gaps fit into four basic categories: relevant
technologies such as tools for reproducibility or requirements on
lifecycle and versioning management for software and data science
artifacts have not been adopted; related human factors problems
defining what it means for people to understand system provenance and
its implication for their goals and tasks have not been solved; there
are not yet accepted standards or even shared best practices for test
and evaluation of software systems for critical applications in many
domains; and there is often insufficient oversight and review of the
behavior of computing systems or capacity for disaffected persons to
seek redress. Some work has aimed at each of these gaps---traceability
systematizes the project of these disparate threads into the
operationalization of a single ethical principle for computing
systems.

Finally, a note about terminology: in this work, we explicitly avoid
the term ``artificial intelligence'', which eludes a rigorous
definition. Instead, we speak of \emph{automation}, or the embodiment
of tasks in technological artifacts (which may or may not be
software-based), and \emph{computing systems}, which use computational
processes, software, and related technologies (e.g., data science and
machine learning) to implement automation. The use of the term
``systems'' here is purposeful -- artifacts on their own (such as
models or programs) cannot exhibit traceability, which is a property
of a tool in a context of use. We can model this by considering
traceability as a property of a system, a set of elements which
interact to produce aggregated behavior by virtue of those
interactions~\cite{mitchell2009complexity,
  martin2020extending}. Although a tool such as a piece of software
can support traceability, because traceability relates the specifics
of the tool's design to the effects of its use, the tool cannot on its
own be said to be traceable. Traceability is a functional property of
the tool in use, and is poorly defined without this context.

\section{Adoption of Traceability}
\label{sec:adoption}
Traceability has been adopted as an explicit principle either by
organizations as a principle for the responsible use of technology or
as policy guidance in official documents. For example, the United
States President's Executive Order 13960~\cite{eo13960} gives nine
principles for the development of ``trustworthy'' AI requires that
agencies of the U.S. Federal Government must make AI systems
``Responsible and Traceable'' when ``designing, developing, acquiring,
and using'' AI. Specifically, that principle states:
\begin{quote}
\textbf{Responsible and traceable.} Agencies shall ensure that human
roles and responsibilities are clearly defined, understood, and
appropriately assigned for the design, development, acquisition, and
use of AI. Agencies shall ensure that AI is used in a manner
consistent with these Principles and the purposes for which each use
of AI is intended. The design, development, acquisition, and use of
AI, as well as relevant inputs and outputs of particular AI
applications, should be well documented and traceable, as appropriate
and to the extent practicable.
\end{quote}
The order is unfortunately light on details, deferring these to
reports demanded of agencies in the weeks following its
release. However, we see that traceability is related to the
documentation of development and acquisition of ``AI'' (a term the
order also leaves undefined).

Almost a year prior to this government-wide policy directive, the
United States Department of Defense adopted traceability as one of its
five ``AI Ethics
Principles''~\cite{dod-official-ethics-principles-2020}. Specifically,
department guidance states the principle:
\begin{quote}
  \textbf{Traceable.} The Department’s AI capabilities will be
  developed and deployed such that relevant personnel possess an
  appropriate understanding of the technology, development processes,
  and operational methods applicable to AI capabilities, including
  with transparent and auditable methodologies, data sources, and
  design procedure and documentation.
\end{quote}
This language originates in a report from a Federal Advisory
Committee, the Defense Innovation Board~\cite{dib2019supporting}. That
study recommends as part of traceability various improvements in
software development discipline, including ``simulation environments,
modeling, automated testing, and validation tools'' but also
improvements to design methodology and assurance that relevant
stakeholders are apprised of development progress. In addition,
traceability is operationalized during system deployment through a
combination of online auditing and careful testing, possibly in
simulated environments. The principle of traceability is also
referenced in documents from the U.S. National Security Commission on
Artificial Intelligence, which adopts the Defense Department language
and approach~\cite{nscai-q2-2020}.

A similar principle that AI systems have ``Documentation of Purpose,
Parameters, Limitations, and Design Outcomes'' can also be found in
the United States Intelligence Community's ``AI Ethics
Framework''~\cite{ic-ai-ethics}. This principle goes further, calling
for documentation stored in a way that is ``accessible to all
potential consumers'' of the technology as well as ``how to verify and
validate'' it. This principle fits it a larger framework that also
demands ``Accounting for Builds, Versions, and Evolutions of an AI''
as well as documentation of ``the test methodology, results, and
changes made based on the test results''. Overall, although the IC
framework does not explicitly use the term ``traceability'', it
clearly espouses the concepts this term signifies in other policy
documents.

The U.S. Federal Data Strategy echoes language from an earlier
Executive Order 13859~\cite{eo13859}, ``Maintaining American
Leadership in Artificial Intelligence'', which calls for
U.S. government agencies to ``Enhance access to high-quality and
\emph{fully traceable} Federal data, models, and computing resources
to increase the value of such resources for AI R\&D, while maintaining
safety, security, privacy, and confidentiality protections consistent
with applicable laws and policies [\emph{emphasis added}].'' This
meaning is expanded in the Federal Data Strategy ``2020 Action
Plan''~\cite{federal-data-strategy-2020-action-plan} to cover:
\begin{itemize}
\item Expanding access to government data and enhancing its quality;
\item Improving guidance around maintaining inventory of data and
  models of that data;
\item Developing standard metadata and formats for identified assets
  to facilitate a government-wide data inventory; and
\item Establishing pilot projects to demonstrate this traceability.
\end{itemize}
Here, traceability is focused on the provenance of software systems
and the decisions made during their creation. However, unlike
transparency, which is often framed as a burden placed on system
creators and controllers, traceability is described as an enabling
value, providing a route to new and more capable systems and a way to
tie access to data and other resources to the provenance of systems.

Traceability is by no means limited to U.S. policy documents,
however. The E.U. High Level Expert Group's ``Ethics Guidelines for
Trustworthy AI''~\cite{eu-hleg-trustworthy-2020} calls for
traceability as a part of its broader theme of transparency, saying:
\begin{quote}
  The data sets and the processes that yield the AI system's decision
  [\ldots] should be documented to the best possible standard to allow
  for traceability and an increase in transparency. This also applies
  to the decisions made by the AI system. This enables identification
  of the reasons why an AI-decision was erroneous which, in turn,
  could help prevent future mistakes. Traceability facilitates
  auditability as well as explainability.
\end{quote}
Again, we see that traceability is tied explicitly to improvements in
the development process, such as documenting design methodology, but
also to improvements in test and validation as well as examining
outcomes driven by the system.

Traceability also appears in international policy guidance. For
example, the Organization for Economic Cooperation and Development
(OECD) states in its ``Recommendation of the Council on Artificial
Intelligence'' that AI systems should have ``Robustness, security, and
safety'' and ``[t]o this end, AI actors should ensure traceability,
including in relation to datasets, processes and decisions made during
the AI system lifecycle, to enable analysis of the AI system’s
outcomes and responses to inquiry, appropriate to the context and
consistent with the state of art''~\cite{oecd-ai-guidelines}. In this
formulation, traceability is explicitly called out for its information
security value, and it is also made clear that operationalizing
traceability must be done in a way that is ``appropriate to the
context'' and which ``enable[s] analysis of the [\ldots]
outcome''. However, we see again a focus on making clear the reasons
for a system's design and the origins of its components as well as the
tools and datasets those components rely on.

Beyond policy documents from major Western superpowers and related
transnational coordinating institutions, traceability also appears in
China's ``Governance Principles'' for ``Responsible
AI''~\cite{china-responsible-ai}. Here it comes in somewhat different
guise, attached to a principle that AI must be ``Secure/Safe and
controllable'', although this principle also addresses issues of
transparency and provenance (here described in terms of
``tamper-resistance'') as the above principles do. Of particular
interest is that the document calls not for traceability as a
requirement but rather says that ``AI systems should [\ldots]
gradually achieve auditability, supervisability, traceability, and
trustworthiness.'' These principles should be viewed in the context of
the earlier Beijing AI Principles, which similarly aspire to
traceability without claiming it as a requirement. Thus, non-Western
conceptions of traceability are similar in substance if perhaps
different in emphasis, calling for ``various parties related to AI
development'' to ``form AI security assessment and management
capabilities'' while describing transparency, accountability, and
traceability all in aspirational terms rather than as requirements.

Across a variety of global policy documents, then, we see that
traceability has emerged as a key requirement for the responsible use
of software systems. This property entails systems where the design
methodology, underlying data sources, and problem definitions are
clearly documented and released to stakeholders (a kind of structured
transparency of the system's structure and development). Additionally,
traceability requires connecting this transparency to outcomes and
behaviors of the system, encompassing auditability of the
system-in-operation as well as the testability of the system during
both development and operation. Further, traceability seeks to relate
disclosed information to the problem of whom to hold responsible for
these behaviors in cases both of failure and success, providing a link
between transparency and disclosure of system provenance and
determinations of
accountability~\cite{kroll2020accountability,wieringa2020account}. An
expansive requirement, traceability lies at the core of system hazard
mitigation and risk management decisions by system controllers.

\subsection{Values Served by Traceability}
To understand the requirements demanded by this conception of
traceability, we must explore the goals articulated by the documents
which espouse it. Traceability is an expansive concept, serving many
values both concrete and abstract.

Although traceability is often described as a kind of transparency, it
does not speak directly to the question of what systems exist or what
the scope of their operations and outputs is, a classical goal of
transparency requirements in policy~\cite{gellman2017fair,
  bruening2019considering}. Instead, as noted in
Section~\ref{sec:intro}, traceability ties the reasons a system works
as it does to its actual operation, supporting audit and interrogation
into the structure and function of a system, thereby serving to
operationalize accountability. However, in requiring structured
disclosure about systems, traceability does serve to enhance
transparency where it exists, providing a link between transparency's
instrumental operation and the values it serves and showing to what
end that transparency is useful. Additionally, traceability provides a
path to understanding a system's integrity in contexts where a
system's supply chain may be in doubt both for reasons of complexity
or for reasons of security. A system which is robustly traceable is
less likely to suffer manipulation by an adversary, as the provenance
of the system's components and their arrangement is made plain to
affected stakeholders. Thus, if an adversary were to substitute or
modify components of or inputs to the system for some or all
decisions, robust traceability would require that this manipulation
become visible to affected parties. And in systems where the origin of
a decision may be complex, robust traceability requires that a
mechanistic justification for that decision be producible on demand,
supporting both system governance and the contestability of system
outputs~\cite{mulligan2019shaping}. Relatedly, a traceable system must
be understandable to the humans intended to trace its operation, so
traceability encompasses aspects of explainability and is at least
partially a human factors question. Finally, traceability serves to
make plain the reasons behind failures, showing where investigators
and analysts can interrogate a system once an undesired behavior
occurs and relating design choices and operational facts to specific
outcomes.

\section{Requirements for Traceability}
\label{sec:reqs}
In this section, we expand on the notion of traceability as adopted in
policy to understand what it requires in technology. Our explication
of requirements is driven by the conceptualization of traceability and
the goals it serves in the policy documents described in
Section~\ref{sec:adoption}. These requirements should be viewed as a
lower bound: traceability requires doing these at a minimum, but may
require other activities depending on the application context.

\subsection{Design Transparency}
The primary demand of traceability is that the design choices made by
system designers be made available to system stakeholders affected by
the system's operation. This could be accomplished via transparency
such as making system source documentation, code, or data
available~\cite{citron2008open} or through more abstracted disclosures
such as impact assessments~\cite{ainow2018impact,
  selbst2017disparate}.

Many proposed approaches would provide standardized disclosures in
support of transparency or traceability~\cite{about-ml}. These include
data sheets~\cite{gebru2018datasheets}, fact
sheets~\cite{arnold2019factsheets}, data
statements~\cite{bender2018data}, data nutrition
labels~\cite{holland2018dataset}, model
cards~\cite{mitchell2019model}, and other standardized disclosure
formats. But disclosure alone does not provide traceability, and while
traceability requires disclosure, they must not be equated. Indeed, if
traceability is more akin to an operationalization of accountability
than of transparency, it may be said that while such tools improve
traceability, they do so only to the extent that they reflect and
enable a broader process of assessment, test, and evaluation.

Design proceeds from requirements. Because traceability asks that
stakeholders be able to understand why design decisions were taken,
traceability also requires that requirements be disclosed as part of
transparency about design. Requirements reflect the way a particular
system's goal was articulated and approached by designers, the
critical aspect of \emph{problem formulation} not otherwise subject to
investigation by persons affected by a computing system and often
invisible in even required disclosures of artifacts like code, data,
or documentation~\cite{passi2019problem}. When requirements are not
specified formally, they should at least be described in an
unambiguous natural language form as relied on during
development. Thus, systems that result from an exploratory process
(e.g., many models derived from standard data science practices) must
be augmented with descriptions of the governance attendant to that
exploration, which controls why particular avenues were or were not
explored and how the fruits of that exploration were considered to be
worthy of application in a particular context.

Transparency of design not only provides a window into how systems
work, it provides visibility into where design choices were taken that
have significant impact. For example, software is used to automate the
work of lab technicians who analyze forensic evidence, but it is
unclear how determinations made by that software are made, a problem
which has spawned much litigation~\cite{kwong2017algorithm}. One tool,
New York City's Forensic Statistical Tool, has essentially fallen out
of use after journalists raised questions around its accuracy, leading
to a court-ordered release of its source code and subsequent public
scrutiny~\cite{kirchner2017fst}. Many commercial offerings remain
trade secrets, despite being used in criminal proceedings regularly. As
a hypothetical example, imagine software which measures a parameter of
a forensic sample and performs differing analysis based on whether the
measured value was above or below a given threshold. The very
existence of this threshold and the process by which it was determined
may be unknown without sufficient traceability. And without that
understanding, the suitability of the analysis, the thresholding that
triggered it, nor the value of that threshold can be challenged or
reviewed. Thus, sufficient traceability means raising decisions about
how parameters internal to the design of a system are set so they are
legible to those outside the design process.

\subsubsection{Testing}
In some descriptions, traceability is established by the ability to
test or audit for particular outcomes. Testing of software systems is
an entire discipline with a long
history~\cite{gelperin1988growth,myers2011art}, but it can help
support external interrogations of a system's behavior under specific
conditions~\cite{desai2018trust, kroll2018fallacy}. Traceability
requires developers and those involved in testing and system
evaluation to minimize the gap between what is known to system
developers through their knowledge of design decisions and
actualization of those decisions through a test and evaluation regime
and what is known to outside stakeholders who do not see the
development or the output of developmental testing. Because the
halting problem prevents testing external to development from being
formally sound~\cite{rice1953classes}, minimizing this gap necessarily
requires disclosing information about the design as well as
information about the system's performance under test.

Thus, traceable systems must have and release information about robust
test and evaluation plans. Further, such systems must be designed to
be testable during development and operation, and ideally to be
testable by outsiders as well as developers. This is driven by the
close relationship between traceability and auditability.

\subsection{Reproducibility}
Related to the requirement to minimize the gap between the view of
developers and other insiders and stakeholders outside the process of
creating or operating the system is the issue of reproducing a
system's behavior so it can be evaluated for correctness under the
stated set of requirements. If a system's behavior cannot be
reproduced by a developer, it cannot be made plain to an outside
stakeholder. A practical issue is that even disclosures of source code
and data (or the use of fully open-source code and data) cannot be
related to compiled software or trained models unless that compilation
or training can be precisely reproduced, which is possible using an
emerging set of tools and software
practices~\cite{reproducible-builds}. More careful reasoning about
compilation can enable efficient updating~\cite{nikitin2017chainiac}
or make verification of software function straightforward from
transparency~\cite{zesty}.

More broadly, it should be possible to reproduce even abstract
conclusions from data or any reported experimental results that claim
scientific authority~\cite{stodden2014best,
  stodden2014implementing}. Without this, an external stakeholder
aiming to verify why and how a system was designed a particular way or
what a system did will be unable to.

A related risk is that loss or modification of development information
(code, data, built components) will lead to a situation where the
system-as-deployed no longer relates to supporting information that
might be disclosed, possibly subsequently. Thus, robust
reproducibility of both artifacts and conclusions must be a
requirement for any traceable system.

\subsection{Operational Recordkeeping}
Beyond traceability at the design stage, traceability is understood to
apply to system operations. Thus, traceable systems must keep records
of their behaviors. But what records must they keep, how and how long
must those records be maintained or retained, and what must those
records show? These questions are highly contextual, and mapping the
associated concrete requirements demands the engagement of
subject-matter experts able to understand both the application area
and the technology intervening on it. Abstractly the requirement must
be that records support sufficient inquiry by external stakeholders or
oversight entities~\cite{kroll2017penn, wieringa2020account,
  kroll2020accountability}.

A further problem is relating the contents of records kept during
operation to a system's observed behavior. For systems which are fully
reproducible, this is possible in principle through transparency or
through the intervention of trusted oversight entities which can
receive disclosures under seal (e.g., law enforcement agencies,
courts, or adversarial parties in litigation). However, it is possible
to use tools from cryptography to bind the contents of records to the
computations performed on those records to make this relation both
more efficient to establish and possible to establish more broadly
even when portions of the retained information must remain secret or
private~\cite{kroll2015accountable}. For example, cryptocurrencies
such as ZeroCash contain protocols to convince the receiver of a
payment that the digital currency the payment represents has not
previously been spent, without revealing the payment's origin or full
history, simulating the privacy properties of physical
cash~\cite{sasson2014zerocash}.

An alternative model is to vest recordkeeping and data access in a
trusted third party such as an oversight entity or an independent data
trust~\cite{young2019beyond}. Such an entity can make assertions about
records and determine the equities of their release on a case-specific
basis.

Finally, the maintenance of records of routine system activity along
with records of failures and near-miss incidents provides a foundation
upon which an understanding of the system's patterns of function can
be built, enabling the development of interventions that improve
outcomes, safety, and overall system function and
reliability~\cite{cook1998complex, barach2000reporting,
  leveson2016engineering}.

\subsection{Human Understanding}
As noted in Section~\ref{sec:adoption}, the principle of traceability
is often justified by the need for humans to understand the decision
rules embedded in the system. Principles call for ``appropriate
understanding of the technology'' and transparency which ``enables
identification of the reasons an AI-decision was
erroneous''. Achieving these abstract principles requires not only
transparency about system function, but careful system engineering to
ensure that systems do not confuse their human
operators~\cite{leveson1997analyzing}, obscure paths of accountability
for accidents where affordances in the design of technology lead human
operators to make judgments that lead to
accidents~\cite{elish2015praise, elish2019moral}, or confuse the loci
of agency and power within a system~\cite{bainbridge1983ironies}.

This requires thinking beyond the current focus on explainability and
causality in AI~\cite{halpern2005causes-ii, doshi2017accountability,
  doshi2017towards, guidotti2018survey, molnar2018interpretable} and a
push toward tying explanations and disclosures to shifts in
power~\cite{miller2017explanation, selbst2017meaningful,
  edwards2017slave, edwards2017enslaving, selbst2018intuitive,
  selbst2019fairness, kalluri2020power}. Such thinking has been a
robust component of research in human
factors~\cite{salvendy2012handbook, sendak2020human}, and developing
appropriate strategies for establishing when and why and how much
humans understand technology remains an important research domain
necessary to enable robust human-machine
collaboration~\cite{lin2018chi, rader2018chi}. Traceability thus
requires that systems be transparent not just about their function but
about whether that function is appropriately communicated to operators
and other affected humans. This is also important in the context of failure analysis, as many accidents result from inappropriate modeling of machines by human operators, of humans by machines and their designers, or at the point of handoff between human and machine~\cite{bainbridge1983ironies,mulligan2020concept}

\subsection{Auditability}
Another component of the traceability principles as stated is that
they support the auditability of systems both before they are fielded
and during operation. This has several meanings.

First, systems must maintain sufficient records during development and
operation that their creation can be reliably established and
reproduced. This requirement is largely encapsulated in the
reproducibility and operational/developmental recordkeeping
requirements listed above.

Beyond requiring that evidence of how a system operated be
established, auditability requires that this evidence be amenable to
review and critique of the system's operation, as well as comparison
of the fidelity of that evidence to reality. Such assessments can be
qualitative or quantitative in nature, and could happen during
development or once a system is fielded. In the accounting literature,
an \emph{audit} compares recorded evidence (``accounting'' for a
system's behavior) to reality to determine whether that evidence is
reliable; alternatively, an \emph{assessment} is the ascription of
some value to that evidence or a judgment about the meaning of that
evidence~\cite{espeland2007accountability}. Scholarly critiques have
noted that understanding the ethical and social implications of
computing systems as simple as sorting~\cite{sandvig2015seeing} and as
complex as news curation~\cite{sandvig2014auditing} require
assessment, but describe this requirement as one of auditing.

Although the term ``audit'' is widely used in policy and principles
documents to describe what is needed (and performed) for computing
systems~\cite{chin1982auditing, blocki2013audit, hannak2013measuring,
  hannak2014measuring, sandvig2014auditing, bashir2016recommended,
  chen2017observing, kearns2017gerrymandering, kim2017auditing,
  elazari2018algorithmbugbounties, reyes2018won,
  chen2018investigating, raji2020closing}, the task referenced by the
traceability principles is more in line with
assessment~\cite{jagadeesan2009towards}. Likely, this is due to the
history in computer security of the use of ``audit methods'' to assess
the security state of a system~\cite{lunt1988automated, habra1992asax,
  bishop1995standard, colbert1996comparison}, to control disclosure of
sensitive information in databases~\cite{helman1993statistical,
  kenthapadi2005simulatable, nabar2006towards, dwork2006calibrating,
  nabar2008survey}, and to establish security policy compliance of the
behavior of system infrastructure~\cite{haeberlen2007peerreview,
  haeberlen2010accountable, haeberlen2010case}. Practical applications
often proceed by applying standard assessment models that use well
defined checklists and controls to achieve desired
outcomes~\cite{nist-cyber-framework-2018,
  nist-privacy-framework-2020}. Other work looks to connect ``audit''
data to sociotechnical goals like the correct operation of an election
to establish an assessment of the entire sociotechnical
system~\cite{waters2004building, adida2008helios,
  hall2010election}. However, some practitioners have criticized the
gap between audit (in a compliance sense) and the assessment of
security derived from it~\cite{bellovin2006brittleness,
  clark2018compliance}. Still others have begun to seek data on the
validity of audit approaches for building assessable metrics of
abstract properties such as security~\cite{de2020scram}.

Scholars from the social sciences have been critical of this
over-quantification of auditing, and the collapse of auditing and
assessment into purely quantitative
methods~\cite{carruthers1991accounting, porter1992quantification,
  espeland2007accountability}. It is notable that widely used
governmental auditing standards allow for both quantitative methods
and qualitative assessment~\cite{gagas-2018}. The collapse of the word
``audit'' onto system assessments in this context is likely due to the
naming of a particular discrimination detection methodology, the
``audit study'', in which similarly qualified test subjects differing
only in (perceived) race or gender or other protected characteristic
under study are subjected to the same process to test for facial
evidence of discrimination in the process~\cite{siegelman1993urban},
an approach which has been and must be updated to include studies of
the behavior of automated systems~\cite{ajunwa2020auditing,
  cfpb-race-imputation}.

Impact assessments are also proffered as a requirement for the
governance of computing systems, and such assessments could guide
requirements for auditability of the system during development or
after it is fielded~\cite{selbst2017disparate, ainow2018impact}. As
with other transparency tools, impact assessments are valuable insofar
as they enable traceability, and considering the requirements of
traceability can help establish the scope of appropriate impact
assessment.

Finally, the extent to which a system can be audited effectively is a
question both of design and governance. The system must have
interfaces that enable audit and also policies that allow effective
uses of those interfaces. In bureaucracies, this is driven by
freedom-of-information laws in addition to standards that define the
sorts of engagements with auditors the system expects. To assess
whether something like an online advertising system admits a
sufficient level of auditing, we need the system to report out
information auditors want, for the system's controller to allow (or to
be compelled) that information to be given to auditors, and for there
to be standards around how that auditing will take place. This is to
say nothing of the problem of finding a sufficient number of skilled
auditors who can apply such standards, a problem which exists even in
domains where standards are clear.

\section{Supporting Traceability}
\label{sec:tech}
In order to meet the requirements laid out in Section~\ref{sec:reqs}
and to support traceability in built systems, we need technical,
organizational, and policy-level tools. Many such tools exist, but
many more do not. Rarely are these tools brought together in an
assemblage that resembles anything like the traceability principles
espoused by many organizations and summarized in
Section~\ref{sec:adoption} or the requirements unpacked from these
principles in Section~\ref{sec:reqs}. And yet, building robustly
accountable computing systems requires embodying this principle in its
full power.

In this section, we summarize known tools and relate their
capabilities and limitations to the requirements laid out in
Section~\ref{sec:reqs}. Our focus is primarily on technical tools
here, though as noted many nontechnical tools are also necessary. It
is likely this cross-functional nature of operationalizing
traceability (and, indeed, any ethical principle) in technology that
makes doing so such a challenge.

\subsection{Development Methodology}
The history of system development, especially software system
development, is littered with
failures~\cite{ewusi2003software}---failures of the system to meet its
stated requirements, failures of the system to function once
delivered, or failures of the system to be developed to the point of
operability at all. These risks have long been recognized---Brooks
famously pointed out in 1975, based on experience developing IBM's
System 360 mainframe operating system, that ``adding manpower to a late
software project makes it later.''~\cite{brooks1995mythical}.

Early development methodologies flowed linearly from requirements to
specifications to deliverables---a ``waterfall'' process, where the
system's components flow along a defined path to
delivery~\cite{petersen2009waterfall}. But all too often, diversions
along this path or changes in the environment by the time a system is
delivered mean that the system does not meet the stated requirements
upon delivery~\cite{brooks1995mythical}. This has led to the development of
various ``iterative'' modalities of software delivery, such as the
principle of ``iterative enhancement''~\cite{basil1975iterative}, the
Agile movement~\cite{beck2001manifesto}, or test-driven
development~\cite{beck2003test}. In all cases, the goal is to feed
back outputs of the development process into future rounds of
development for continuous, incremental improvement and
learning. The result is meant to be a product
which is closer in its delivered form to envisaged requirements.

However, although iterative development often leads to better outcomes
faster, issues can arise from the way the problem to be solved is
defined initially~\cite{argyris1977double} and these methods may be
based on unfounded assumptions in many
cases~\cite{turk2005assumptions}. These methods have also been
critiqued for attending too much to problems and bugs identified early
in a project's history (as the visibility and tracking of issues
demands their rectification and continuous improvement) as well as for
creating a model unsuitable for high-assurance tasks (as the project
is under continuous revision, there is not an obvious point at which
to verify or validate its functionality and these assessments can
become problematic in a continuous-assessment
setting)~\cite{boehm2002get, mcbreen2002questioning}.

Yet there has been some progress towards traceability in software
development. Discipline around version control of digital artifacts
like code, data sets, and model products has been enhanced by the
adoption of versioning-focused
methodologies~\cite{bass2015devops,carter2016data} and the
introduction of powerful, general digital object versioning
systems~\cite{loeliger2012version}. Although the problem of keeping
digital archives of any sort is a major
challenge~\cite{hedstrom1997digital, rothenberg1999avoiding},
standards exist for creating trustworthy repositories and auditing
their reliability~\cite{iso16363}, especially for high-assurance
applications like space data systems~\cite{ccsds2011audit}. Coupled
with careful testing of desired invariants and augmented by disclosure
of decision points in the development methodology, these tools can be
extended to methods for trustworthy system
development~\cite{brundage2020toward}.

\subsection{Reproducibility and Provenance}
Many tools exist to support the reproducibility of created software
artifacts and models of data sets~\cite{reproducible-builds,
  boettiger2015introduction}. This is critical to scientific
research~\cite{stodden2014best, stodden2014implementing, geiger2017},
and to the practice of software development and data science in
industry both for reasons of maintaining clarity of system
function~\cite{bass2015devops} and security~\cite{sbom}. Despite the
importance of this function, software projects and especially data
science and machine learning products are rarely able to precisely
reproduce their work to a sufficient level to provide reliability, let
alone traceability~\cite{warden2018reproducibility}.

Another relevant genre of tooling and research effort concerns the
maintenance of system-level metadata which can be used to establish
data and artifact \emph{provenance}~\cite{muniswamy2006provenance,
  moreau2008provenance, buneman2001provenance,
  muniswamy2010provenance, herschel2017survey,
  perez2018systematic}. Work in this area out of the operating systems
and database research communities has led to efficient systems that
can be transparently layered into existing scientific
workflows~\cite{mcphillips2015yesworkflow, ludascher2016brief} or
standard infrastructure components~\cite{muniswamy2010provenance}.

These tools support the requirements of reproducibility and can be
used along with development methodology and discipline to support
requirements about documenting and substantiating the origin of data
and components. Unlike other aspects of traceability, this is an area
with well developed, field-tested tooling, and solutions which can be
brought immediately into practice for clear gains, both for
traceability and for broader benefits (e.g., the management of these
artifacts; capabilities to share results within communities of
practice or research).

\subsection{Design and Structure}
The stated principles of traceability tie system records to system
understanding and ultimately to the way that systems embody
values. There is a long history of research into the way design
reflects values, and the analytical framework of this line of research
gives a foundational framework for making both of these
links~\cite{moor1985computer, star1994steps, friedman1996value,
  star1996steps, nissenbaum2001, flanagan2005values,
  nissenbaum2005values, friedman2008, ledantec2009,
  gurses2011engineering, knobel2011, irani2014critical,
  jackson2014policy, shilton2014, jafarinaimi2015,
  steinhardt2016breaking, wagenknecht2016, ziewitz2017not,
  mulligan2018saving, shilton2018, zhu2018,
  young2019toward}. Specifically, this line of work notes that systems
must be designed at a fundamental level to embody values and that it
is not enough to add abstraction layers or minor changes at the end of
a design process divorced from values. Tools such as analytic
frameworks for operationalizing values~\cite{burrell2016machine,
  mulligan2016privacy, mulligan2019thing}, techniques for explicitly
capturing certain concrete operationalizations of contested values in
systems~\cite{albarghouthi2016fairness, bonchi2016bias,
  chouldechova2017fair, albarghouthi2019fairness, beutel2019putting,
  wong2019bringing}, and mechanisms for observing the ultimate
value-sensitivity of the resulting system~\cite{lum2016predict,
  lipton2017does, buolamwini2018gender, raghavan2020mitigating} all
arise from this rich and deep vein of work and provide approaches to
capturing the value of traceability in real systems. We hope the
requirements of Section~\ref{sec:reqs} can serve as a similar analytic
framework for the principle of traceability, and that the techniques
summarized in this section can be useful in making such requirements
actionable in real systems.

One well-studied value that can be operationalized through design is
privacy. Privacy-by-design is a principle recognized both by
technologists and lawyers, academics and
practitioners~\cite{cavoukian2009privacy,cavoukian2011privacy,
  cccpbdworkshop1, cccpbdworkshop2, cccpbdworkshop3}. The exact
operationalization of privacy can be a contested, or even an
essentially contested point~\cite{mulligan2016privacy}. Many systems
purport to protect privacy through such design approaches, generally
operationalizing privacy as the restriction of some information from
some parties in the system~\cite{camenisch2006balancing,
  barth2007privacy, balasch2010pretp}. Some scholars have argued that
privacy-by-design leads toward a narrow view of what privacy is and
away from issues of power and control of
information~\cite{dwork2013not}. Legal obligations for privacy-by-design may
hypothetically limit the number and severity of privacy incidents in
practical systems~\cite{rubinstein2013privacy}. Empirical studies have demonstrated
that attitudes toward privacy and compliance with data protection law
vary drastically by sector and by country~\cite{bamberger2010privacy,
  bamberger2015privacy}.

Of course, design means little if the ultimate implementation does not
comport with the design. Translating values-sensitive design to
values-embodying implementation remains a key open challenge, as does
assuring a relationship between specification and
implementation. Closing this specification-implementation gap is a
core competency of the traditional computer science field of software
\emph{verification}~\cite{boehm1984verifying, appel2011verified}. A
related problem is that of \emph{validation}, ensuring that a given
specification captures the intended value. The discipline of
requirements engineering has developed formal and semi-formal
approaches to the latter problem~\cite{breaux2006towards,
  gordon2013assessing}.

Capturing principles in the design of systems that include software
and computers is not an idea that originated with values. Indeed, some
of the earliest large computing systems derived their designs from core
principles, such as the Internet's ``end-to-end''
principle~\cite{gillespie2006engineering}. This focus on design
principles even under an imagined ``value free'' ideal remains to this
day in the design of Internet platforms such as ISPs, social media and
forum websites, and hosting providers~\cite{gillespie2010politics}.

Traceability demands that the design of systems be visible to affected
stakeholders. Another approach to legible design, beyond disclosure of
design desiderata, is the use of transparent and inclusive design
processes, a topic which has received much attention in the technical
literature recently, but which has a longer history in political
science~\cite{rosner2014designing, wolf2018, young2019beyond,
  young2019toward, lee2019webuildai, lee2019procedural,
  lee2020human}. Such community-driven design can lead to better
overall decision-making affected stakeholders also find more
acceptable, even given
tradeoffs~\cite{chouldechova2018case}. Ethnographic work has found
substantially similar attitudes toward decision-making which is fully
embodied in a system across disparate
domains~\cite{christin2017algorithms}. Survey-driven human factors
work has similarly discovered that participatory design can lead to
more acceptance of system-driven
outcomes~\cite{binns2018reducing}. Design toolkits which combine
participation in design with explicit consideration of equity and also
drive toward auditability provide concrete tools for advancing the
state of play in values-oriented system development, realizing the
goal of values-driven design in practical system implementation and
fielding~\cite{katell2019algorithmic}. Research questions remain,
however, around the efficacy of such tools in actually preventing
harms such as discrimination or the amassing of technologically
centralized and entrenched power. One approach under active scrutiny
is the idea of designing systems for \emph{contestability}, both the
ability to appeal incorrect decisions and the ability to understand
the mechanism of decisions well enough to determine when decision
guidance should be disregarded or
overridden~\cite{mulligan2019shaping}.

Finally, the design of systems must take into account the role of
humans within those systems. Systems include more than just technical
components and rules-driven automation~\cite{desai2018trust}, but also
decisions by human decision-makers who can exercise
discretion. Assuring that discretion within systems is exercised
transparently and at appropriate times requires careful attention to
design, including requirements on when humans create records or review
existing records for accuracy or policy
compliance~\cite{ellison2007ceremony}. Humans may at times prefer
decisions which are driven by discretion or negotiation rather than
purely algorithmic decisions~\cite{lee2017algorithmic}. In general,
automation increases the capabilities of humans by taking over certain
tasks and freeing human efforts for strategic optimization and
additional productive work. But in doing so, automation also robs
involved humans of situational awareness and expertise in the detailed
operation of the system. Thus, humans coupled to automation are both
more critical to the outcomes driven by the system and less able to
control those outcomes~\cite{bainbridge1983ironies}.

\subsection{Structured Logs}
As traceability requires recordkeeping both at the development stage
and the operational stage, we must consider the classic tradeoff in
computer science between direct policy enforcement via the design of
and compliance with specifications and the detection of policy
violations via recordkeeping and
review~\cite{breaux2006enforceability, jagadeesan2009towards,
  pearson2009accountability, kuesters2010accountability,
  datta2014privacy}. Several authors have proposed tools for
\emph{verified computation}, which provides a proof of its execution
that can be reviewed later to ensure particular computations took
place as claimed~\cite{ben-sasson, ben2013snarks, braun2013verifying,
  vu2013hybrid, ben2014succinct, ben2015secure}. Others have also
proposed extending such technology to create structured logs in a
commit-and-prove style to provide explicit technological demonstration
to a skeptical stakeholder or oversight entity of the
values-appropriateness of a computation or its procedural fairness and
regularity~\cite{kroll2015accountable, kroll2017penn}. Structured
logging with disclosable side-conditions is a well-studied problem in
computing, with associated properties ranging from simple integrity
checking~\cite{merkle-tree, crosby2009efficient} to full real-world
systems for proving the nonexistence of security-critical digital
objects such as the certificates used for Internet
security~\cite{laurie2013certificate}. A related line of work is the
vast field of tracing in operating systems and distributed
systems~\cite{mace2015pivot}.

We observed above that traceability is often conceptualized as a
transparency principle, but it is more akin to a principle for
enabling accountability. Many authors have called explicitly for the
use of structured recordkeeping tools to enhance accountability and
have built the conceptual framework tying such recordkeeping to
improved accountability~\cite{nissenbaum1996accountability,
  nature-editorial, wieringa2020account}.

Empirical studies of trust in data science projects have shown that
the sort of recordkeeping called for under traceability requirements
may enhance the credibility of data science work products both within
and beyond the teams creating them~\cite{passi2018}. Along with
traditional software engineering and project management discipline
(lifecycles, versioning tools), tools for reproducibility and
recordkeeping comprise a set of currently available techniques with
the capacity to improve traceability in real-world applications but
which are currently underused. Wider adoption of these techniques
could improve not only traceability, but thereby the broader
governance of computing systems generally.

\subsection{Structured Transparency Disclosures}
The sort of developmental and operational recordkeeping imagined in
the prior section is often shorthanded into demands for better system
documentation, often in a structured format such as a
checklist~\cite{gawande2009checklist, gebru2018datasheets,
  mitchell2019model, arnold2019factsheets, bender2018data,
  holland2018dataset, about-ml}. Although documentation of computing
systems does support traceability, there is little in the way of
research establishing its effectiveness at communicating within
organizations or actually mitigating harms. Further, there is often a
substantial practical gap between the state of the
system-as-documented and the state of the system-as-realized. It is
important to remember that documentation is generated by a healthy
governance process, but as a side effect. Documentation does not, on
its own, engender complete or useful governance of systems.

This gap is highlighted in the operationalization of data protection
law. To maintain ``adequacy'' and enable legal transfers of data
between the European Union, where the General Data Protection
Regulation applies to all data processing, and the United States,
which has a considerably more flexible legal regime for data
protection based on sector-specific regulation, the Privacy Shield framework
was set up to replace a prior Safe Harbor
agreement~\cite{bruening2019considering}. However, both structures
have now been found to be insufficient to protect fundamental rights
guaranteed under EU law. Despite this, analysis done when these
frameworks were in force examines transparency and governance rights
in the US sector-specific approach (e.g., in laws like the Fair Credit
Reporting Act and the Equal Credit Opportunity Act) and finds them as
powerful if not more powerful than the analogous rights within the
GDPR~\cite{bodea2018euautomated, ec2018privacyshield}.

\section{Gaps and Needs}
Although many technologies exist which support traceability and many
organizations aspire to embody this principle in practice or even
claim to achieve it at a sufficient level, substantial gaps exist at
the basic technical level. This is to say nothing of gaps in
organizational governance or in the remainder of the sociotechnical
control structure---no technical control can be more capable than the
implementing organization's willingness to take action based on
it~\cite{vaughan1996challenger, leveson2016engineering}.

\subsubsection{Adoption}
As noted in a few places in Section~\ref{sec:tech}, there exist tools
and technologies that could be deployed immediately in service of
better traceability, and yet they are not used. Understanding why this
is will be critical to advancing this principle. And because this
principle is straightforwardly recognizable and potentially amenable
to assessment and measurement more directly than abstract goals such
as fairness or equitability, understanding the barriers to
operationalizing this principle bears on the feasibility and success
of the entire values-in-design project.

Of specific note is the lack of methodologies and standards around the
use of tools for the reproducibility of data science products or
machine learning models. This is despite substantial progress having
been made at integrating such tools into common modeling
frameworks---even with the capability to checkpoint models or reset
the seeds of pseudorandom generators, this is rarely done in practice
and the best guidance exists in the form of folkloric knowledge
scattered across blogs, answers on question answering sites, and forum
posts. This despite the fact that the frameworks could be designed to
afford reproducibility as a default, or at least an easily and
straightforwardly achievable property of implementations. Yet tools
like data science notebooks lean towards interactive environments in
service of the ``democratization'' of powerful tools, but in ways that
make it hard to establish what, precisely, resultant insights actually
mean~\cite{jacobs2018measurement}. And even rectifying this dismal
situation would not address the problem of framework dependencies and
the potential for numerical instability, bugs in core algorithms, or
differences in behavior sufficient to change the output of models in
practice~\cite{selsam2017developing}.

\subsubsection{Human Factors and Understanding}
Traceability claims explicitly to act in service of better
understanding of the decision-making processes at work in an automated
system. Yet while there is a massive literature on machine learning
explainability~\cite{halpern2005causes-i, doshi2017accountability,
  doshi2017towards, edwards2017slave, guidotti2018survey,
  molnar2018interpretable}, there is comparatively little work on the
human factors of how such explanation methods serve
understanding~\cite{lin2018chi, rader2018chi} nor much work that
explicitly relates the value of generated explanations to a theory or
philosophy of what explanations should
achieve~\cite{miller2017explanation, selbst2017meaningful,
  edwards2017slave, edwards2017enslaving, selbst2018intuitive,
  selbst2019fairness, kalluri2020power}. For example, in one
framework, Miller argues that good explanations must be causal,
contrastive, selective, and social and that they act as both a product
conveying information and a process of inquiry between the explainer
and the explainee~\cite{miller2017explanation}. Yet few techniques
from the machine learning literature demonstrably have more than one
of these properties, and precious little work examines such systems
\emph{in situ} to determine whether they improve
outcomes~\cite{lou2012intelligible, sendak2020human}.

\subsubsection{Test and Evaluation, Standards}
In a similar vein, although much work considers specific robustness
and generalization properties, or responses to bias or adversarially
generated input, work on test and evaluation methodologies for
data-driven automation is nascent, and work on test and evaluation of
software lags the capabilities of test and evaluation for physical
systems despite decades of research. Investigating how to assess
algorithmic systems generally leads to more questions than answers.

At present, there is no generally accepted standard for establishing
what evidence an automated decision-making system should produce so
that its outputs can be considered traceable. This is in contrast to
other complex systems, such as corporate finance and risk
management~\cite{sr11-7}, nuclear safety~\cite{rees2009hostages}, or
aviation safety~\cite{huang2009aviation}. And while standards are only
part of a governance approach~\cite{leveson2016engineering}, they can
help build the organizational structures necessary to assess and
investigate unintended behaviors before they happen. Updates to
accounting or audit standards, in both the public and private sectors,
would make the assessment of traceability substantially more
straightforward. Further, investigating the effectiveness of such
standards in furthering the goals of traceability (as measured by,
say, perspectives on the performance of a
system~\cite{veale2017logics, binns2018reducing}) would provide useful
benchmarks for those charged with the test and evaluation of practical
systems. Understanding the operationalization of this principle,
seemingly amenable to testing and assessment more than other,
contested principles dealing with fairness, bias, and equity, would
demonstrate a path to operationalizing ethical principles in the
development of practical systems, and is thus of paramount importance.

\subsection{Oversight, Governance, and Responsibility}
Closing the gaps between the lofty goals of principles documents and
the behavior of algorithmic systems in practice is the key research
challenge of the values-oriented design movement and research
community~\cite{metcalf2019owning, mittelstadt2019ai,
  mittelstadt2019principles, fjeld2020principled}. It is in particular
imperative that the size and nature of any such gap be legible both to
affected stakeholders and to competent oversight authorities in order
to support effective governance and
accountability~\cite{mulgan2000accountability, mulgan2003holding,
  kroll2017penn, kroll2018fallacy}. Operationalizing
traceability provides an excellent opportunity for progress in this
critical research domain. Our investigation opens related research
questions, such as why available tools are ignored even when their use
would drive clear benefit both for the function of development teams,
the function of governance structures, and the embodiment of human
values.

\section{Conclusion}
Traceability is a widely demanded principle, appearing in a variety of
AI policy and AI ethics principles documents adopted around the world,
notably in several important pieces of government-issued guidance, in
coordinating documents promulgated by transnational organizations such
as the United Nations and the OECD, and in the guidance of the
European Union's High Level Expert Group on building Trustworthy and
Responsible AI. Although it is often conceptualized in these documents
as a principle which operationalizes transparency as a way to achieve
governance, in reality that governance is achieved by enhancements to
accountability and enhanced capabilities of both affected parties and
competent authorities to notice when systems are going wrong and
rectify the issue. Traceability serves to demonstrate when and why
transparency is valuable, connecting the desire for disclosures about
how a system functions to consumption of that disclosure for a defined
purpose.

Traceability is an excellent principle for driving system assessment,
as it is seemingly more concretely realizable and recognizable
compared to other goals like equitability or nondiscrimination. Like
transparency, it is an instrumental principle that serves ends beyond
itself (including other ethical principles such as fairness,
accountability, or governability). While it is possible to see such
instruments as a goal unto themselves, their primary role in the
operationalization of policies for ethical computing systems is to
enable more abstract assessments.

That said, the concreteness of traceability does not equate to ease of
realization in practical systems. We conclude via our analysis of
operationalizing this principle that substantial gaps exist between
the requirements we have identified and the tools presently available
to meet those requirements. A major gap is simply adoption: tools
which could improve traceability remain unused despite decades of
development and the existence of mature realizations. Understanding
why these tools go un-adopted remains an important open research
question for the governance of computing systems. Other gaps are more
serious and require new research to close: establishing an appropriate
human factors understanding for computing systems that work in tandem
with humans challenges traceability as much as it challenges any other
question of ethical computing; the lack of accepted standards or even
widely used best practices for the assessment of computing system
ethics remains a barrier; and a lack of practical governance,
oversight, and review limits how well robust traceability can support
meaningful assignments of responsibility for the behaviors of
computing systems or assessments of those systems' fidelity to
normative governance goals.

Traceability serves as a foundation for other goals in aligning the
behavior of automated systems to human values and in governing those
systems to make that alignment plain to anyone potentially affected or
harmed by the operation of those systems or the social outcomes they
drive. Only by making systems traceable can we hold them accountable
and ensure they comport with applicable, contextually appropriate
social, political, and legal norms.

\begin{acks}
The author wishes to thank the anonymous reviewers for their helpful
comments and insights, and is grateful for feedback from and
discussions with colleagues including CDR Ed Jatho, USN; Abigail
Jacobs; and Andrew Smart.

This work was sponsored by a grant from the Naval Postgraduate
School's Research Initiation Program for new faculty. Views expressed
are those of the author and not of the Naval Postgraduate School, the
Department of the Navy, the Department of Defense, or the United
States Government.
\end{acks}

\bibliographystyle{ACM-Reference-Format}
\bibliography{nps}


\begin{thebibliography}{250}


\ifx \showCODEN    \undefined \def \showCODEN     #1{\unskip}     \fi
\ifx \showDOI      \undefined \def \showDOI       #1{#1}\fi
\ifx \showISBNx    \undefined \def \showISBNx     #1{\unskip}     \fi
\ifx \showISBNxiii \undefined \def \showISBNxiii  #1{\unskip}     \fi
\ifx \showISSN     \undefined \def \showISSN      #1{\unskip}     \fi
\ifx \showLCCN     \undefined \def \showLCCN      #1{\unskip}     \fi
\ifx \shownote     \undefined \def \shownote      #1{#1}          \fi
\ifx \showarticletitle \undefined \def \showarticletitle #1{#1}   \fi
\ifx \showURL      \undefined \def \showURL       {\relax}        \fi
\providecommand\bibfield[2]{#2}
\providecommand\bibinfo[2]{#2}
\providecommand\natexlab[1]{#1}
\providecommand\showeprint[2][]{arXiv:#2}

\bibitem[\protect\citeauthoryear{Abdul, Vermeulen, Wang, Lim, and
  Kankanhalli}{Abdul et~al\mbox{.}}{2018}]%
        {lin2018chi}
\bibfield{author}{\bibinfo{person}{Ashraf Abdul}, \bibinfo{person}{Jo
  Vermeulen}, \bibinfo{person}{Danding Wang}, \bibinfo{person}{Brian~Y. Lim},
  {and} \bibinfo{person}{Mohan Kankanhalli}.} \bibinfo{year}{2018}\natexlab{}.
\newblock \showarticletitle{Trends and Trajectories for Explainable,
  Accountable and Intelligible Systems}.
\newblock \bibinfo{journal}{\emph{Proceedings of the International Conference
  on Human Factors in Computer Systems (CHI)}} (\bibinfo{year}{2018}).
\newblock


\bibitem[\protect\citeauthoryear{Adida}{Adida}{2008}]%
        {adida2008helios}
\bibfield{author}{\bibinfo{person}{Ben Adida}.}
  \bibinfo{year}{2008}\natexlab{}.
\newblock \showarticletitle{Helios: Web-based Open-Audit Voting.}. In
  \bibinfo{booktitle}{\emph{USENIX Security Symposium}},
  Vol.~\bibinfo{volume}{17}. \bibinfo{pages}{335--348}.
\newblock


\bibitem[\protect\citeauthoryear{Ajunwa}{Ajunwa}{2021}]%
        {ajunwa2020auditing}
\bibfield{author}{\bibinfo{person}{Ifeoma Ajunwa}.}
  \bibinfo{year}{2021}\natexlab{}.
\newblock \showarticletitle{The Auditing Imperative for Automated Hiring}.
\newblock \bibinfo{journal}{\emph{Harvard Journal of Law \& Technology}}
  \bibinfo{volume}{34} (\bibinfo{year}{2021}).
\newblock


\bibitem[\protect\citeauthoryear{Albarghouthi, D'Antoni, Drews, and
  Nori}{Albarghouthi et~al\mbox{.}}{2016}]%
        {albarghouthi2016fairness}
\bibfield{author}{\bibinfo{person}{Aws Albarghouthi}, \bibinfo{person}{Loris
  D'Antoni}, \bibinfo{person}{Samuel Drews}, {and} \bibinfo{person}{Aditya
  Nori}.} \bibinfo{year}{2016}\natexlab{}.
\newblock \showarticletitle{Fairness as a program property}.
\newblock \bibinfo{journal}{\emph{arXiv preprint arXiv:1610.06067}}
  (\bibinfo{year}{2016}).
\newblock


\bibitem[\protect\citeauthoryear{Albarghouthi and Vinitsky}{Albarghouthi and
  Vinitsky}{2019}]%
        {albarghouthi2019fairness}
\bibfield{author}{\bibinfo{person}{Aws Albarghouthi} {and}
  \bibinfo{person}{Samuel Vinitsky}.} \bibinfo{year}{2019}\natexlab{}.
\newblock \showarticletitle{Fairness-aware programming}. In
  \bibinfo{booktitle}{\emph{Conference on Fairness, Accountability, and
  Transparency}}. ACM, \bibinfo{pages}{211--219}.
\newblock


\bibitem[\protect\citeauthoryear{{Amit Elazari Bar On}}{{Amit Elazari Bar
  On}}{2018}]%
        {elazari2018algorithmbugbounties}
\bibfield{author}{\bibinfo{person}{{Amit Elazari Bar On}}.}
  \bibinfo{year}{2018}\natexlab{}.
\newblock \showarticletitle{We Need Bug Bounties for Bad Algorithms}.
\newblock
  \bibinfo{howpublished}{\url{https://www.vice.com/en/article/8xkyj3/we-need-bug-bounties-for-bad-algorithms}}.
\newblock \bibinfo{journal}{\emph{{Vice News}}} (\bibinfo{date}{3 May}
  \bibinfo{year}{2018}).
\newblock


\bibitem[\protect\citeauthoryear{Appel}{Appel}{2011}]%
        {appel2011verified}
\bibfield{author}{\bibinfo{person}{Andrew~W Appel}.}
  \bibinfo{year}{2011}\natexlab{}.
\newblock \showarticletitle{Verified software toolchain}. In
  \bibinfo{booktitle}{\emph{European Symposium on Programming}}. Springer,
  \bibinfo{pages}{1--17}.
\newblock


\bibitem[\protect\citeauthoryear{Argyraki, Maniatis, Irzak, and
  Shenker}{Argyraki et~al\mbox{.}}{2007}]%
        {argyraki2007accountability}
\bibfield{author}{\bibinfo{person}{K Argyraki}, \bibinfo{person}{P Maniatis},
  \bibinfo{person}{O Irzak}, {and} \bibinfo{person}{S Shenker}.}
  \bibinfo{year}{2007}\natexlab{}.
\newblock \showarticletitle{An accountability interface for the Internet}. In
  \bibinfo{booktitle}{\emph{Proc. 14th ICNP}}.
\newblock


\bibitem[\protect\citeauthoryear{Argyris}{Argyris}{1977}]%
        {argyris1977double}
\bibfield{author}{\bibinfo{person}{Chris Argyris}.}
  \bibinfo{year}{1977}\natexlab{}.
\newblock \showarticletitle{Double loop learning in organizations}.
\newblock \bibinfo{journal}{\emph{Harvard business review}}
  \bibinfo{volume}{55}, \bibinfo{number}{5} (\bibinfo{year}{1977}),
  \bibinfo{pages}{115--125}.
\newblock


\bibitem[\protect\citeauthoryear{Arnold, Bellamy, Hind, Houde, Mehta,
  Mojsilovi{\'c}, Nair, Ramamurthy, Olteanu, Piorkowski, et~al\mbox{.}}{Arnold
  et~al\mbox{.}}{2019}]%
        {arnold2019factsheets}
\bibfield{author}{\bibinfo{person}{Matthew Arnold}, \bibinfo{person}{Rachel~KE
  Bellamy}, \bibinfo{person}{Michael Hind}, \bibinfo{person}{Stephanie Houde},
  \bibinfo{person}{Sameep Mehta}, \bibinfo{person}{A Mojsilovi{\'c}},
  \bibinfo{person}{Ravi Nair}, \bibinfo{person}{K~Natesan Ramamurthy},
  \bibinfo{person}{Alexandra Olteanu}, \bibinfo{person}{David Piorkowski},
  {et~al\mbox{.}}} \bibinfo{year}{2019}\natexlab{}.
\newblock \showarticletitle{FactSheets: Increasing trust in AI services through
  supplier's declarations of conformity}.
\newblock \bibinfo{journal}{\emph{IBM Journal of Research and Development}}
  \bibinfo{volume}{63}, \bibinfo{number}{4/5} (\bibinfo{year}{2019}),
  \bibinfo{pages}{6--1}.
\newblock


\bibitem[\protect\citeauthoryear{Backes, Druschel, Haeberlen, and Unruh}{Backes
  et~al\mbox{.}}{2009}]%
        {csar}
\bibfield{author}{\bibinfo{person}{Michael Backes}, \bibinfo{person}{Peter
  Druschel}, \bibinfo{person}{Andreas Haeberlen}, {and}
  \bibinfo{person}{Dominique Unruh}.} \bibinfo{year}{2009}\natexlab{}.
\newblock \showarticletitle{{CSAR:} A Practical and Provable Technique to Make
  Randomized Systems Accountable}.
\newblock \bibinfo{journal}{\emph{Proc. NDSS}} (\bibinfo{year}{2009}).
\newblock


\bibitem[\protect\citeauthoryear{Bainbridge}{Bainbridge}{1983}]%
        {bainbridge1983ironies}
\bibfield{author}{\bibinfo{person}{Lisanne Bainbridge}.}
  \bibinfo{year}{1983}\natexlab{}.
\newblock \showarticletitle{Ironies of automation}.
\newblock In \bibinfo{booktitle}{\emph{Analysis, design and evaluation of
  man--machine systems}}. \bibinfo{publisher}{Elsevier},
  \bibinfo{pages}{129--135}.
\newblock


\bibitem[\protect\citeauthoryear{Balasch, Rial, Troncoso, Preneel, Verbauwhede,
  and Geuens}{Balasch et~al\mbox{.}}{2010}]%
        {balasch2010pretp}
\bibfield{author}{\bibinfo{person}{Josep Balasch}, \bibinfo{person}{Alfredo
  Rial}, \bibinfo{person}{Carmela Troncoso}, \bibinfo{person}{Bart Preneel},
  \bibinfo{person}{Ingrid Verbauwhede}, {and} \bibinfo{person}{Christophe
  Geuens}.} \bibinfo{year}{2010}\natexlab{}.
\newblock \showarticletitle{PrETP: Privacy-Preserving Electronic Toll
  Pricing.}. In \bibinfo{booktitle}{\emph{USENIX Security Symposium}}.
  \bibinfo{pages}{63--78}.
\newblock


\bibitem[\protect\citeauthoryear{Bamberger and Mulligan}{Bamberger and
  Mulligan}{2010}]%
        {bamberger2010privacy}
\bibfield{author}{\bibinfo{person}{Kenneth~A Bamberger} {and}
  \bibinfo{person}{Deirdre~K Mulligan}.} \bibinfo{year}{2010}\natexlab{}.
\newblock \showarticletitle{Privacy on the Books and on the Ground}.
\newblock \bibinfo{journal}{\emph{Stan. L. Rev.}}  \bibinfo{volume}{63}
  (\bibinfo{year}{2010}), \bibinfo{pages}{247}.
\newblock


\bibitem[\protect\citeauthoryear{Bamberger and Mulligan}{Bamberger and
  Mulligan}{2015}]%
        {bamberger2015privacy}
\bibfield{author}{\bibinfo{person}{Kenneth~A Bamberger} {and}
  \bibinfo{person}{Deirdre~K Mulligan}.} \bibinfo{year}{2015}\natexlab{}.
\newblock \bibinfo{booktitle}{\emph{Privacy on the ground: driving corporate
  behavior in the United States and Europe}}.
\newblock \bibinfo{publisher}{MIT Press}.
\newblock


\bibitem[\protect\citeauthoryear{Barach and Small}{Barach and Small}{2000}]%
        {barach2000reporting}
\bibfield{author}{\bibinfo{person}{Paul Barach} {and}
  \bibinfo{person}{Stephen~D Small}.} \bibinfo{year}{2000}\natexlab{}.
\newblock \showarticletitle{Reporting and preventing medical mishaps: lessons
  from non-medical near miss reporting systems}.
\newblock \bibinfo{journal}{\emph{Bmj}} \bibinfo{volume}{320},
  \bibinfo{number}{7237} (\bibinfo{year}{2000}), \bibinfo{pages}{759--763}.
\newblock


\bibitem[\protect\citeauthoryear{Barth, Mitchell, Datta, and Sundaram}{Barth
  et~al\mbox{.}}{2007}]%
        {barth2007privacy}
\bibfield{author}{\bibinfo{person}{Adam Barth}, \bibinfo{person}{John~C
  Mitchell}, \bibinfo{person}{Anupam Datta}, {and} \bibinfo{person}{Sharada
  Sundaram}.} \bibinfo{year}{2007}\natexlab{}.
\newblock \showarticletitle{Privacy and utility in business processes}. In
  \bibinfo{booktitle}{\emph{Computer Security Foundations Symposium, 2007.
  CSF'07. 20th IEEE}}. IEEE, \bibinfo{pages}{279--294}.
\newblock


\bibitem[\protect\citeauthoryear{Bashir, Arshad, and Wilson}{Bashir
  et~al\mbox{.}}{2016}]%
        {bashir2016recommended}
\bibfield{author}{\bibinfo{person}{Muhammad~Ahmad Bashir},
  \bibinfo{person}{Sajjad Arshad}, {and} \bibinfo{person}{Christo Wilson}.}
  \bibinfo{year}{2016}\natexlab{}.
\newblock \showarticletitle{Recommended For You: A First Look at Content
  Recommendation Networks}. In \bibinfo{booktitle}{\emph{Proceedings of the
  2016 Internet Measurement Conference}}. ACM.
\newblock


\bibitem[\protect\citeauthoryear{Basil and Turner}{Basil and Turner}{1975}]%
        {basil1975iterative}
\bibfield{author}{\bibinfo{person}{Victor~R Basil} {and}
  \bibinfo{person}{Albert~J Turner}.} \bibinfo{year}{1975}\natexlab{}.
\newblock \showarticletitle{Iterative enhancement: A practical technique for
  software development}.
\newblock \bibinfo{journal}{\emph{IEEE Transactions on Software Engineering}}
  \bibinfo{number}{4} (\bibinfo{year}{1975}), \bibinfo{pages}{390--396}.
\newblock


\bibitem[\protect\citeauthoryear{Bass, Weber, and Zhu}{Bass
  et~al\mbox{.}}{2015}]%
        {bass2015devops}
\bibfield{author}{\bibinfo{person}{Len Bass}, \bibinfo{person}{Ingo Weber},
  {and} \bibinfo{person}{Liming Zhu}.} \bibinfo{year}{2015}\natexlab{}.
\newblock \bibinfo{booktitle}{\emph{DevOps: A software architect's
  perspective}}.
\newblock \bibinfo{publisher}{Addison-Wesley Professional}.
\newblock


\bibitem[\protect\citeauthoryear{Beck}{Beck}{2003}]%
        {beck2003test}
\bibfield{author}{\bibinfo{person}{Kent Beck}.}
  \bibinfo{year}{2003}\natexlab{}.
\newblock \bibinfo{booktitle}{\emph{Test-driven development: by example}}.
\newblock \bibinfo{publisher}{Addison-Wesley Prof.}
\newblock


\bibitem[\protect\citeauthoryear{Beck, Beedle, Van~Bennekum, Cockburn,
  Cunningham, et~al\mbox{.}}{Beck et~al\mbox{.}}{[n.d.]}]%
        {beck2001manifesto}
\bibfield{author}{\bibinfo{person}{Kent Beck}, \bibinfo{person}{Mike Beedle},
  \bibinfo{person}{Arie Van~Bennekum}, \bibinfo{person}{Alistair Cockburn},
  \bibinfo{person}{Ward Cunningham}, {et~al\mbox{.}}}
  \bibinfo{year}{[n.d.]}\natexlab{}.
\newblock \showarticletitle{Manifesto for agile software development, 2001}.
\newblock  (\bibinfo{year}{[n.\,d.]}).
\newblock


\bibitem[\protect\citeauthoryear{Bellovin}{Bellovin}{2006}]%
        {bellovin2006brittleness}
\bibfield{author}{\bibinfo{person}{Steven~M Bellovin}.}
  \bibinfo{year}{2006}\natexlab{}.
\newblock \showarticletitle{On the brittleness of software and the
  infeasibility of security metrics}.
\newblock \bibinfo{journal}{\emph{IEEE Annals of the History of Computing}}
  \bibinfo{volume}{4}, \bibinfo{number}{04} (\bibinfo{year}{2006}),
  \bibinfo{pages}{96--96}.
\newblock


\bibitem[\protect\citeauthoryear{Ben-Sasson, Chiesa, Genkin, and
  Tromer}{Ben-Sasson et~al\mbox{.}}{2012}]%
        {ben-sasson}
\bibfield{author}{\bibinfo{person}{Eli Ben-Sasson}, \bibinfo{person}{Alessandro
  Chiesa}, \bibinfo{person}{Daniel Genkin}, {and} \bibinfo{person}{Eran
  Tromer}.} \bibinfo{year}{2012}\natexlab{}.
\newblock \bibinfo{booktitle}{\emph{On the Concrete-Efficiency Threshold of
  Probabilistically-Checkable Proofs}}.
\newblock \bibinfo{type}{{T}echnical {R}eport}.
  \bibinfo{institution}{Electronic Colloquium on Computational Complexity}.
\newblock
\newblock
\shownote{\url{http://eccc.hpi-web.de/report/2012/045/}.}


\bibitem[\protect\citeauthoryear{Ben-Sasson, Chiesa, Genkin, Tromer, and
  Virza}{Ben-Sasson et~al\mbox{.}}{2013}]%
        {ben2013snarks}
\bibfield{author}{\bibinfo{person}{Eli Ben-Sasson}, \bibinfo{person}{Alessandro
  Chiesa}, \bibinfo{person}{Daniel Genkin}, \bibinfo{person}{Eran Tromer},
  {and} \bibinfo{person}{Madars Virza}.} \bibinfo{year}{2013}\natexlab{}.
\newblock \showarticletitle{{SNARKs for C}: Verifying program executions
  succinctly and in zero knowledge}.
\newblock \bibinfo{journal}{\emph{CRYPTO}} (\bibinfo{year}{2013}).
\newblock


\bibitem[\protect\citeauthoryear{Ben-Sasson, Chiesa, Green, Tromer, and
  Virza}{Ben-Sasson et~al\mbox{.}}{2015}]%
        {ben2015secure}
\bibfield{author}{\bibinfo{person}{Eli Ben-Sasson}, \bibinfo{person}{Alessandro
  Chiesa}, \bibinfo{person}{Matthew Green}, \bibinfo{person}{Eran Tromer},
  {and} \bibinfo{person}{Madars Virza}.} \bibinfo{year}{2015}\natexlab{}.
\newblock \showarticletitle{Secure sampling of public parameters for succinct
  zero knowledge proofs}. In \bibinfo{booktitle}{\emph{IEEE Symposium on
  Security and Privacy}}.
\newblock


\bibitem[\protect\citeauthoryear{Ben-Sasson, Chiesa, Tromer, and
  Virza}{Ben-Sasson et~al\mbox{.}}{2014}]%
        {ben2014succinct}
\bibfield{author}{\bibinfo{person}{Eli Ben-Sasson}, \bibinfo{person}{Alessandro
  Chiesa}, \bibinfo{person}{Eran Tromer}, {and} \bibinfo{person}{Madars
  Virza}.} \bibinfo{year}{2014}\natexlab{}.
\newblock \showarticletitle{Succinct non-interactive zero knowledge for a von
  Neumann architecture}. In \bibinfo{booktitle}{\emph{USENIX Security}}.
  \bibinfo{pages}{781--796}.
\newblock


\bibitem[\protect\citeauthoryear{Bender and Friedman}{Bender and
  Friedman}{2018}]%
        {bender2018data}
\bibfield{author}{\bibinfo{person}{Emily~M Bender} {and} \bibinfo{person}{Batya
  Friedman}.} \bibinfo{year}{2018}\natexlab{}.
\newblock \showarticletitle{Data statements for natural language processing:
  Toward mitigating system bias and enabling better science}.
\newblock \bibinfo{journal}{\emph{Transactions of the Association for
  Computational Linguistics}}  \bibinfo{volume}{6} (\bibinfo{year}{2018}),
  \bibinfo{pages}{587--604}.
\newblock


\bibitem[\protect\citeauthoryear{Beutel, Chen, Doshi, Qian, Woodruff, Luu,
  Kreitmann, Bischof, and Chi}{Beutel et~al\mbox{.}}{2019}]%
        {beutel2019putting}
\bibfield{author}{\bibinfo{person}{Alex Beutel}, \bibinfo{person}{Jilin Chen},
  \bibinfo{person}{Tulsee Doshi}, \bibinfo{person}{Hai Qian},
  \bibinfo{person}{Allison Woodruff}, \bibinfo{person}{Christine Luu},
  \bibinfo{person}{Pierre Kreitmann}, \bibinfo{person}{Jonathan Bischof}, {and}
  \bibinfo{person}{Ed~H Chi}.} \bibinfo{year}{2019}\natexlab{}.
\newblock \showarticletitle{Putting Fairness Principles into Practice:
  Challenges, Metrics, and Improvements}.
\newblock \bibinfo{journal}{\emph{arXiv preprint arXiv:1901.04562}}
  (\bibinfo{year}{2019}).
\newblock


\bibitem[\protect\citeauthoryear{Binns, Van~Kleek, Veale, Lyngs, Zhao, and
  Shadbolt}{Binns et~al\mbox{.}}{2018}]%
        {binns2018reducing}
\bibfield{author}{\bibinfo{person}{Reuben Binns}, \bibinfo{person}{Max
  Van~Kleek}, \bibinfo{person}{Michael Veale}, \bibinfo{person}{Ulrik Lyngs},
  \bibinfo{person}{Jun Zhao}, {and} \bibinfo{person}{Nigel Shadbolt}.}
  \bibinfo{year}{2018}\natexlab{}.
\newblock \showarticletitle{'It's Reducing a Human Being to a Percentage':
  Perceptions of Justice in Algorithmic Decisions}. In
  \bibinfo{booktitle}{\emph{Proceedings of the 2018 CHI Conference on Human
  Factors in Computing Systems}}. ACM, \bibinfo{pages}{377}.
\newblock


\bibitem[\protect\citeauthoryear{Bishop}{Bishop}{1995}]%
        {bishop1995standard}
\bibfield{author}{\bibinfo{person}{Matt Bishop}.}
  \bibinfo{year}{1995}\natexlab{}.
\newblock \showarticletitle{A standard audit trail format}. In
  \bibinfo{booktitle}{\emph{Proceedings of the 18th National Information
  Systems Security Conference}}. \bibinfo{pages}{136--145}.
\newblock


\bibitem[\protect\citeauthoryear{Blocki, Christin, Datta, Procaccia, and
  Sinha}{Blocki et~al\mbox{.}}{2013}]%
        {blocki2013audit}
\bibfield{author}{\bibinfo{person}{Jeremiah Blocki}, \bibinfo{person}{Nicolas
  Christin}, \bibinfo{person}{Anupam Datta}, \bibinfo{person}{Ariel~D
  Procaccia}, {and} \bibinfo{person}{Arunesh Sinha}.}
  \bibinfo{year}{2013}\natexlab{}.
\newblock \showarticletitle{Audit games}. In
  \bibinfo{booktitle}{\emph{Proceedings of the Twenty-Third international joint
  conference on Artificial Intelligence}}. AAAI Press, \bibinfo{pages}{41--47}.
\newblock


\bibitem[\protect\citeauthoryear{Bodea, Karanikolova, Mulligan, and
  Makagon}{Bodea et~al\mbox{.}}{2018a}]%
        {bodea2018euautomated}
\bibfield{author}{\bibinfo{person}{Gabriela Bodea}, \bibinfo{person}{Kristina
  Karanikolova}, \bibinfo{person}{Deirdre~K. Mulligan}, {and}
  \bibinfo{person}{Jael Makagon}.} \bibinfo{year}{2018}\natexlab{a}.
\newblock \bibinfo{booktitle}{\emph{{Automated decision-making on the basis of
  personal data that has been transferred from the EU to companies certified
  under the EU-U.S. Privacy Shield: Fact-finding and assessment of safeguards
  provided by U.S. law}}}.
\newblock \bibinfo{type}{{T}echnical {R}eport}. \bibinfo{institution}{European
  Commission}.
\newblock
\urldef\tempurl%
\url{https://ec.europa.eu/info/sites/info/files/independent_study_on_automated_decision-making.pdf}
\showURL{%
\tempurl}


\bibitem[\protect\citeauthoryear{Bodea, Karanikolova, Mulligan, and
  Makagon}{Bodea et~al\mbox{.}}{2018b}]%
        {ec2018privacyshield}
\bibfield{author}{\bibinfo{person}{Gabriela Bodea}, \bibinfo{person}{Kristina
  Karanikolova}, \bibinfo{person}{Deirdre~K. Mulligan}, {and}
  \bibinfo{person}{Jael Makagon}.} \bibinfo{year}{2018}\natexlab{b}.
\newblock \bibinfo{title}{Automated decision-making on the basis of personal
  data that has been transferred from the EU to companies certified under the
  EU-U.S. Privacy Shield. Fact-finding and assessment of safeguards provided by
  U.S. law}.
\newblock \bibinfo{howpublished}{Independent Report promulgated by the European
  Commission, Directorate-General for Justice and Consumers}.
\newblock
\urldef\tempurl%
\url{https://ec.europa.eu/info/sites/info/files/independent_study_on_automated_decision-making.pdf}
\showURL{%
\tempurl}


\bibitem[\protect\citeauthoryear{Boehm}{Boehm}{2002}]%
        {boehm2002get}
\bibfield{author}{\bibinfo{person}{Barry Boehm}.}
  \bibinfo{year}{2002}\natexlab{}.
\newblock \showarticletitle{Get ready for agile methods, with care}.
\newblock \bibinfo{journal}{\emph{Computer}} \bibinfo{volume}{35},
  \bibinfo{number}{1} (\bibinfo{year}{2002}), \bibinfo{pages}{64--69}.
\newblock


\bibitem[\protect\citeauthoryear{Boehm}{Boehm}{1984}]%
        {boehm1984verifying}
\bibfield{author}{\bibinfo{person}{Barry~W. Boehm}.}
  \bibinfo{year}{1984}\natexlab{}.
\newblock \showarticletitle{Verifying and validating software requirements and
  design specifications}.
\newblock \bibinfo{journal}{\emph{IEEE software}} \bibinfo{volume}{1},
  \bibinfo{number}{1} (\bibinfo{year}{1984}), \bibinfo{pages}{75}.
\newblock


\bibitem[\protect\citeauthoryear{Boettiger}{Boettiger}{2015}]%
        {boettiger2015introduction}
\bibfield{author}{\bibinfo{person}{Carl Boettiger}.}
  \bibinfo{year}{2015}\natexlab{}.
\newblock \showarticletitle{An introduction to Docker for reproducible
  research}.
\newblock \bibinfo{journal}{\emph{ACM SIGOPS Operating Systems Review}}
  \bibinfo{volume}{49}, \bibinfo{number}{1} (\bibinfo{year}{2015}),
  \bibinfo{pages}{71--79}.
\newblock


\bibitem[\protect\citeauthoryear{Bonchi, Castillo, and Hajian}{Bonchi
  et~al\mbox{.}}{2016}]%
        {bonchi2016bias}
\bibfield{author}{\bibinfo{person}{Francisco Bonchi}, \bibinfo{person}{Carlos
  Castillo}, {and} \bibinfo{person}{Sara Hajian}.}
  \bibinfo{year}{2016}\natexlab{}.
\newblock \bibinfo{title}{Algorithmic bias: from discrimination discovery to
  fairness-aware data mining}.
\newblock \bibinfo{howpublished}{Knowledge Discovery and Data Mining, Tutorials
  Track}.
\newblock


\bibitem[\protect\citeauthoryear{Braun, Feldman, Ren, Setty, Blumberg, and
  Walfish}{Braun et~al\mbox{.}}{2013}]%
        {braun2013verifying}
\bibfield{author}{\bibinfo{person}{Benjamin Braun}, \bibinfo{person}{Ariel~J
  Feldman}, \bibinfo{person}{Zuocheng Ren}, \bibinfo{person}{Srinath Setty},
  \bibinfo{person}{Andrew~J Blumberg}, {and} \bibinfo{person}{Michael
  Walfish}.} \bibinfo{year}{2013}\natexlab{}.
\newblock \showarticletitle{Verifying computations with state}. In
  \bibinfo{booktitle}{\emph{Twenty-Fourth ACM Symposium on Operating Systems
  Principles}}. ACM, \bibinfo{pages}{341--357}.
\newblock


\bibitem[\protect\citeauthoryear{Breaux, Ant{\'o}n, Karat, and Karat}{Breaux
  et~al\mbox{.}}{2006a}]%
        {breaux2006enforceability}
\bibfield{author}{\bibinfo{person}{Travis~D Breaux}, \bibinfo{person}{Annie~I
  Ant{\'o}n}, \bibinfo{person}{C-M Karat}, {and} \bibinfo{person}{John Karat}.}
  \bibinfo{year}{2006}\natexlab{a}.
\newblock \showarticletitle{Enforceability vs. accountability in electronic
  policies}. In \bibinfo{booktitle}{\emph{Policies for Distributed Systems and
  Networks, 2006. Policy 2006. Seventh IEEE International Workshop on}}. IEEE,
  \bibinfo{pages}{4--pp}.
\newblock


\bibitem[\protect\citeauthoryear{Breaux, Ant{\'o}n, and Spafford}{Breaux
  et~al\mbox{.}}{2009}]%
        {breaux2009distributed}
\bibfield{author}{\bibinfo{person}{Travis~D Breaux}, \bibinfo{person}{Annie~I
  Ant{\'o}n}, {and} \bibinfo{person}{Eugene~H Spafford}.}
  \bibinfo{year}{2009}\natexlab{}.
\newblock \showarticletitle{A distributed requirements management framework for
  legal compliance and accountability}.
\newblock \bibinfo{journal}{\emph{computers \& security}} \bibinfo{volume}{28},
  \bibinfo{number}{1-2} (\bibinfo{year}{2009}), \bibinfo{pages}{8--17}.
\newblock


\bibitem[\protect\citeauthoryear{Breaux, Vail, and Anton}{Breaux
  et~al\mbox{.}}{2006b}]%
        {breaux2006towards}
\bibfield{author}{\bibinfo{person}{Travis~D Breaux}, \bibinfo{person}{Matthew~W
  Vail}, {and} \bibinfo{person}{Annie~I Anton}.}
  \bibinfo{year}{2006}\natexlab{b}.
\newblock \showarticletitle{Towards regulatory compliance: Extracting rights
  and obligations to align requirements with regulations}. In
  \bibinfo{booktitle}{\emph{14th IEEE International Requirements Engineering
  Conference (RE'06)}}. IEEE, \bibinfo{pages}{49--58}.
\newblock


\bibitem[\protect\citeauthoryear{Brooks~Jr}{Brooks~Jr}{1995}]%
        {brooks1995mythical}
\bibfield{author}{\bibinfo{person}{Frederick~P Brooks~Jr}.}
  \bibinfo{year}{1995}\natexlab{}.
\newblock \bibinfo{booktitle}{\emph{The mythical man-month: essays on software
  engineering}}.
\newblock \bibinfo{publisher}{Addison-Wesley}.
\newblock


\bibitem[\protect\citeauthoryear{Bruening and Kroll}{Bruening and
  Kroll}{2019}]%
        {bruening2019considering}
\bibfield{author}{\bibinfo{person}{Paula Bruening} {and}
  \bibinfo{person}{Joshua~A. Kroll}.} \bibinfo{year}{2019}\natexlab{}.
\newblock \showarticletitle{Considering Transparency: AI, Data Science, and the
  GDPR}.
\newblock \bibinfo{journal}{\emph{Computers, Privacy, and Data Protection}}
  (\bibinfo{year}{2019}).
\newblock


\bibitem[\protect\citeauthoryear{Brundage, Avin, Wang, Belfield, Krueger,
  Hadfield, Khlaaf, Yang, Toner, Fong, et~al\mbox{.}}{Brundage
  et~al\mbox{.}}{2020}]%
        {brundage2020toward}
\bibfield{author}{\bibinfo{person}{Miles Brundage}, \bibinfo{person}{Shahar
  Avin}, \bibinfo{person}{Jasmine Wang}, \bibinfo{person}{Haydn Belfield},
  \bibinfo{person}{Gretchen Krueger}, \bibinfo{person}{Gillian Hadfield},
  \bibinfo{person}{Heidy Khlaaf}, \bibinfo{person}{Jingying Yang},
  \bibinfo{person}{Helen Toner}, \bibinfo{person}{Ruth Fong}, {et~al\mbox{.}}}
  \bibinfo{year}{2020}\natexlab{}.
\newblock \showarticletitle{Toward trustworthy AI development: mechanisms for
  supporting verifiable claims}.
\newblock \bibinfo{journal}{\emph{OpenAI Technical Report, arXiv preprint
  arXiv:2004.07213}} (\bibinfo{year}{2020}).
\newblock


\bibitem[\protect\citeauthoryear{Buneman, Khanna, and Wang-Chiew}{Buneman
  et~al\mbox{.}}{2001}]%
        {buneman2001provenance}
\bibfield{author}{\bibinfo{person}{Peter Buneman}, \bibinfo{person}{Sanjeev
  Khanna}, {and} \bibinfo{person}{Tan Wang-Chiew}.}
  \bibinfo{year}{2001}\natexlab{}.
\newblock \showarticletitle{Why and where: A characterization of data
  provenance}.
\newblock In \bibinfo{booktitle}{\emph{Database Theory—ICDT 2001}}.
  \bibinfo{publisher}{Springer}.
\newblock


\bibitem[\protect\citeauthoryear{Buolamwini and Gebru}{Buolamwini and
  Gebru}{2018}]%
        {buolamwini2018gender}
\bibfield{author}{\bibinfo{person}{Joy Buolamwini} {and}
  \bibinfo{person}{Timnit Gebru}.} \bibinfo{year}{2018}\natexlab{}.
\newblock \showarticletitle{Gender shades: Intersectional accuracy disparities
  in commercial gender classification}. In \bibinfo{booktitle}{\emph{Conference
  on Fairness, Accountability and Transparency}}. \bibinfo{pages}{77--91}.
\newblock


\bibitem[\protect\citeauthoryear{Burrell}{Burrell}{2016}]%
        {burrell2016machine}
\bibfield{author}{\bibinfo{person}{Jenna Burrell}.}
  \bibinfo{year}{2016}\natexlab{}.
\newblock \showarticletitle{How the machine `thinks': Understanding opacity in
  machine learning algorithms}.
\newblock \bibinfo{journal}{\emph{Big Data \& Society}} \bibinfo{volume}{3},
  \bibinfo{number}{1} (\bibinfo{year}{2016}).
\newblock


\bibitem[\protect\citeauthoryear{Camenisch, Hohenberger, and
  Lysyanskaya}{Camenisch et~al\mbox{.}}{2006}]%
        {camenisch2006balancing}
\bibfield{author}{\bibinfo{person}{Jan Camenisch}, \bibinfo{person}{Susan
  Hohenberger}, {and} \bibinfo{person}{Anna Lysyanskaya}.}
  \bibinfo{year}{2006}\natexlab{}.
\newblock \showarticletitle{Balancing accountability and privacy using e-cash}.
\newblock In \bibinfo{booktitle}{\emph{Security and Cryptography for
  Networks}}. \bibinfo{publisher}{Springer}, \bibinfo{pages}{141--155}.
\newblock


\bibitem[\protect\citeauthoryear{Carruthers and Espeland}{Carruthers and
  Espeland}{1991}]%
        {carruthers1991accounting}
\bibfield{author}{\bibinfo{person}{Bruce~G Carruthers} {and}
  \bibinfo{person}{Wendy~Nelson Espeland}.} \bibinfo{year}{1991}\natexlab{}.
\newblock \showarticletitle{Accounting for rationality: Double-entry
  bookkeeping and the rhetoric of economic rationality}.
\newblock \bibinfo{journal}{\emph{American journal of sociology}}
  \bibinfo{volume}{97}, \bibinfo{number}{1} (\bibinfo{year}{1991}),
  \bibinfo{pages}{31--69}.
\newblock


\bibitem[\protect\citeauthoryear{Carter and Sholler}{Carter and
  Sholler}{2016}]%
        {carter2016data}
\bibfield{author}{\bibinfo{person}{Daniel Carter} {and} \bibinfo{person}{Dan
  Sholler}.} \bibinfo{year}{2016}\natexlab{}.
\newblock \showarticletitle{Data science on the ground: Hype, criticism, and
  everyday work}.
\newblock \bibinfo{journal}{\emph{Journal of the Association for Information
  Science and Technology}} \bibinfo{volume}{67}, \bibinfo{number}{10}
  (\bibinfo{year}{2016}), \bibinfo{pages}{2309--2319}.
\newblock


\bibitem[\protect\citeauthoryear{Cavoukian}{Cavoukian}{2011}]%
        {cavoukian2011privacy}
\bibfield{author}{\bibinfo{person}{Ann Cavoukian}.}
  \bibinfo{year}{2011}\natexlab{}.
\newblock \showarticletitle{Privacy by design in law, policy and practice}.
\newblock \bibinfo{journal}{\emph{A white paper for regulators, decision-makers
  and policy-makers}} (\bibinfo{year}{2011}).
\newblock


\bibitem[\protect\citeauthoryear{Cavoukian et~al\mbox{.}}{Cavoukian
  et~al\mbox{.}}{2009}]%
        {cavoukian2009privacy}
\bibfield{author}{\bibinfo{person}{Ann Cavoukian} {et~al\mbox{.}}}
  \bibinfo{year}{2009}\natexlab{}.
\newblock \showarticletitle{Privacy by design: The 7 foundational principles}.
\newblock \bibinfo{journal}{\emph{Information and Privacy Commissioner of
  Ontario, Canada}}  \bibinfo{volume}{5} (\bibinfo{year}{2009}).
\newblock


\bibitem[\protect\citeauthoryear{Chen, Ma, Hann{\'a}k, and Wilson}{Chen
  et~al\mbox{.}}{2018}]%
        {chen2018investigating}
\bibfield{author}{\bibinfo{person}{Le Chen}, \bibinfo{person}{Ruijun Ma},
  \bibinfo{person}{Anik{\'o} Hann{\'a}k}, {and} \bibinfo{person}{Christo
  Wilson}.} \bibinfo{year}{2018}\natexlab{}.
\newblock \showarticletitle{Investigating the Impact of Gender on Rank in
  Resume Search Engines}. In \bibinfo{booktitle}{\emph{Proceedings of the 2018
  CHI Conference on Human Factors in Computing Systems}}. ACM.
\newblock


\bibitem[\protect\citeauthoryear{Chen and Wilson}{Chen and Wilson}{2017}]%
        {chen2017observing}
\bibfield{author}{\bibinfo{person}{Le Chen} {and} \bibinfo{person}{Christo
  Wilson}.} \bibinfo{year}{2017}\natexlab{}.
\newblock \showarticletitle{Observing algorithmic marketplaces in-the-wild}.
\newblock \bibinfo{journal}{\emph{ACM SIGecom Exchanges}} \bibinfo{volume}{15},
  \bibinfo{number}{2} (\bibinfo{year}{2017}), \bibinfo{pages}{34--39}.
\newblock


\bibitem[\protect\citeauthoryear{Chin and Ozsoyoglu}{Chin and
  Ozsoyoglu}{1982}]%
        {chin1982auditing}
\bibfield{author}{\bibinfo{person}{Francis~Y Chin} {and}
  \bibinfo{person}{Gultekin Ozsoyoglu}.} \bibinfo{year}{1982}\natexlab{}.
\newblock \showarticletitle{Auditing and inference control in statistical
  databases}.
\newblock \bibinfo{journal}{\emph{IEEE Transactions on Software Engineering}}
  (\bibinfo{year}{1982}), \bibinfo{pages}{574--582}.
\newblock


\bibitem[\protect\citeauthoryear{Chouldechova}{Chouldechova}{2017}]%
        {chouldechova2017fair}
\bibfield{author}{\bibinfo{person}{Alexandra Chouldechova}.}
  \bibinfo{year}{2017}\natexlab{}.
\newblock \showarticletitle{Fair prediction with disparate impact: A study of
  bias in recidivism prediction instruments}.
\newblock \bibinfo{journal}{\emph{Big data}} \bibinfo{volume}{5},
  \bibinfo{number}{2} (\bibinfo{year}{2017}), \bibinfo{pages}{153--163}.
\newblock


\bibitem[\protect\citeauthoryear{Chouldechova, Benavides-Prado, Fialko, and
  Vaithianathan}{Chouldechova et~al\mbox{.}}{2018}]%
        {chouldechova2018case}
\bibfield{author}{\bibinfo{person}{Alexandra Chouldechova},
  \bibinfo{person}{Diana Benavides-Prado}, \bibinfo{person}{Oleksandr Fialko},
  {and} \bibinfo{person}{Rhema Vaithianathan}.}
  \bibinfo{year}{2018}\natexlab{}.
\newblock \showarticletitle{A case study of algorithm-assisted decision making
  in child maltreatment hotline screening decisions}. In
  \bibinfo{booktitle}{\emph{Conference on Fairness, Accountability and
  Transparency}}. \bibinfo{pages}{134--148}.
\newblock


\bibitem[\protect\citeauthoryear{Christin}{Christin}{2017}]%
        {christin2017algorithms}
\bibfield{author}{\bibinfo{person}{Ang{\`e}le Christin}.}
  \bibinfo{year}{2017}\natexlab{}.
\newblock \showarticletitle{Algorithms in practice: Comparing web journalism
  and criminal justice}.
\newblock \bibinfo{journal}{\emph{Big Data \& Society}} \bibinfo{volume}{4},
  \bibinfo{number}{2} (\bibinfo{year}{2017}).
\newblock


\bibitem[\protect\citeauthoryear{Citron}{Citron}{2007}]%
        {citron2007technological}
\bibfield{author}{\bibinfo{person}{Danielle Citron}.}
  \bibinfo{year}{2007}\natexlab{}.
\newblock \showarticletitle{Technological due process}.
\newblock \bibinfo{journal}{\emph{Washington University Law Review}}
  \bibinfo{volume}{85} (\bibinfo{year}{2007}), \bibinfo{pages}{1249--1313}.
\newblock


\bibitem[\protect\citeauthoryear{Citron}{Citron}{2008}]%
        {citron2008open}
\bibfield{author}{\bibinfo{person}{Danielle~Keats Citron}.}
  \bibinfo{year}{2008}\natexlab{}.
\newblock \showarticletitle{Open code governance}. In
  \bibinfo{booktitle}{\emph{University of Chicago Legal Forum}}.
  \bibinfo{pages}{355--387}.
\newblock


\bibitem[\protect\citeauthoryear{Citron and Calo}{Citron and Calo}{2020}]%
        {citron2020automated}
\bibfield{author}{\bibinfo{person}{Danielle~K Citron} {and}
  \bibinfo{person}{Ryan Calo}.} \bibinfo{year}{2020}\natexlab{}.
\newblock \showarticletitle{The Automated Administrative State: A Crisis of
  Legitimacy}.
\newblock \bibinfo{journal}{\emph{Emory Law Journal}} (\bibinfo{year}{2020}).
\newblock


\bibitem[\protect\citeauthoryear{Clark}{Clark}{2018}]%
        {clark2018compliance}
\bibfield{author}{\bibinfo{person}{Robert Clark}.}
  \bibinfo{year}{2018}\natexlab{}.
\newblock \showarticletitle{{Compliance != Security (Except When It Might
  Be)}}. In \bibinfo{booktitle}{\emph{Enigma 2018}}.
\newblock


\bibitem[\protect\citeauthoryear{Colbert and Bowen}{Colbert and Bowen}{1996}]%
        {colbert1996comparison}
\bibfield{author}{\bibinfo{person}{Janet~L Colbert} {and} \bibinfo{person}{Paul
  Bowen}.} \bibinfo{year}{1996}\natexlab{}.
\newblock \showarticletitle{A Comparison of Internal Control: COBIT, SAC, COSO
  and SAS}.
\newblock \bibinfo{journal}{\emph{IS Audit and Control Journal}}
  (\bibinfo{year}{1996}), \bibinfo{pages}{26--35}.
\newblock


\bibitem[\protect\citeauthoryear{{Computing Community Consortium}}{{Computing
  Community Consortium}}{2015a}]%
        {cccpbdworkshop3}
\bibfield{author}{\bibinfo{person}{{Computing Community Consortium}}.}
  \bibinfo{year}{2015}\natexlab{a}.
\newblock \bibinfo{title}{Privacy by Design---Engineering Privacy: Workshop 3
  Report}.
\newblock
  \bibinfo{howpublished}{\url{https://cra.org/ccc/wp-content/uploads/sites/2/2015/12/PbD3-Workshop-Report-v2.pdf}}.
\newblock


\bibitem[\protect\citeauthoryear{{Computing Community Consortium}}{{Computing
  Community Consortium}}{2015b}]%
        {cccpbdworkshop2}
\bibfield{author}{\bibinfo{person}{{Computing Community Consortium}}.}
  \bibinfo{year}{2015}\natexlab{b}.
\newblock \bibinfo{title}{Privacy by Design---Privacy Enabling Design: Workshop
  2 Report}.
\newblock
  \bibinfo{howpublished}{\url{https://cra.org/ccc/wp-content/uploads/sites/2/2015/05/PbD2-Report-v5.pdf}}.
\newblock


\bibitem[\protect\citeauthoryear{{Computing Community Consortium}}{{Computing
  Community Consortium}}{2015c}]%
        {cccpbdworkshop1}
\bibfield{author}{\bibinfo{person}{{Computing Community Consortium}}.}
  \bibinfo{year}{2015}\natexlab{c}.
\newblock \bibinfo{title}{Privacy by Design-State of Research and Practice:
  Workshop 1 Report}.
\newblock
  \bibinfo{howpublished}{\url{https://cra.org/ccc/wp-content/uploads/sites/2/2015/02/PbD-Workshop-1-Report-.pdf}}.
\newblock


\bibitem[\protect\citeauthoryear{{Consultative Committee for Space Data
  Systems}}{{Consultative Committee for Space Data Systems}}{2011}]%
        {ccsds2011audit}
\bibfield{author}{\bibinfo{person}{{Consultative Committee for Space Data
  Systems}}.} \bibinfo{year}{2011}\natexlab{}.
\newblock \bibinfo{title}{Audit and certification of Trustworthy Digital
  Repositories, Recommended Practice}.
\newblock
\newblock


\bibitem[\protect\citeauthoryear{{Consumer Financial Protection
  Bureau}}{{Consumer Financial Protection Bureau}}{2014}]%
        {cfpb-race-imputation}
\bibfield{author}{\bibinfo{person}{{Consumer Financial Protection Bureau}}.}
  \bibinfo{year}{2014}\natexlab{}.
\newblock \bibinfo{title}{{Using Publicly Available Information to Proxy for
  Unidentified Race and Ethnicity: A Methodology and Assessment}}.
\newblock
  \bibinfo{howpublished}{\url{https://www.consumerfinance.gov/data-research/research-reports/using-publicly-available-information-to-proxy-for-unidentified-race-and-ethnicity/}}.
\newblock


\bibitem[\protect\citeauthoryear{Cook}{Cook}{1998}]%
        {cook1998complex}
\bibfield{author}{\bibinfo{person}{Richard~I Cook}.}
  \bibinfo{year}{1998}\natexlab{}.
\newblock \showarticletitle{How complex systems fail}.
\newblock \bibinfo{journal}{\emph{Cognitive Technologies Laboratory, University
  of Chicago. Chicago IL}} (\bibinfo{year}{1998}).
\newblock


\bibitem[\protect\citeauthoryear{Crosby and Wallach}{Crosby and
  Wallach}{2009}]%
        {crosby2009efficient}
\bibfield{author}{\bibinfo{person}{Scott~A Crosby} {and} \bibinfo{person}{Dan~S
  Wallach}.} \bibinfo{year}{2009}\natexlab{}.
\newblock \showarticletitle{Efficient Data Structures For Tamper-Evident
  Logging.}. In \bibinfo{booktitle}{\emph{USENIX Security Symposium}}.
  \bibinfo{pages}{317--334}.
\newblock


\bibitem[\protect\citeauthoryear{Datta}{Datta}{2014}]%
        {datta2014privacy}
\bibfield{author}{\bibinfo{person}{Anupam Datta}.}
  \bibinfo{year}{2014}\natexlab{}.
\newblock \showarticletitle{Privacy through Accountability: A Computer Science
  Perspective}.
\newblock In \bibinfo{booktitle}{\emph{Distributed Computing and Internet
  Technology}}. \bibinfo{publisher}{Springer}, \bibinfo{pages}{43--49}.
\newblock


\bibitem[\protect\citeauthoryear{de~Castro, Lo, Reynolds, Susan,
  Vaikuntanathan, Weitzner, and Zhang}{de~Castro et~al\mbox{.}}{2020}]%
        {de2020scram}
\bibfield{author}{\bibinfo{person}{Leo de Castro}, \bibinfo{person}{Andrew~W
  Lo}, \bibinfo{person}{Taylor Reynolds}, \bibinfo{person}{Fransisca Susan},
  \bibinfo{person}{Vinod Vaikuntanathan}, \bibinfo{person}{Daniel Weitzner},
  {and} \bibinfo{person}{Nicolas Zhang}.} \bibinfo{year}{2020}\natexlab{}.
\newblock \showarticletitle{SCRAM: A Platform for Securely Measuring Cyber
  Risk}.
\newblock \bibinfo{journal}{\emph{Harvard Data Science Review}}
  (\bibinfo{year}{2020}).
\newblock


\bibitem[\protect\citeauthoryear{{Defense Innovation Board}}{{Defense
  Innovation Board}}{2019}]%
        {dib2019supporting}
\bibfield{author}{\bibinfo{person}{{Defense Innovation Board}}.}
  \bibinfo{year}{2019}\natexlab{}.
\newblock \bibinfo{title}{{AI Principles: Recommendations on the Ethical Use of
  Artificial Intelligence by the Department of Defense: Supporting Document}}.
\newblock \bibinfo{howpublished}{\url{https://innovation.defense.gov/ai/}}.
\newblock


\bibitem[\protect\citeauthoryear{Desai and Kroll}{Desai and Kroll}{2018}]%
        {desai2018trust}
\bibfield{author}{\bibinfo{person}{Deven Desai} {and}
  \bibinfo{person}{Joshua~A. Kroll}.} \bibinfo{year}{2018}\natexlab{}.
\newblock \showarticletitle{Trust but Verify: A Guide to Algorithms and the
  Law}.
\newblock \bibinfo{journal}{\emph{Harvard J. of Law and Tech.}}
  \bibinfo{volume}{31}, \bibinfo{number}{1} (\bibinfo{year}{2018}).
\newblock


\bibitem[\protect\citeauthoryear{{Division of Banking Supervision and
  Regulation}}{{Division of Banking Supervision and Regulation}}{2011}]%
        {sr11-7}
\bibfield{author}{\bibinfo{person}{{Division of Banking Supervision and
  Regulation}}.} \bibinfo{year}{2011}\natexlab{}.
\newblock \bibinfo{title}{SR 11-7: Guidance on Model Risk Management}.
\newblock
\newblock


\bibitem[\protect\citeauthoryear{Doshi-Velez and Kim}{Doshi-Velez and
  Kim}{2017}]%
        {doshi2017towards}
\bibfield{author}{\bibinfo{person}{Finale Doshi-Velez} {and}
  \bibinfo{person}{Been Kim}.} \bibinfo{year}{2017}\natexlab{}.
\newblock \showarticletitle{Towards a rigorous science of interpretable machine
  learning}.
\newblock \bibinfo{journal}{\emph{arXiv preprint arXiv:1702.08608}}
  (\bibinfo{year}{2017}).
\newblock


\bibitem[\protect\citeauthoryear{Doshi-Velez, Kortz, Budish, Bavitz, Gershman,
  O'Brien, Schieber, Waldo, Weinberger, and Wood}{Doshi-Velez
  et~al\mbox{.}}{2017}]%
        {doshi2017accountability}
\bibfield{author}{\bibinfo{person}{Finale Doshi-Velez}, \bibinfo{person}{Mason
  Kortz}, \bibinfo{person}{Ryan Budish}, \bibinfo{person}{Chris Bavitz},
  \bibinfo{person}{Sam Gershman}, \bibinfo{person}{David O'Brien},
  \bibinfo{person}{Stuart Schieber}, \bibinfo{person}{James Waldo},
  \bibinfo{person}{David Weinberger}, {and} \bibinfo{person}{Alexandra Wood}.}
  \bibinfo{year}{2017}\natexlab{}.
\newblock \showarticletitle{Accountability of {AI} Under the Law: The Role of
  Explanation}.
\newblock \bibinfo{journal}{\emph{arXiv preprint arXiv:1711.01134}}
  (\bibinfo{year}{2017}).
\newblock


\bibitem[\protect\citeauthoryear{Dwork, McSherry, Nissim, and Smith}{Dwork
  et~al\mbox{.}}{2006}]%
        {dwork2006calibrating}
\bibfield{author}{\bibinfo{person}{Cynthia Dwork}, \bibinfo{person}{Frank
  McSherry}, \bibinfo{person}{Kobbi Nissim}, {and} \bibinfo{person}{Adam
  Smith}.} \bibinfo{year}{2006}\natexlab{}.
\newblock \showarticletitle{Calibrating noise to sensitivity in private data
  analysis}. In \bibinfo{booktitle}{\emph{Theory of cryptography conference}}.
  Springer, \bibinfo{pages}{265--284}.
\newblock


\bibitem[\protect\citeauthoryear{Dwork and Mulligan}{Dwork and
  Mulligan}{2013}]%
        {dwork2013not}
\bibfield{author}{\bibinfo{person}{Cynthia Dwork} {and}
  \bibinfo{person}{Deirdre~K Mulligan}.} \bibinfo{year}{2013}\natexlab{}.
\newblock \showarticletitle{It's not privacy, and it's not fair}.
\newblock \bibinfo{journal}{\emph{Stanford Law Review Online}}
  \bibinfo{volume}{66} (\bibinfo{year}{2013}), \bibinfo{pages}{35}.
\newblock


\bibitem[\protect\citeauthoryear{Editorial}{Editorial}{2016}]%
        {nature-editorial}
\bibfield{author}{\bibinfo{person}{Editorial}.}
  \bibinfo{year}{2016}\natexlab{}.
\newblock \showarticletitle{More Accountability for Big Data Algorithms}.
\newblock \bibinfo{journal}{\emph{Nature}}  \bibinfo{volume}{537}
  (\bibinfo{year}{2016}).
\newblock
Issue 7621.


\bibitem[\protect\citeauthoryear{Edwards and Veale}{Edwards and Veale}{2017a}]%
        {edwards2017enslaving}
\bibfield{author}{\bibinfo{person}{Lilian Edwards} {and}
  \bibinfo{person}{Michael Veale}.} \bibinfo{year}{2017}\natexlab{a}.
\newblock \showarticletitle{Enslaving the Algorithm: From a ``Right to an
  Explanation'' to a ``Right to Better Decisions''?}
\newblock \bibinfo{journal}{\emph{IEEE Security and Privacy}}
  (\bibinfo{year}{2017}).
\newblock


\bibitem[\protect\citeauthoryear{Edwards and Veale}{Edwards and Veale}{2017b}]%
        {edwards2017slave}
\bibfield{author}{\bibinfo{person}{Lilian Edwards} {and}
  \bibinfo{person}{Michael Veale}.} \bibinfo{year}{2017}\natexlab{b}.
\newblock \showarticletitle{Slave to the Algorithm? Why a ``Right to
  Explanation'' is Probably Not the Remedy You are Looking for}.
\newblock \bibinfo{journal}{\emph{Duke Technology Law Journal}}
  \bibinfo{volume}{16}, \bibinfo{number}{1} (\bibinfo{year}{2017}),
  \bibinfo{pages}{18--84}.
\newblock


\bibitem[\protect\citeauthoryear{Elish and Hwang}{Elish and Hwang}{2015}]%
        {elish2015praise}
\bibfield{author}{\bibinfo{person}{MC Elish} {and} \bibinfo{person}{Tim
  Hwang}.} \bibinfo{year}{2015}\natexlab{}.
\newblock \showarticletitle{Praise the Machine! Punish the Human! The
  contradictory history of accountability in automated aviation}.
\newblock \bibinfo{journal}{\emph{Data \& Society Report}}
  (\bibinfo{year}{2015}).
\newblock


\bibitem[\protect\citeauthoryear{Elish}{Elish}{2019}]%
        {elish2019moral}
\bibfield{author}{\bibinfo{person}{Madeleine~Clare Elish}.}
  \bibinfo{year}{2019}\natexlab{}.
\newblock \showarticletitle{Moral crumple zones: Cautionary tales in
  human-robot interaction}.
\newblock \bibinfo{journal}{\emph{Engaging Science, Technology, and Society}}
  \bibinfo{volume}{5} (\bibinfo{year}{2019}), \bibinfo{pages}{40--60}.
\newblock


\bibitem[\protect\citeauthoryear{Ellison}{Ellison}{2007}]%
        {ellison2007ceremony}
\bibfield{author}{\bibinfo{person}{Carl Ellison}.}
  \bibinfo{year}{2007}\natexlab{}.
\newblock \showarticletitle{Ceremony Design and Analysis}.
\newblock \bibinfo{journal}{\emph{IACR eprint archive}}  \bibinfo{volume}{399}
  (\bibinfo{year}{2007}).
\newblock
\newblock
\shownote{\url{http://eprint.iacr.org/2007/399.pdf}.}


\bibitem[\protect\citeauthoryear{Ellison, Goodenough, Weinstock, and
  Woody}{Ellison et~al\mbox{.}}{2010}]%
        {ellison2010evaluating}
\bibfield{author}{\bibinfo{person}{Robert~J Ellison}, \bibinfo{person}{John~B
  Goodenough}, \bibinfo{person}{Charles~B Weinstock}, {and}
  \bibinfo{person}{Carol Woody}.} \bibinfo{year}{2010}\natexlab{}.
\newblock \bibinfo{booktitle}{\emph{Evaluating and mitigating software supply
  chain security risks}}.
\newblock \bibinfo{type}{{T}echnical {R}eport}.
  \bibinfo{institution}{Carnegie-Mellon University Software Engineering
  Institute}.
\newblock


\bibitem[\protect\citeauthoryear{Espeland and Vannebo}{Espeland and
  Vannebo}{2007}]%
        {espeland2007accountability}
\bibfield{author}{\bibinfo{person}{Wendy~Nelson Espeland} {and}
  \bibinfo{person}{Berit~Irene Vannebo}.} \bibinfo{year}{2007}\natexlab{}.
\newblock \showarticletitle{Accountability, quantification, and law}.
\newblock \bibinfo{journal}{\emph{Annu. Rev. Law Soc. Sci.}}
  \bibinfo{volume}{3} (\bibinfo{year}{2007}), \bibinfo{pages}{21--43}.
\newblock


\bibitem[\protect\citeauthoryear{{European Commission Independent High-Level
  Expert Group on Artificial Intelligence}}{{European Commission Independent
  High-Level Expert Group on Artificial Intelligence}}{2020}]%
        {eu-hleg-trustworthy-2020}
\bibfield{author}{\bibinfo{person}{{European Commission Independent High-Level
  Expert Group on Artificial Intelligence}}.} \bibinfo{year}{2020}\natexlab{}.
\newblock \bibinfo{title}{Ethics Guidelines for Trustworthy AI}.
\newblock
  \bibinfo{howpublished}{\url{https://ec.europa.eu/digital-single-market/en/news/ethics-guidelines-trustworthy-ai}}.
\newblock


\bibitem[\protect\citeauthoryear{Ewusi-Mensah}{Ewusi-Mensah}{2003}]%
        {ewusi2003software}
\bibfield{author}{\bibinfo{person}{Kweku Ewusi-Mensah}.}
  \bibinfo{year}{2003}\natexlab{}.
\newblock \bibinfo{booktitle}{\emph{Software development failures}}.
\newblock \bibinfo{publisher}{Mit Press}.
\newblock


\bibitem[\protect\citeauthoryear{{Executive Office of the President of the
  United States}}{{Executive Office of the President of the United
  States}}{2019}]%
        {eo13859}
\bibfield{author}{\bibinfo{person}{{Executive Office of the President of the
  United States}}.} \bibinfo{year}{2019}\natexlab{}.
\newblock \bibinfo{title}{{Executive Order 13859: Maintaining American
  Leadership in Artificial Intelligence}}.
\newblock
  \bibinfo{howpublished}{\url{https://www.govinfo.gov/app/details/DCPD-201900073}}.
\newblock


\bibitem[\protect\citeauthoryear{{Executive Office of the President of the
  United States}}{{Executive Office of the President of the United
  States}}{2020}]%
        {eo13960}
\bibfield{author}{\bibinfo{person}{{Executive Office of the President of the
  United States}}.} \bibinfo{year}{2020}\natexlab{}.
\newblock \bibinfo{title}{{Executive Order 13960: Promoting the Use of
  Trustworthy Artificial Intelligence in the Federal Government}}.
\newblock
  \bibinfo{howpublished}{\url{https://www.govinfo.gov/app/details/DCPD-202000870}}.
\newblock


\bibitem[\protect\citeauthoryear{Feigenbaum, Jaggard, Wright, and
  Xiao}{Feigenbaum et~al\mbox{.}}{2012}]%
        {feigenbaum2012systematizing}
\bibfield{author}{\bibinfo{person}{Joan Feigenbaum}, \bibinfo{person}{Aaron~D
  Jaggard}, \bibinfo{person}{Rebecca~N Wright}, {and} \bibinfo{person}{Hongda
  Xiao}.} \bibinfo{year}{2012}\natexlab{}.
\newblock \bibinfo{booktitle}{\emph{Systematizing “Accountability” in
  Computer Science (Version of Feb. 17, 2012)}}.
\newblock \bibinfo{type}{{T}echnical {R}eport} YALEU/DCS/TR-1452.
  \bibinfo{institution}{Yale University, New Haven, CT}.
\newblock


\bibitem[\protect\citeauthoryear{Fjeld, Achten, Hilligoss, Nagy, and
  Srikumar}{Fjeld et~al\mbox{.}}{2020}]%
        {fjeld2020principled}
\bibfield{author}{\bibinfo{person}{Jessica Fjeld}, \bibinfo{person}{Nele
  Achten}, \bibinfo{person}{Hannah Hilligoss}, \bibinfo{person}{Adam Nagy},
  {and} \bibinfo{person}{Madhulika Srikumar}.} \bibinfo{year}{2020}\natexlab{}.
\newblock \showarticletitle{Principled artificial intelligence: Mapping
  consensus in ethical and rights-based approaches to principles for AI}.
\newblock \bibinfo{journal}{\emph{Berkman Klein Center Research Publication}}
  \bibinfo{number}{2020-1} (\bibinfo{year}{2020}).
\newblock


\bibitem[\protect\citeauthoryear{Flanagan, Howe, and Nissenbaum}{Flanagan
  et~al\mbox{.}}{2005}]%
        {flanagan2005values}
\bibfield{author}{\bibinfo{person}{Mary Flanagan}, \bibinfo{person}{Daniel~C.
  Howe}, {and} \bibinfo{person}{Helen Nissenbaum}.}
  \bibinfo{year}{2005}\natexlab{}.
\newblock \showarticletitle{Values at Play: Design Tradeoffs in
  Socially-oriented Game Design}. In \bibinfo{booktitle}{\emph{Proceedings of
  the SIGCHI Conference on Human Factors in Computing Systems}}
  \emph{(\bibinfo{series}{CHI '05})}. \bibinfo{publisher}{ACM},
  \bibinfo{address}{New York, NY, USA}, \bibinfo{pages}{751--760}.
\newblock
\showISBNx{1-58113-998-5}


\bibitem[\protect\citeauthoryear{Floridi}{Floridi}{2019}]%
        {floridi2019establishing}
\bibfield{author}{\bibinfo{person}{Luciano Floridi}.}
  \bibinfo{year}{2019}\natexlab{}.
\newblock \showarticletitle{Establishing the rules for building trustworthy
  AI}.
\newblock \bibinfo{journal}{\emph{Nature Machine Intelligence}}
  \bibinfo{volume}{1}, \bibinfo{number}{6} (\bibinfo{year}{2019}),
  \bibinfo{pages}{261--262}.
\newblock


\bibitem[\protect\citeauthoryear{Friedman}{Friedman}{1996}]%
        {friedman1996value}
\bibfield{author}{\bibinfo{person}{Batya Friedman}.}
  \bibinfo{year}{1996}\natexlab{}.
\newblock \showarticletitle{Value-sensitive design}.
\newblock \bibinfo{journal}{\emph{interactions}} \bibinfo{volume}{3},
  \bibinfo{number}{6} (\bibinfo{year}{1996}), \bibinfo{pages}{16--23}.
\newblock


\bibitem[\protect\citeauthoryear{Friedman, Kahn, and Borning}{Friedman
  et~al\mbox{.}}{2008}]%
        {friedman2008}
\bibfield{author}{\bibinfo{person}{Batya Friedman}, \bibinfo{person}{Peter~H.
  Kahn}, {and} \bibinfo{person}{Alan Borning}.}
  \bibinfo{year}{2008}\natexlab{}.
\newblock \showarticletitle{{Value Sensitive Design and Information Systems}}.
\newblock In \bibinfo{booktitle}{\emph{The Handbook of Information and Computer
  Ethics}}, \bibfield{editor}{\bibinfo{person}{Kenneth~Einar Himma} {and}
  \bibinfo{person}{Herman~T. Tavani}} (Eds.). \bibinfo{publisher}{John Wiley
  {\&} Sons, Inc.}, Chapter~4, \bibinfo{pages}{69--101}.
\newblock
\showISBNx{9400778430}
\showISSN{10725520}


\bibitem[\protect\citeauthoryear{Friedman, Thomas, Grudin, Nass, Nissenbaum,
  Schlager, and Shneiderman}{Friedman et~al\mbox{.}}{1999}]%
        {friedman1999trust}
\bibfield{author}{\bibinfo{person}{Batya Friedman}, \bibinfo{person}{John~C
  Thomas}, \bibinfo{person}{Jonathan Grudin}, \bibinfo{person}{Clifford Nass},
  \bibinfo{person}{Helen Nissenbaum}, \bibinfo{person}{Mark Schlager}, {and}
  \bibinfo{person}{Ben Shneiderman}.} \bibinfo{year}{1999}\natexlab{}.
\newblock \showarticletitle{Trust me, I'm accountable: trust and accountability
  online}. In \bibinfo{booktitle}{\emph{CHI'99 Extended Abstracts on Human
  Factors in Computing Systems}}. ACM, \bibinfo{pages}{79--80}.
\newblock


\bibitem[\protect\citeauthoryear{Gawande}{Gawande}{2009}]%
        {gawande2009checklist}
\bibfield{author}{\bibinfo{person}{Atul Gawande}.}
  \bibinfo{year}{2009}\natexlab{}.
\newblock \bibinfo{booktitle}{\emph{{The Checklist Manifesto: How to Get Things
  Right}}}.
\newblock \bibinfo{publisher}{Metropolitan Books}.
\newblock


\bibitem[\protect\citeauthoryear{Gebru, Morgenstern, Vecchione, Vaughan,
  Wallach, Daum{\'e}~III, and Crawford}{Gebru et~al\mbox{.}}{2018}]%
        {gebru2018datasheets}
\bibfield{author}{\bibinfo{person}{Timnit Gebru}, \bibinfo{person}{Jamie
  Morgenstern}, \bibinfo{person}{Briana Vecchione},
  \bibinfo{person}{Jennifer~Wortman Vaughan}, \bibinfo{person}{Hanna Wallach},
  \bibinfo{person}{Hal Daum{\'e}~III}, {and} \bibinfo{person}{Kate Crawford}.}
  \bibinfo{year}{2018}\natexlab{}.
\newblock \showarticletitle{Datasheets for datasets}.
\newblock \bibinfo{journal}{\emph{arXiv preprint arXiv:1803.09010}}
  (\bibinfo{year}{2018}).
\newblock


\bibitem[\protect\citeauthoryear{Geiger and Halfaker}{Geiger and
  Halfaker}{2017}]%
        {geiger2017}
\bibfield{author}{\bibinfo{person}{R.~Stuart Geiger} {and}
  \bibinfo{person}{Aaron Halfaker}.} \bibinfo{year}{2017}\natexlab{}.
\newblock \showarticletitle{{Operationalizing Conflict and Cooperation between
  Automated Software Agents in Wikipedia: A Replication and Expansion of "Even
  Good Bots Fight"}}.
\newblock \bibinfo{journal}{\emph{Proceedings of the ACM on Human-Computer
  Interaction}} \bibinfo{volume}{1}, \bibinfo{number}{CSCW}
  (\bibinfo{date}{dec} \bibinfo{year}{2017}), \bibinfo{pages}{1--33}.
\newblock
\showISSN{25730142}


\bibitem[\protect\citeauthoryear{Gellman}{Gellman}{2017}]%
        {gellman2017fair}
\bibfield{author}{\bibinfo{person}{Robert Gellman}.}
  \bibinfo{year}{2017}\natexlab{}.
\newblock \showarticletitle{Fair information practices: A basic history}.
\newblock \bibinfo{journal}{\emph{SSRN, 2415020}} (\bibinfo{year}{2017}).
\newblock


\bibitem[\protect\citeauthoryear{Gelperin and Hetzel}{Gelperin and
  Hetzel}{1988}]%
        {gelperin1988growth}
\bibfield{author}{\bibinfo{person}{David Gelperin} {and} \bibinfo{person}{Bill
  Hetzel}.} \bibinfo{year}{1988}\natexlab{}.
\newblock \showarticletitle{The growth of software testing}.
\newblock \bibinfo{journal}{\emph{Commun. ACM}} \bibinfo{volume}{31},
  \bibinfo{number}{6} (\bibinfo{year}{1988}), \bibinfo{pages}{687--695}.
\newblock


\bibitem[\protect\citeauthoryear{Gillespie}{Gillespie}{2006}]%
        {gillespie2006engineering}
\bibfield{author}{\bibinfo{person}{Tarleton Gillespie}.}
  \bibinfo{year}{2006}\natexlab{}.
\newblock \showarticletitle{Engineering a Principle: ‘End-to-End’in the
  Design of the Internet}.
\newblock \bibinfo{journal}{\emph{Social Studies of Science}}
  \bibinfo{volume}{36}, \bibinfo{number}{3} (\bibinfo{year}{2006}),
  \bibinfo{pages}{427--457}.
\newblock


\bibitem[\protect\citeauthoryear{Gillespie}{Gillespie}{2010}]%
        {gillespie2010politics}
\bibfield{author}{\bibinfo{person}{Tarleton Gillespie}.}
  \bibinfo{year}{2010}\natexlab{}.
\newblock \showarticletitle{The politics of ‘platforms’}.
\newblock \bibinfo{journal}{\emph{New media \& society}} \bibinfo{volume}{12},
  \bibinfo{number}{3} (\bibinfo{year}{2010}), \bibinfo{pages}{347--364}.
\newblock


\bibitem[\protect\citeauthoryear{Gordon and Breaux}{Gordon and Breaux}{2013}]%
        {gordon2013assessing}
\bibfield{author}{\bibinfo{person}{David~G Gordon} {and}
  \bibinfo{person}{Travis~D Breaux}.} \bibinfo{year}{2013}\natexlab{}.
\newblock \showarticletitle{Assessing regulatory change through legal
  requirements coverage modeling}. In \bibinfo{booktitle}{\emph{2013 21st IEEE
  International Requirements Engineering Conference (RE)}}. IEEE,
  \bibinfo{pages}{145--154}.
\newblock


\bibitem[\protect\citeauthoryear{{Government Accountability
  Office}}{{Government Accountability Office}}{2018}]%
        {gagas-2018}
\bibfield{author}{\bibinfo{person}{{Government Accountability Office}}.}
  \bibinfo{year}{2018}\natexlab{}.
\newblock \bibinfo{title}{Government Auditing Standards, 2018 Revision}.
\newblock
\newblock


\bibitem[\protect\citeauthoryear{Guidotti, Monreale, Turini, Pedreschi, and
  Giannotti}{Guidotti et~al\mbox{.}}{2018}]%
        {guidotti2018survey}
\bibfield{author}{\bibinfo{person}{Riccardo Guidotti}, \bibinfo{person}{Anna
  Monreale}, \bibinfo{person}{Franco Turini}, \bibinfo{person}{Dino Pedreschi},
  {and} \bibinfo{person}{Fosca Giannotti}.} \bibinfo{year}{2018}\natexlab{}.
\newblock \showarticletitle{A Survey Of Methods For Explaining Black Box
  Models}.
\newblock \bibinfo{journal}{\emph{arXiv preprint arXiv:1802.01933}}
  (\bibinfo{year}{2018}).
\newblock


\bibitem[\protect\citeauthoryear{G{\"u}rses, Troncoso, and Diaz}{G{\"u}rses
  et~al\mbox{.}}{2011}]%
        {gurses2011engineering}
\bibfield{author}{\bibinfo{person}{Seda G{\"u}rses}, \bibinfo{person}{Carmela
  Troncoso}, {and} \bibinfo{person}{Claudia Diaz}.}
  \bibinfo{year}{2011}\natexlab{}.
\newblock \showarticletitle{Engineering privacy by design}.
\newblock \bibinfo{journal}{\emph{Conference on Computers, Privacy, and Data
  Protection}} (\bibinfo{year}{2011}).
\newblock


\bibitem[\protect\citeauthoryear{Habra, Le~Charlier, Mounji, and Mathieu}{Habra
  et~al\mbox{.}}{1992}]%
        {habra1992asax}
\bibfield{author}{\bibinfo{person}{Naji Habra}, \bibinfo{person}{Baudouin
  Le~Charlier}, \bibinfo{person}{Abdelaziz Mounji}, {and}
  \bibinfo{person}{Isabelle Mathieu}.} \bibinfo{year}{1992}\natexlab{}.
\newblock \showarticletitle{ASAX: Software architecture and rule-based language
  for universal audit trail analysis}.
\newblock In \bibinfo{booktitle}{\emph{Computer Security—ESORICS 92}}.
  \bibinfo{publisher}{Springer}, \bibinfo{pages}{435--450}.
\newblock


\bibitem[\protect\citeauthoryear{Haeberlen}{Haeberlen}{2010}]%
        {haeberlen2010case}
\bibfield{author}{\bibinfo{person}{Andreas Haeberlen}.}
  \bibinfo{year}{2010}\natexlab{}.
\newblock \showarticletitle{A case for the accountable cloud}.
\newblock \bibinfo{journal}{\emph{ACM SIGOPS Operating Systems Review}}
  \bibinfo{volume}{44}, \bibinfo{number}{2} (\bibinfo{year}{2010}),
  \bibinfo{pages}{52--57}.
\newblock


\bibitem[\protect\citeauthoryear{Haeberlen, Aditya, Rodrigues, and
  Druschel}{Haeberlen et~al\mbox{.}}{2010}]%
        {haeberlen2010accountable}
\bibfield{author}{\bibinfo{person}{Andreas Haeberlen},
  \bibinfo{person}{Paarijaat Aditya}, \bibinfo{person}{Rodrigo Rodrigues},
  {and} \bibinfo{person}{Peter Druschel}.} \bibinfo{year}{2010}\natexlab{}.
\newblock \showarticletitle{Accountable Virtual Machines.}. In
  \bibinfo{booktitle}{\emph{OSDI}}. \bibinfo{pages}{119--134}.
\newblock


\bibitem[\protect\citeauthoryear{Haeberlen, Kouznetsov, and Druschel}{Haeberlen
  et~al\mbox{.}}{2007}]%
        {haeberlen2007peerreview}
\bibfield{author}{\bibinfo{person}{Andreas Haeberlen}, \bibinfo{person}{Petr
  Kouznetsov}, {and} \bibinfo{person}{Peter Druschel}.}
  \bibinfo{year}{2007}\natexlab{}.
\newblock \showarticletitle{PeerReview: Practical accountability for
  distributed systems}. In \bibinfo{booktitle}{\emph{ACM SIGOPS Operating
  Systems Review}}, Vol.~\bibinfo{volume}{41:6}. ACM,
  \bibinfo{pages}{175--188}.
\newblock


\bibitem[\protect\citeauthoryear{Hall}{Hall}{2010}]%
        {hall2010election}
\bibfield{author}{\bibinfo{person}{Joseph~Lorenzo Hall}.}
  \bibinfo{year}{2010}\natexlab{}.
\newblock \bibinfo{title}{Election Auditing Bibliography}.
\newblock
\newblock
\newblock
\shownote{\url{https://josephhall.org/papers/auditing_biblio.pdf}.}


\bibitem[\protect\citeauthoryear{Halpern and Pearl}{Halpern and Pearl}{2005a}]%
        {halpern2005causes-i}
\bibfield{author}{\bibinfo{person}{Joseph~Y Halpern} {and}
  \bibinfo{person}{Judea Pearl}.} \bibinfo{year}{2005}\natexlab{a}.
\newblock \showarticletitle{Causes and explanations: A structural-model
  approach. Part I: Causes}.
\newblock \bibinfo{journal}{\emph{The British journal for the philosophy of
  science}} \bibinfo{volume}{56}, \bibinfo{number}{4} (\bibinfo{year}{2005}),
  \bibinfo{pages}{843--887}.
\newblock


\bibitem[\protect\citeauthoryear{Halpern and Pearl}{Halpern and Pearl}{2005b}]%
        {halpern2005causes-ii}
\bibfield{author}{\bibinfo{person}{Joseph~Y Halpern} {and}
  \bibinfo{person}{Judea Pearl}.} \bibinfo{year}{2005}\natexlab{b}.
\newblock \showarticletitle{Causes and explanations: A structural-model
  approach. Part II: Explanations}.
\newblock \bibinfo{journal}{\emph{The British Journal for the Philosophy of
  Science}} \bibinfo{volume}{56}, \bibinfo{number}{4} (\bibinfo{year}{2005}),
  \bibinfo{pages}{889--911}.
\newblock


\bibitem[\protect\citeauthoryear{Hannak, Sapiezynski, Molavi~Kakhki,
  Krishnamurthy, Lazer, Mislove, and Wilson}{Hannak et~al\mbox{.}}{2013}]%
        {hannak2013measuring}
\bibfield{author}{\bibinfo{person}{Aniko Hannak}, \bibinfo{person}{Piotr
  Sapiezynski}, \bibinfo{person}{Arash Molavi~Kakhki},
  \bibinfo{person}{Balachander Krishnamurthy}, \bibinfo{person}{David Lazer},
  \bibinfo{person}{Alan Mislove}, {and} \bibinfo{person}{Christo Wilson}.}
  \bibinfo{year}{2013}\natexlab{}.
\newblock \showarticletitle{Measuring personalization of web search}. In
  \bibinfo{booktitle}{\emph{Proceedings of the 22nd international conference on
  World Wide Web}}. ACM.
\newblock


\bibitem[\protect\citeauthoryear{Hannak, Soeller, Lazer, Mislove, and
  Wilson}{Hannak et~al\mbox{.}}{[n.d.]}]%
        {hannak2014measuring}
\bibfield{author}{\bibinfo{person}{Aniko Hannak}, \bibinfo{person}{Gary
  Soeller}, \bibinfo{person}{David Lazer}, \bibinfo{person}{Alan Mislove},
  {and} \bibinfo{person}{Christo Wilson}.} \bibinfo{year}{[n.d.]}\natexlab{}.
\newblock \showarticletitle{Measuring price discrimination and steering on
  e-commerce web sites}. In \bibinfo{booktitle}{\emph{Proceedings of the 2014
  conference on internet measurement conference}}. ACM,
  \bibinfo{pages}{305--318}.
\newblock


\bibitem[\protect\citeauthoryear{Hedstrom}{Hedstrom}{1997}]%
        {hedstrom1997digital}
\bibfield{author}{\bibinfo{person}{Margaret Hedstrom}.}
  \bibinfo{year}{1997}\natexlab{}.
\newblock \showarticletitle{Digital preservation: a time bomb for digital
  libraries}.
\newblock \bibinfo{journal}{\emph{Computers and the Humanities}}
  \bibinfo{volume}{31}, \bibinfo{number}{3} (\bibinfo{year}{1997}),
  \bibinfo{pages}{189}.
\newblock


\bibitem[\protect\citeauthoryear{Helman and Liepins}{Helman and
  Liepins}{1993}]%
        {helman1993statistical}
\bibfield{author}{\bibinfo{person}{Paul Helman} {and} \bibinfo{person}{Gunar
  Liepins}.} \bibinfo{year}{1993}\natexlab{}.
\newblock \showarticletitle{Statistical foundations of audit trail analysis for
  the detection of computer misuse}.
\newblock \bibinfo{journal}{\emph{Software Engineering, IEEE Transactions on}}
  \bibinfo{volume}{19}, \bibinfo{number}{9} (\bibinfo{year}{1993}),
  \bibinfo{pages}{886--901}.
\newblock


\bibitem[\protect\citeauthoryear{Herschel, Diestelk{\"a}mper, and
  Lahmar}{Herschel et~al\mbox{.}}{2017}]%
        {herschel2017survey}
\bibfield{author}{\bibinfo{person}{Melanie Herschel}, \bibinfo{person}{Ralf
  Diestelk{\"a}mper}, {and} \bibinfo{person}{Houssem~Ben Lahmar}.}
  \bibinfo{year}{2017}\natexlab{}.
\newblock \showarticletitle{A survey on provenance: {What for? What form? What
  from?}}
\newblock \bibinfo{journal}{\emph{The VLDB Journal}} \bibinfo{volume}{26},
  \bibinfo{number}{6} (\bibinfo{year}{2017}), \bibinfo{pages}{881--906}.
\newblock


\bibitem[\protect\citeauthoryear{Holland, Hosny, Newman, Joseph, and
  Chmielinski}{Holland et~al\mbox{.}}{2018}]%
        {holland2018dataset}
\bibfield{author}{\bibinfo{person}{Sarah Holland}, \bibinfo{person}{Ahmed
  Hosny}, \bibinfo{person}{Sarah Newman}, \bibinfo{person}{Joshua Joseph},
  {and} \bibinfo{person}{Kasia Chmielinski}.} \bibinfo{year}{2018}\natexlab{}.
\newblock \showarticletitle{The dataset nutrition label: A framework to drive
  higher data quality standards}.
\newblock \bibinfo{journal}{\emph{arXiv preprint arXiv:1805.03677}}
  (\bibinfo{year}{2018}).
\newblock


\bibitem[\protect\citeauthoryear{Huang}{Huang}{2009}]%
        {huang2009aviation}
\bibfield{author}{\bibinfo{person}{Jiefang Huang}.}
  \bibinfo{year}{2009}\natexlab{}.
\newblock \bibinfo{booktitle}{\emph{Aviation safety through the rule of law:
  ICAO's mechanisms and practices}}.
\newblock \bibinfo{publisher}{Kluwer Law International BV}.
\newblock


\bibitem[\protect\citeauthoryear{Irani and Silberman}{Irani and
  Silberman}{2014}]%
        {irani2014critical}
\bibfield{author}{\bibinfo{person}{Lilly Irani} {and} \bibinfo{person}{M.~Six
  Silberman}.} \bibinfo{year}{2014}\natexlab{}.
\newblock \showarticletitle{{From critical design to critical infrastructure}}.
\newblock \bibinfo{journal}{\emph{Interactions}} \bibinfo{volume}{21},
  \bibinfo{number}{4} (\bibinfo{year}{2014}), \bibinfo{pages}{32--35}.
\newblock
\showISSN{10725520}


\bibitem[\protect\citeauthoryear{ISO}{ISO}{2012}]%
        {iso16363}
\bibfield{author}{\bibinfo{person}{ISO}.} \bibinfo{year}{2012}\natexlab{}.
\newblock \bibinfo{title}{Certification of Trustworthy Digital Repositories}.
\newblock
\newblock


\bibitem[\protect\citeauthoryear{Jabbra and Dwivedi}{Jabbra and
  Dwivedi}{1989}]%
        {jabbra1989public}
\bibfield{editor}{\bibinfo{person}{J.G. Jabbra} {and} \bibinfo{person}{O.P.
  Dwivedi}} (Eds.). \bibinfo{year}{1989}\natexlab{}.
\newblock \bibinfo{booktitle}{\emph{Public Service Accountability: A
  Comparative Perspective}}.
\newblock \bibinfo{publisher}{Kumarian Press}.
\newblock
\showISBNx{9780931816413}
\showLCCN{88013220}


\bibitem[\protect\citeauthoryear{Jackson, Gillespie, and Payette}{Jackson
  et~al\mbox{.}}{2014}]%
        {jackson2014policy}
\bibfield{author}{\bibinfo{person}{Steven~J. Jackson},
  \bibinfo{person}{Tarleton Gillespie}, {and} \bibinfo{person}{Sandy Payette}.}
  \bibinfo{year}{2014}\natexlab{}.
\newblock \showarticletitle{{The Policy Knot: Re-integrating Policy, Practice
  and Design in CSCW Studies of Social Computing}}. In
  \bibinfo{booktitle}{\emph{Proceedings of the 17th ACM Conference on Computer
  Supported Cooperative Work \& Social Computing}} \emph{(\bibinfo{series}{CSCW
  '14})}. \bibinfo{publisher}{ACM}, \bibinfo{address}{New York, NY, USA},
  \bibinfo{pages}{588--602}.
\newblock
\showISBNx{978-1-4503-2540-0}


\bibitem[\protect\citeauthoryear{Jacobs and Wallach}{Jacobs and
  Wallach}{2021}]%
        {jacobs2018measurement}
\bibfield{author}{\bibinfo{person}{Abigail~Z. Jacobs} {and}
  \bibinfo{person}{Hanna Wallach}.} \bibinfo{year}{2021}\natexlab{}.
\newblock \showarticletitle{Measurement and Fairness}. In
  \bibinfo{booktitle}{\emph{ACM Conference on Fairness, Accountability, and
  Transparency}}.
\newblock


\bibitem[\protect\citeauthoryear{JafariNaimi, Nathan, and
  Hargraves}{JafariNaimi et~al\mbox{.}}{2015}]%
        {jafarinaimi2015}
\bibfield{author}{\bibinfo{person}{Nassim JafariNaimi}, \bibinfo{person}{Lisa
  Nathan}, {and} \bibinfo{person}{Ian Hargraves}.}
  \bibinfo{year}{2015}\natexlab{}.
\newblock \showarticletitle{{Values as Hypotheses: Design, Inquiry, and the
  Service of Values}}.
\newblock \bibinfo{journal}{\emph{Design Issues}} \bibinfo{volume}{31},
  \bibinfo{number}{4} (\bibinfo{date}{oct} \bibinfo{year}{2015}),
  \bibinfo{pages}{91--104}.
\newblock
\showISBNx{13978-3-923859-82-5}
\showISSN{0747-9360}


\bibitem[\protect\citeauthoryear{Jagadeesan, Jeffrey, Pitcher, and
  Riely}{Jagadeesan et~al\mbox{.}}{2009}]%
        {jagadeesan2009towards}
\bibfield{author}{\bibinfo{person}{Radha Jagadeesan}, \bibinfo{person}{Alan
  Jeffrey}, \bibinfo{person}{Corin Pitcher}, {and} \bibinfo{person}{James
  Riely}.} \bibinfo{year}{2009}\natexlab{}.
\newblock \showarticletitle{Towards a theory of accountability and audit}.
\newblock In \bibinfo{booktitle}{\emph{Computer Security--ESORICS 2009}}.
  \bibinfo{publisher}{Springer}, \bibinfo{pages}{152--167}.
\newblock


\bibitem[\protect\citeauthoryear{{Ka-Ping Yee}}{{Ka-Ping Yee}}{2007}]%
        {zesty}
\bibfield{author}{\bibinfo{person}{{Ka-Ping Yee}}.}
  \bibinfo{year}{2007}\natexlab{}.
\newblock \emph{\bibinfo{title}{{Building Reliable Voting Machine Software}}}.
\newblock \bibinfo{thesistype}{Ph.D. Dissertation}.
  \bibinfo{school}{{University of California}}.
\newblock
\newblock
\shownote{\url{http://zesty.ca/pubs/yee-phd.pdf}.}


\bibitem[\protect\citeauthoryear{Kalluri}{Kalluri}{2020}]%
        {kalluri2020power}
\bibfield{author}{\bibinfo{person}{Pratyusha Kalluri}.}
  \bibinfo{year}{2020}\natexlab{}.
\newblock \showarticletitle{Don't ask if artificial intelligence is good or
  fair, ask how it shifts power.}
\newblock \bibinfo{journal}{\emph{Nature}} \bibinfo{volume}{583},
  \bibinfo{number}{7815} (\bibinfo{year}{2020}), \bibinfo{pages}{169--169}.
\newblock


\bibitem[\protect\citeauthoryear{Katell, Young, Herman, Dailey, Tam, Guetler,
  Binz, Raz, and Krafft}{Katell et~al\mbox{.}}{2019}]%
        {katell2019algorithmic}
\bibfield{author}{\bibinfo{person}{Michael Katell}, \bibinfo{person}{Meg
  Young}, \bibinfo{person}{Bernease Herman}, \bibinfo{person}{Dharma Dailey},
  \bibinfo{person}{Aaron Tam}, \bibinfo{person}{Vivian Guetler},
  \bibinfo{person}{Corinne Binz}, \bibinfo{person}{Daniella Raz}, {and}
  \bibinfo{person}{PM Krafft}.} \bibinfo{year}{2019}\natexlab{}.
\newblock \showarticletitle{An Algorithmic Equity Toolkit for Technology Audits
  by Community Advocates and Activists}.
\newblock \bibinfo{journal}{\emph{arXiv preprint arXiv:1912.02943}}
  (\bibinfo{year}{2019}).
\newblock


\bibitem[\protect\citeauthoryear{Kearns, Neel, Roth, and Wu}{Kearns
  et~al\mbox{.}}{2017}]%
        {kearns2017gerrymandering}
\bibfield{author}{\bibinfo{person}{Michael Kearns}, \bibinfo{person}{Seth
  Neel}, \bibinfo{person}{Aaron Roth}, {and} \bibinfo{person}{Zhiwei~Steven
  Wu}.} \bibinfo{year}{2017}\natexlab{}.
\newblock \showarticletitle{Preventing Fairness Gerrymandering: Auditing and
  Learning for Subgroup Fairness}.
\newblock \bibinfo{journal}{\emph{arXiv preprint arXiv:1711.05144}}
  (\bibinfo{year}{2017}).
\newblock


\bibitem[\protect\citeauthoryear{Kenthapadi, Mishra, and Nissim}{Kenthapadi
  et~al\mbox{.}}{2005}]%
        {kenthapadi2005simulatable}
\bibfield{author}{\bibinfo{person}{Krishnaram Kenthapadi},
  \bibinfo{person}{Nina Mishra}, {and} \bibinfo{person}{Kobbi Nissim}.}
  \bibinfo{year}{2005}\natexlab{}.
\newblock \showarticletitle{Simulatable auditing}. In
  \bibinfo{booktitle}{\emph{Proceedings of the twenty-fourth ACM
  SIGMOD-SIGACT-SIGART symposium on Principles of database systems}}. ACM,
  \bibinfo{pages}{118--127}.
\newblock


\bibitem[\protect\citeauthoryear{Kim}{Kim}{2017}]%
        {kim2017auditing}
\bibfield{author}{\bibinfo{person}{Pauline Kim}.}
  \bibinfo{year}{2017}\natexlab{}.
\newblock \showarticletitle{Auditing Algorithms for Discrimination}.
\newblock \bibinfo{journal}{\emph{University of Pennsylvania Law Review
  Online}} \bibinfo{volume}{166}, \bibinfo{number}{189} (\bibinfo{year}{2017}).
\newblock


\bibitem[\protect\citeauthoryear{Kirchner}{Kirchner}{2017}]%
        {kirchner2017fst}
\bibfield{author}{\bibinfo{person}{Lauren Kirchner}.}
  \bibinfo{year}{2017}\natexlab{}.
\newblock \showarticletitle{Thousands of Criminal Cases in New York Relied on
  Disputed DNA Testing Techniques}.
\newblock \bibinfo{journal}{\emph{{ProPublica}}} (\bibinfo{date}{4 Sept.}
  \bibinfo{year}{2017}).
\newblock


\bibitem[\protect\citeauthoryear{Knobel and Bowker}{Knobel and Bowker}{2011}]%
        {knobel2011}
\bibfield{author}{\bibinfo{person}{Cory Knobel} {and}
  \bibinfo{person}{Geoffrey~C. Bowker}.} \bibinfo{year}{2011}\natexlab{}.
\newblock \showarticletitle{{Values in design}}.
\newblock \bibinfo{journal}{\emph{Commun. ACM}}  \bibinfo{volume}{54}
  (\bibinfo{year}{2011}), \bibinfo{pages}{26}.
\newblock
\showISSN{00010782}


\bibitem[\protect\citeauthoryear{Kroll}{Kroll}{2015}]%
        {kroll2015accountable}
\bibfield{author}{\bibinfo{person}{Joshua~A. Kroll}.}
  \bibinfo{year}{2015}\natexlab{}.
\newblock \emph{\bibinfo{title}{Accountable Algorithms}}.
\newblock \bibinfo{thesistype}{Ph.D. Dissertation}. \bibinfo{school}{Princeton
  University}.
\newblock


\bibitem[\protect\citeauthoryear{Kroll}{Kroll}{2018}]%
        {kroll2018fallacy}
\bibfield{author}{\bibinfo{person}{Joshua~A Kroll}.}
  \bibinfo{year}{2018}\natexlab{}.
\newblock \showarticletitle{The fallacy of inscrutability}.
\newblock \bibinfo{journal}{\emph{Phil. Trans. R. Soc. A}}
  \bibinfo{volume}{376}, \bibinfo{number}{2133} (\bibinfo{year}{2018}), 14.
\newblock


\bibitem[\protect\citeauthoryear{Kroll}{Kroll}{2020}]%
        {kroll2020accountability}
\bibfield{author}{\bibinfo{person}{Joshua~A. Kroll}.}
  \bibinfo{year}{2020}\natexlab{}.
\newblock \showarticletitle{Accountability in Computer Systems}.
\newblock In \bibinfo{booktitle}{\emph{The Oxford Handbook of the Ethics of
  Artificial Intelligence}}, \bibfield{editor}{\bibinfo{person}{Markus Dubber},
  \bibinfo{person}{Frank Pasquale}, {and} \bibinfo{person}{Sunit Das}} (Eds.).
  \bibinfo{publisher}{Oxford University Press}, \bibinfo{address}{Oxford, UK},
  \bibinfo{pages}{181--196}.
\newblock


\bibitem[\protect\citeauthoryear{Kroll, Huey, Barocas, Felten, Reidenberg,
  Robinson, and Yu.}{Kroll et~al\mbox{.}}{2017}]%
        {kroll2017penn}
\bibfield{author}{\bibinfo{person}{Joshua~A. Kroll}, \bibinfo{person}{Joanna
  Huey}, \bibinfo{person}{Solon Barocas}, \bibinfo{person}{Edward~W. Felten},
  \bibinfo{person}{Joel~R. Reidenberg}, \bibinfo{person}{David~G. Robinson},
  {and} \bibinfo{person}{Harlan Yu.}} \bibinfo{year}{2017}\natexlab{}.
\newblock \showarticletitle{Accountable Algorithms}.
\newblock \bibinfo{journal}{\emph{University of Pennsylvania Law Review (to
  appear)}}  \bibinfo{volume}{165} (\bibinfo{year}{2017}),
  \bibinfo{pages}{633--705}.
\newblock
Issue 3.


\bibitem[\protect\citeauthoryear{K{\"u}nnemann, Esiyok, and
  Backes}{K{\"u}nnemann et~al\mbox{.}}{2019}]%
        {kunnemann2019automated}
\bibfield{author}{\bibinfo{person}{Robert K{\"u}nnemann},
  \bibinfo{person}{Ilkan Esiyok}, {and} \bibinfo{person}{Michael Backes}.}
  \bibinfo{year}{2019}\natexlab{}.
\newblock \showarticletitle{Automated Verification of Accountability in
  Security Protocols}. In \bibinfo{booktitle}{\emph{2019 IEEE 32nd Computer
  Security Foundations Symposium (CSF)}}. IEEE, \bibinfo{pages}{397--39716}.
\newblock


\bibitem[\protect\citeauthoryear{K{\"u}sters, Truderung, and Vogt}{K{\"u}sters
  et~al\mbox{.}}{2010}]%
        {kuesters2010accountability}
\bibfield{author}{\bibinfo{person}{Ralf K{\"u}sters}, \bibinfo{person}{Tomasz
  Truderung}, {and} \bibinfo{person}{Andreas Vogt}.}
  \bibinfo{year}{2010}\natexlab{}.
\newblock \showarticletitle{Accountability: definition and relationship to
  verifiability}. In \bibinfo{booktitle}{\emph{Proc. 17th ACM conf. Computer
  and Communications Security}}. ACM, \bibinfo{pages}{526--535}.
\newblock


\bibitem[\protect\citeauthoryear{Kwong}{Kwong}{2017}]%
        {kwong2017algorithm}
\bibfield{author}{\bibinfo{person}{Katherine Kwong}.}
  \bibinfo{year}{2017}\natexlab{}.
\newblock \showarticletitle{The Algorithm says you did it: The use of Black Box
  Algorithms to analyze complex DNA evidence}.
\newblock \bibinfo{journal}{\emph{Harv. JL \& Tech.}}  \bibinfo{volume}{31}
  (\bibinfo{year}{2017}), \bibinfo{pages}{275}.
\newblock


\bibitem[\protect\citeauthoryear{Lamb, Levson, Rizzolo, and Cascadian}{Lamb
  et~al\mbox{.}}{2020}]%
        {reproducible-builds}
\bibfield{author}{\bibinfo{person}{Chris Lamb}, \bibinfo{person}{Holger
  Levson}, \bibinfo{person}{Mattia Rizzolo}, {and} \bibinfo{person}{Vagrant
  Cascadian}.} \bibinfo{year}{2020}\natexlab{}.
\newblock \bibinfo{title}{Reproducible Builds}.
\newblock \bibinfo{howpublished}{\url{https://reproducible-builds.org}}.
\newblock


\bibitem[\protect\citeauthoryear{Laurie, Langley, and Kasper}{Laurie
  et~al\mbox{.}}{2013}]%
        {laurie2013certificate}
\bibfield{author}{\bibinfo{person}{Ben Laurie}, \bibinfo{person}{Adam Langley},
  {and} \bibinfo{person}{Emilia Kasper}.} \bibinfo{year}{2013}\natexlab{}.
\newblock \bibinfo{booktitle}{\emph{Certificate Transparency}}.
\newblock \bibinfo{type}{{T}echnical {R}eport} RFC 6962.
  \bibinfo{institution}{Internet Engineering Task Force}.
\newblock


\bibitem[\protect\citeauthoryear{{Le Dantec}, Poole, and Wyche}{{Le Dantec}
  et~al\mbox{.}}{2009}]%
        {ledantec2009}
\bibfield{author}{\bibinfo{person}{Christopher~A. {Le Dantec}},
  \bibinfo{person}{Erika~Shehan Poole}, {and} \bibinfo{person}{Susan~P.
  Wyche}.} \bibinfo{year}{2009}\natexlab{}.
\newblock \showarticletitle{{Values as lived experience: Evolving value
  sensitive design in support of value discovery}}. In
  \bibinfo{booktitle}{\emph{Proceedings of the 27th international conference on
  Human factors in computing systems - CHI 09}}. \bibinfo{publisher}{ACM
  Press}, \bibinfo{address}{New York, New York, USA}, \bibinfo{pages}{1141}.
\newblock
\showISBNx{9781605582467}
\showISSN{9781605582467}


\bibitem[\protect\citeauthoryear{Lee and Baykal}{Lee and Baykal}{2017}]%
        {lee2017algorithmic}
\bibfield{author}{\bibinfo{person}{Min~Kyung Lee} {and} \bibinfo{person}{Su
  Baykal}.} \bibinfo{year}{2017}\natexlab{}.
\newblock \showarticletitle{Algorithmic Mediation in Group Decisions: Fairness
  Perceptions of Algorithmically Mediated vs. Discussion-Based Social
  Division}. In \bibinfo{booktitle}{\emph{Proceedings of the 2017 ACM
  Conference on Computer Supported Cooperative Work and Social Computing}}
  \emph{(\bibinfo{series}{CSCW '17})}. \bibinfo{publisher}{ACM},
  \bibinfo{address}{New York, NY, USA}, \bibinfo{pages}{1035--1048}.
\newblock
\showISBNx{978-1-4503-4335-0}


\bibitem[\protect\citeauthoryear{Lee, Grgi{\'c}-Hla{\v{c}}a, Tschantz, Binns,
  Weller, Carney, and Inkpen}{Lee et~al\mbox{.}}{2020}]%
        {lee2020human}
\bibfield{author}{\bibinfo{person}{Min~Kyung Lee}, \bibinfo{person}{Nina
  Grgi{\'c}-Hla{\v{c}}a}, \bibinfo{person}{Michael~Carl Tschantz},
  \bibinfo{person}{Reuben Binns}, \bibinfo{person}{Adrian Weller},
  \bibinfo{person}{Michelle Carney}, {and} \bibinfo{person}{Kori Inkpen}.}
  \bibinfo{year}{2020}\natexlab{}.
\newblock \showarticletitle{Human-Centered Approaches to Fair and Responsible
  AI}. In \bibinfo{booktitle}{\emph{Extended Abstracts of the 2020 CHI
  Conference on Human Factors in Computing Systems}}. \bibinfo{pages}{1--8}.
\newblock


\bibitem[\protect\citeauthoryear{Lee, Jain, Cha, Ojha, and Kusbit}{Lee
  et~al\mbox{.}}{2019a}]%
        {lee2019procedural}
\bibfield{author}{\bibinfo{person}{Min~Kyung Lee}, \bibinfo{person}{Anuraag
  Jain}, \bibinfo{person}{Hea~Jin Cha}, \bibinfo{person}{Shashank Ojha}, {and}
  \bibinfo{person}{Daniel Kusbit}.} \bibinfo{year}{2019}\natexlab{a}.
\newblock \showarticletitle{Procedural justice in algorithmic fairness:
  Leveraging transparency and outcome control for fair algorithmic mediation}.
\newblock \bibinfo{journal}{\emph{Proceedings of the ACM on Human-Computer
  Interaction}} \bibinfo{volume}{3}, \bibinfo{number}{CSCW}
  (\bibinfo{year}{2019}), \bibinfo{pages}{1--26}.
\newblock


\bibitem[\protect\citeauthoryear{Lee, Kusbit, Kahng, Kim, Yuan, Chan, See,
  Noothigattu, Lee, Psomas, et~al\mbox{.}}{Lee et~al\mbox{.}}{2019b}]%
        {lee2019webuildai}
\bibfield{author}{\bibinfo{person}{Min~Kyung Lee}, \bibinfo{person}{Daniel
  Kusbit}, \bibinfo{person}{Anson Kahng}, \bibinfo{person}{Ji~Tae Kim},
  \bibinfo{person}{Xinran Yuan}, \bibinfo{person}{Allissa Chan},
  \bibinfo{person}{Daniel See}, \bibinfo{person}{Ritesh Noothigattu},
  \bibinfo{person}{Siheon Lee}, \bibinfo{person}{Alexandros Psomas},
  {et~al\mbox{.}}} \bibinfo{year}{2019}\natexlab{b}.
\newblock \showarticletitle{WeBuildAI: Participatory framework for algorithmic
  governance}.
\newblock \bibinfo{journal}{\emph{Proceedings of the ACM on Human-Computer
  Interaction}} \bibinfo{volume}{3}, \bibinfo{number}{CSCW}
  (\bibinfo{year}{2019}), \bibinfo{pages}{1--35}.
\newblock


\bibitem[\protect\citeauthoryear{{Leveraging Data as a Strategic Asset
  Cross-Agency Priority Team}}{{Leveraging Data as a Strategic Asset
  Cross-Agency Priority Team}}{2020}]%
        {federal-data-strategy-2020-action-plan}
\bibfield{author}{\bibinfo{person}{{Leveraging Data as a Strategic Asset
  Cross-Agency Priority Team}}.} \bibinfo{year}{2020}\natexlab{}.
\newblock \bibinfo{title}{{Federal Data Strategy: 2020 Action Plan}}.
\newblock \bibinfo{howpublished}{\url{https://strategy.data.gov/action-plan/}}.
\newblock


\bibitem[\protect\citeauthoryear{Leveson, Pinnel, Sandys, Koga, and
  Reese}{Leveson et~al\mbox{.}}{1997}]%
        {leveson1997analyzing}
\bibfield{author}{\bibinfo{person}{Nancy Leveson}, \bibinfo{person}{L~Denise
  Pinnel}, \bibinfo{person}{Sean~David Sandys}, \bibinfo{person}{Shuichi Koga},
  {and} \bibinfo{person}{Jon~Damon Reese}.} \bibinfo{year}{1997}\natexlab{}.
\newblock \showarticletitle{Analyzing software specifications for mode
  confusion potential}. In \bibinfo{booktitle}{\emph{Proceedings of a workshop
  on human error and system development}}. Glasgow Accident Analysis Group,
  \bibinfo{pages}{132--146}.
\newblock


\bibitem[\protect\citeauthoryear{Leveson}{Leveson}{2016}]%
        {leveson2016engineering}
\bibfield{author}{\bibinfo{person}{Nancy~G Leveson}.}
  \bibinfo{year}{2016}\natexlab{}.
\newblock \bibinfo{booktitle}{\emph{Engineering a safer world: Systems thinking
  applied to safety}}.
\newblock \bibinfo{publisher}{The MIT Press}.
\newblock


\bibitem[\protect\citeauthoryear{Lipton, Chouldechova, and McAuley}{Lipton
  et~al\mbox{.}}{2017}]%
        {lipton2017does}
\bibfield{author}{\bibinfo{person}{Zachary~C Lipton},
  \bibinfo{person}{Alexandra Chouldechova}, {and} \bibinfo{person}{Julian
  McAuley}.} \bibinfo{year}{2017}\natexlab{}.
\newblock \showarticletitle{Does mitigating ML's disparate impact require
  disparate treatment?}
\newblock \bibinfo{journal}{\emph{arXiv preprint arXiv:1711.07076}}
  (\bibinfo{year}{2017}).
\newblock


\bibitem[\protect\citeauthoryear{Loeliger and McCullough}{Loeliger and
  McCullough}{2012}]%
        {loeliger2012version}
\bibfield{author}{\bibinfo{person}{Jon Loeliger} {and} \bibinfo{person}{Matthew
  McCullough}.} \bibinfo{year}{2012}\natexlab{}.
\newblock \bibinfo{booktitle}{\emph{Version Control with Git: Powerful tools
  and techniques for collaborative software development}}.
\newblock \bibinfo{publisher}{O'Reilly Media}.
\newblock


\bibitem[\protect\citeauthoryear{Lou, Caruana, and Gehrke}{Lou
  et~al\mbox{.}}{2012}]%
        {lou2012intelligible}
\bibfield{author}{\bibinfo{person}{Yin Lou}, \bibinfo{person}{Rich Caruana},
  {and} \bibinfo{person}{Johannes Gehrke}.} \bibinfo{year}{2012}\natexlab{}.
\newblock \showarticletitle{Intelligible models for classification and
  regression}. In \bibinfo{booktitle}{\emph{Proceedings of the 18th ACM SIGKDD
  international conference on Knowledge discovery and data mining}}.
  \bibinfo{pages}{150--158}.
\newblock


\bibitem[\protect\citeauthoryear{Lud{\"a}scher}{Lud{\"a}scher}{2016}]%
        {ludascher2016brief}
\bibfield{author}{\bibinfo{person}{Bertram Lud{\"a}scher}.}
  \bibinfo{year}{2016}\natexlab{}.
\newblock \showarticletitle{A brief tour through provenance in scientific
  workflows and databases}.
\newblock In \bibinfo{booktitle}{\emph{Building Trust in Information}}.
  \bibinfo{publisher}{Springer}, \bibinfo{pages}{103--126}.
\newblock


\bibitem[\protect\citeauthoryear{Lum and Isaac}{Lum and Isaac}{2016}]%
        {lum2016predict}
\bibfield{author}{\bibinfo{person}{Kristian Lum} {and} \bibinfo{person}{William
  Isaac}.} \bibinfo{year}{2016}\natexlab{}.
\newblock \showarticletitle{To predict and serve?}
\newblock \bibinfo{journal}{\emph{Significance}} \bibinfo{volume}{13},
  \bibinfo{number}{5} (\bibinfo{year}{2016}), \bibinfo{pages}{14--19}.
\newblock


\bibitem[\protect\citeauthoryear{Lunt}{Lunt}{1988}]%
        {lunt1988automated}
\bibfield{author}{\bibinfo{person}{Teresa~F Lunt}.}
  \bibinfo{year}{1988}\natexlab{}.
\newblock \showarticletitle{Automated audit trail analysis and intrusion
  detection: A survey}. In \bibinfo{booktitle}{\emph{In Proceedings of the 11th
  National Computer Security Conference}}.
\newblock


\bibitem[\protect\citeauthoryear{Mace, Roelke, and Fonseca}{Mace
  et~al\mbox{.}}{2015}]%
        {mace2015pivot}
\bibfield{author}{\bibinfo{person}{Jonathan Mace}, \bibinfo{person}{Ryan
  Roelke}, {and} \bibinfo{person}{Rodrigo Fonseca}.}
  \bibinfo{year}{2015}\natexlab{}.
\newblock \showarticletitle{Pivot tracing: Dynamic causal monitoring for
  distributed systems}. In \bibinfo{booktitle}{\emph{Proceedings of the 25th
  Symposium on Operating Systems Principles}}. \bibinfo{pages}{378--393}.
\newblock


\bibitem[\protect\citeauthoryear{Martin~Jr, Prabhakaran, Kuhlberg, Smart, and
  Isaac}{Martin~Jr et~al\mbox{.}}{2020}]%
        {martin2020extending}
\bibfield{author}{\bibinfo{person}{Donald Martin~Jr},
  \bibinfo{person}{Vinodkumar Prabhakaran}, \bibinfo{person}{Jill Kuhlberg},
  \bibinfo{person}{Andrew Smart}, {and} \bibinfo{person}{William~S Isaac}.}
  \bibinfo{year}{2020}\natexlab{}.
\newblock \showarticletitle{Extending the machine learning abstraction
  boundary: A Complex systems approach to incorporate societal context}.
\newblock \bibinfo{journal}{\emph{arXiv preprint arXiv:2006.09663}}
  (\bibinfo{year}{2020}).
\newblock


\bibitem[\protect\citeauthoryear{McBreen and Foreword By-Beck}{McBreen and
  Foreword By-Beck}{2002}]%
        {mcbreen2002questioning}
\bibfield{author}{\bibinfo{person}{Pete McBreen} {and} \bibinfo{person}{Kent
  Foreword By-Beck}.} \bibinfo{year}{2002}\natexlab{}.
\newblock \bibinfo{booktitle}{\emph{Questioning extreme programming}}.
\newblock \bibinfo{publisher}{Addison-Wesley Longman Publishing Co., Inc.}
\newblock


\bibitem[\protect\citeauthoryear{McPhillips, Song, Kolisnik, Aulenbach,
  Belhajjame, Bocinsky, Cao, Chirigati, Dey, Freire, Huntzinger, Jones, Koop,
  Missier, Schildhauer, Schwalm, Wei, Cheney, Bieda, and Ludascher}{McPhillips
  et~al\mbox{.}}{2015}]%
        {mcphillips2015yesworkflow}
\bibfield{author}{\bibinfo{person}{Timothy McPhillips},
  \bibinfo{person}{Tianhong Song}, \bibinfo{person}{Tyler Kolisnik},
  \bibinfo{person}{Steve Aulenbach}, \bibinfo{person}{Khalid Belhajjame},
  \bibinfo{person}{Kyle Bocinsky}, \bibinfo{person}{Yang Cao},
  \bibinfo{person}{Fernando Chirigati}, \bibinfo{person}{Saumen Dey},
  \bibinfo{person}{Juliana Freire}, \bibinfo{person}{Deborah Huntzinger},
  \bibinfo{person}{Christopher Jones}, \bibinfo{person}{David Koop},
  \bibinfo{person}{Paolo Missier}, \bibinfo{person}{Mark Schildhauer},
  \bibinfo{person}{Christopher Schwalm}, \bibinfo{person}{Yaxing Wei},
  \bibinfo{person}{James Cheney}, \bibinfo{person}{Mark Bieda}, {and}
  \bibinfo{person}{Bertram Ludascher}.} \bibinfo{year}{2015}\natexlab{}.
\newblock \showarticletitle{YesWorkflow: a user-oriented, language-independent
  tool for recovering workflow information from scripts}.
\newblock \bibinfo{journal}{\emph{International Data Curation Conference}}
  (\bibinfo{year}{2015}).
\newblock


\bibitem[\protect\citeauthoryear{Merkle}{Merkle}{1987}]%
        {merkle-tree}
\bibfield{author}{\bibinfo{person}{Ralph~C. Merkle}.}
  \bibinfo{year}{1987}\natexlab{}.
\newblock \showarticletitle{A Digital Signature Based on a Conventional
  Encryption Function}.
\newblock \bibinfo{journal}{\emph{CRYPTO}} (\bibinfo{year}{1987}).
\newblock


\bibitem[\protect\citeauthoryear{Metcalf, Moss, et~al\mbox{.}}{Metcalf
  et~al\mbox{.}}{2019}]%
        {metcalf2019owning}
\bibfield{author}{\bibinfo{person}{Jacob Metcalf}, \bibinfo{person}{Emanuel
  Moss}, {et~al\mbox{.}}} \bibinfo{year}{2019}\natexlab{}.
\newblock \showarticletitle{Owning Ethics: Corporate Logics, Silicon Valley,
  and the Institutionalization of Ethics}.
\newblock \bibinfo{journal}{\emph{Social Research: An International Quarterly}}
  \bibinfo{volume}{86}, \bibinfo{number}{2} (\bibinfo{year}{2019}),
  \bibinfo{pages}{449--476}.
\newblock


\bibitem[\protect\citeauthoryear{Miller}{Miller}{2019}]%
        {miller2017explanation}
\bibfield{author}{\bibinfo{person}{Tim Miller}.}
  \bibinfo{year}{2019}\natexlab{}.
\newblock \showarticletitle{Explanation in artificial intelligence: Insights
  from the social sciences}.
\newblock \bibinfo{journal}{\emph{Artificial Intelligence}}
  \bibinfo{volume}{267} (\bibinfo{year}{2019}), \bibinfo{pages}{1--38}.
\newblock


\bibitem[\protect\citeauthoryear{Mitchell}{Mitchell}{2009}]%
        {mitchell2009complexity}
\bibfield{author}{\bibinfo{person}{Melanie Mitchell}.}
  \bibinfo{year}{2009}\natexlab{}.
\newblock \bibinfo{booktitle}{\emph{Complexity: A guided tour}}.
\newblock \bibinfo{publisher}{Oxford University Press},
  \bibinfo{address}{Oxford, UK}.
\newblock


\bibitem[\protect\citeauthoryear{Mitchell, Wu, Zaldivar, Barnes, Vasserman,
  Hutchinson, Spitzer, Raji, and Gebru}{Mitchell et~al\mbox{.}}{2019}]%
        {mitchell2019model}
\bibfield{author}{\bibinfo{person}{Margaret Mitchell}, \bibinfo{person}{Simone
  Wu}, \bibinfo{person}{Andrew Zaldivar}, \bibinfo{person}{Parker Barnes},
  \bibinfo{person}{Lucy Vasserman}, \bibinfo{person}{Ben Hutchinson},
  \bibinfo{person}{Elena Spitzer}, \bibinfo{person}{Inioluwa~Deborah Raji},
  {and} \bibinfo{person}{Timnit Gebru}.} \bibinfo{year}{2019}\natexlab{}.
\newblock \showarticletitle{Model cards for model reporting}. In
  \bibinfo{booktitle}{\emph{Proceedings of the Conference on Fairness,
  Accountability, and Transparency}}. ACM, \bibinfo{pages}{220--229}.
\newblock


\bibitem[\protect\citeauthoryear{Mittelstadt}{Mittelstadt}{2019a}]%
        {mittelstadt2019ai}
\bibfield{author}{\bibinfo{person}{Brent Mittelstadt}.}
  \bibinfo{year}{2019}\natexlab{a}.
\newblock \showarticletitle{AI Ethics--Too principled to fail}.
\newblock \bibinfo{journal}{\emph{arXiv preprint arXiv:1906.06668}}
  (\bibinfo{year}{2019}).
\newblock


\bibitem[\protect\citeauthoryear{Mittelstadt}{Mittelstadt}{2019b}]%
        {mittelstadt2019principles}
\bibfield{author}{\bibinfo{person}{Brent Mittelstadt}.}
  \bibinfo{year}{2019}\natexlab{b}.
\newblock \showarticletitle{Principles alone cannot guarantee ethical AI}.
\newblock \bibinfo{journal}{\emph{Nature Machine Intelligence}}
  (\bibinfo{year}{2019}), \bibinfo{pages}{1--7}.
\newblock


\bibitem[\protect\citeauthoryear{Molnar}{Molnar}{2018}]%
        {molnar2018interpretable}
\bibfield{author}{\bibinfo{person}{Christoph Molnar}.}
  \bibinfo{year}{2018}\natexlab{}.
\newblock \bibinfo{booktitle}{\emph{Interpretable Machine Learning: A Guide for
  Making Black-box Models Explainable}}.
\newblock \bibinfo{publisher}{Online:
  \url{https://christophm.github.io/interpretable-ml-book/}}.
\newblock


\bibitem[\protect\citeauthoryear{Moor}{Moor}{1985}]%
        {moor1985computer}
\bibfield{author}{\bibinfo{person}{James~H Moor}.}
  \bibinfo{year}{1985}\natexlab{}.
\newblock \showarticletitle{What is computer ethics?}
\newblock \bibinfo{journal}{\emph{Metaphilosophy}} \bibinfo{volume}{16},
  \bibinfo{number}{4} (\bibinfo{year}{1985}).
\newblock


\bibitem[\protect\citeauthoryear{Moreau, Groth, Miles, Vazquez-Salceda,
  Ibbotson, Jiang, Munroe, Rana, Schreiber, Tan, et~al\mbox{.}}{Moreau
  et~al\mbox{.}}{2008}]%
        {moreau2008provenance}
\bibfield{author}{\bibinfo{person}{Luc Moreau}, \bibinfo{person}{Paul Groth},
  \bibinfo{person}{Simon Miles}, \bibinfo{person}{Javier Vazquez-Salceda},
  \bibinfo{person}{John Ibbotson}, \bibinfo{person}{Sheng Jiang},
  \bibinfo{person}{Steve Munroe}, \bibinfo{person}{Omer Rana},
  \bibinfo{person}{Andreas Schreiber}, \bibinfo{person}{Victor Tan},
  {et~al\mbox{.}}} \bibinfo{year}{2008}\natexlab{}.
\newblock \showarticletitle{The provenance of electronic data}.
\newblock \bibinfo{journal}{\emph{Commun. ACM}} \bibinfo{volume}{51},
  \bibinfo{number}{4} (\bibinfo{year}{2008}), \bibinfo{pages}{52--58}.
\newblock


\bibitem[\protect\citeauthoryear{Mulgan}{Mulgan}{2000}]%
        {mulgan2000accountability}
\bibfield{author}{\bibinfo{person}{Richard Mulgan}.}
  \bibinfo{year}{2000}\natexlab{}.
\newblock \showarticletitle{‘Accountability’: An ever-expanding concept?}
\newblock \bibinfo{journal}{\emph{Public administration}} \bibinfo{volume}{78},
  \bibinfo{number}{3} (\bibinfo{year}{2000}), \bibinfo{pages}{555--573}.
\newblock


\bibitem[\protect\citeauthoryear{Mulgan}{Mulgan}{2003}]%
        {mulgan2003holding}
\bibfield{author}{\bibinfo{person}{Richard~G Mulgan}.}
  \bibinfo{year}{2003}\natexlab{}.
\newblock \bibinfo{booktitle}{\emph{Holding power to account: accountability in
  modern democracies}}.
\newblock \bibinfo{publisher}{Palgrave Macmillan}.
\newblock


\bibitem[\protect\citeauthoryear{Mulligan and Bamberger}{Mulligan and
  Bamberger}{2018}]%
        {mulligan2018saving}
\bibfield{author}{\bibinfo{person}{Deirdre~K Mulligan} {and}
  \bibinfo{person}{Kenneth~A Bamberger}.} \bibinfo{year}{2018}\natexlab{}.
\newblock \showarticletitle{Saving Governance-by-Design}.
\newblock \bibinfo{journal}{\emph{California Law Review}}
  \bibinfo{volume}{106}, \bibinfo{number}{101} (\bibinfo{year}{2018}).
\newblock


\bibitem[\protect\citeauthoryear{Mulligan, Kluttz, and Kohli}{Mulligan
  et~al\mbox{.}}{2019a}]%
        {mulligan2019shaping}
\bibfield{author}{\bibinfo{person}{Deirdre~K Mulligan}, \bibinfo{person}{Daniel
  Kluttz}, {and} \bibinfo{person}{Nitin Kohli}.}
  \bibinfo{year}{2019}\natexlab{a}.
\newblock \showarticletitle{Shaping Our Tools: Contestability as a Means to
  Promote Responsible Algorithmic Decision Making in the Professions}.
\newblock \bibinfo{journal}{\emph{Preprint available at SSRN 3311894}}
  (\bibinfo{year}{2019}).
\newblock


\bibitem[\protect\citeauthoryear{Mulligan, Koopman, and Doty}{Mulligan
  et~al\mbox{.}}{2016}]%
        {mulligan2016privacy}
\bibfield{author}{\bibinfo{person}{Deirdre~K Mulligan}, \bibinfo{person}{Colin
  Koopman}, {and} \bibinfo{person}{Nick Doty}.}
  \bibinfo{year}{2016}\natexlab{}.
\newblock \showarticletitle{Privacy is an essentially contested concept: a
  multi-dimensional analytic for mapping privacy}.
\newblock \bibinfo{journal}{\emph{Philosophical Transactions of the Royal
  Society A: Mathematical, Physical and Engineering Sciences}}
  \bibinfo{volume}{374}, \bibinfo{number}{2083} (\bibinfo{year}{2016}),
  \bibinfo{pages}{20160118}.
\newblock


\bibitem[\protect\citeauthoryear{Mulligan, Kroll, Kohli, and Wong}{Mulligan
  et~al\mbox{.}}{2019b}]%
        {mulligan2019thing}
\bibfield{author}{\bibinfo{person}{Deirdre~K Mulligan},
  \bibinfo{person}{Joshua~A Kroll}, \bibinfo{person}{Nitin Kohli}, {and}
  \bibinfo{person}{Richmond~Y Wong}.} \bibinfo{year}{2019}\natexlab{b}.
\newblock \showarticletitle{This Thing Called Fairness: Disciplinary Confusion
  Realizing a Value in Technology}.
\newblock \bibinfo{journal}{\emph{Proceedings of the ACM on Human-Computer
  Interaction}} \bibinfo{volume}{3}, \bibinfo{number}{CSCW}
  (\bibinfo{year}{2019}), \bibinfo{pages}{119}.
\newblock


\bibitem[\protect\citeauthoryear{Mulligan and Nissenbaum}{Mulligan and
  Nissenbaum}{2020}]%
        {mulligan2020concept}
\bibfield{author}{\bibinfo{person}{Deirdre~K Mulligan} {and}
  \bibinfo{person}{Helen Nissenbaum}.} \bibinfo{year}{2020}\natexlab{}.
\newblock \showarticletitle{The Concept of Handoff as a Model for Ethical
  Analysis and Design}.
\newblock In \bibinfo{booktitle}{\emph{The Oxford Handbook of Ethics of AI}},
  \bibfield{editor}{\bibinfo{person}{Markus Dubber}, \bibinfo{person}{Frank
  Pasquale}, {and} \bibinfo{person}{Sunit Das}} (Eds.).
  \bibinfo{publisher}{Oxford University Press}, \bibinfo{address}{Oxford, UK}.
\newblock


\bibitem[\protect\citeauthoryear{{Multistakeholder Process on Software
  Component Transparency}}{{Multistakeholder Process on Software Component
  Transparency}}{2019}]%
        {sbom}
\bibfield{author}{\bibinfo{person}{{Multistakeholder Process on Software
  Component Transparency}}.} \bibinfo{year}{2019}\natexlab{}.
\newblock \bibinfo{title}{Framing Software Component Transparency: Establishing
  a Common Software Bill of Material (SBOM)}.
\newblock
  \bibinfo{howpublished}{\url{https://www.ntia.gov/files/ntia/publications/framingsbom_20191112.pdf}}.
\newblock


\bibitem[\protect\citeauthoryear{Muniswamy-Reddy, Holland, Braun, and
  Seltzer}{Muniswamy-Reddy et~al\mbox{.}}{2006}]%
        {muniswamy2006provenance}
\bibfield{author}{\bibinfo{person}{Kiran-Kumar Muniswamy-Reddy},
  \bibinfo{person}{David~A Holland}, \bibinfo{person}{Uri Braun}, {and}
  \bibinfo{person}{Margo~I Seltzer}.} \bibinfo{year}{2006}\natexlab{}.
\newblock \showarticletitle{Provenance-Aware Storage Systems.}. In
  \bibinfo{booktitle}{\emph{USENIX Annual Technical Conference, General
  Track}}. \bibinfo{pages}{43--56}.
\newblock


\bibitem[\protect\citeauthoryear{Muniswamy-Reddy, Macko, and
  Seltzer}{Muniswamy-Reddy et~al\mbox{.}}{2010}]%
        {muniswamy2010provenance}
\bibfield{author}{\bibinfo{person}{Kiran-Kumar Muniswamy-Reddy},
  \bibinfo{person}{Peter Macko}, {and} \bibinfo{person}{Margo~I Seltzer}.}
  \bibinfo{year}{2010}\natexlab{}.
\newblock \showarticletitle{Provenance for the Cloud.}. In
  \bibinfo{booktitle}{\emph{FAST}}, Vol.~\bibinfo{volume}{10}.
  \bibinfo{pages}{15--14}.
\newblock


\bibitem[\protect\citeauthoryear{Myers, Sandler, and Badgett}{Myers
  et~al\mbox{.}}{1979}]%
        {myers2011art}
\bibfield{author}{\bibinfo{person}{Glenford~J Myers}, \bibinfo{person}{Corey
  Sandler}, {and} \bibinfo{person}{Tom Badgett}.} \bibinfo{year}{2011
  (1979)}\natexlab{}.
\newblock \bibinfo{booktitle}{\emph{The art of software testing}}.
\newblock \bibinfo{publisher}{John Wiley \& Sons}.
\newblock


\bibitem[\protect\citeauthoryear{Nabar, Kenthapadi, Mishra, and Motwani}{Nabar
  et~al\mbox{.}}{2008}]%
        {nabar2008survey}
\bibfield{author}{\bibinfo{person}{Shubha~U Nabar}, \bibinfo{person}{Krishnaram
  Kenthapadi}, \bibinfo{person}{Nina Mishra}, {and} \bibinfo{person}{Rajeev
  Motwani}.} \bibinfo{year}{2008}\natexlab{}.
\newblock \showarticletitle{A survey of query auditing techniques for data
  privacy}.
\newblock In \bibinfo{booktitle}{\emph{Privacy-Preserving Data Mining}}.
  \bibinfo{publisher}{Springer}, \bibinfo{pages}{415--431}.
\newblock


\bibitem[\protect\citeauthoryear{Nabar, Marthi, Kenthapadi, Mishra, and
  Motwani}{Nabar et~al\mbox{.}}{2006}]%
        {nabar2006towards}
\bibfield{author}{\bibinfo{person}{Shubha~U Nabar}, \bibinfo{person}{Bhaskara
  Marthi}, \bibinfo{person}{Krishnaram Kenthapadi}, \bibinfo{person}{Nina
  Mishra}, {and} \bibinfo{person}{Rajeev Motwani}.}
  \bibinfo{year}{2006}\natexlab{}.
\newblock \showarticletitle{Towards robustness in query auditing}. In
  \bibinfo{booktitle}{\emph{Proceedings of the 32nd international conference on
  Very large data bases}}. \bibinfo{pages}{151--162}.
\newblock


\bibitem[\protect\citeauthoryear{{National Institute of Standards and
  Technology}}{{National Institute of Standards and Technology}}{2018}]%
        {nist-cyber-framework-2018}
\bibfield{author}{\bibinfo{person}{{National Institute of Standards and
  Technology}}.} \bibinfo{year}{2018}\natexlab{}.
\newblock \bibinfo{title}{Framework for Improving Critical Infrastructure
  Cybersecurity}.
\newblock \bibinfo{howpublished}{\url{https://www.nist.gov/cyberframework}}.
\newblock


\bibitem[\protect\citeauthoryear{{National Institute of Standards and
  Technology}}{{National Institute of Standards and Technology}}{2020}]%
        {nist-privacy-framework-2020}
\bibfield{author}{\bibinfo{person}{{National Institute of Standards and
  Technology}}.} \bibinfo{year}{2020}\natexlab{}.
\newblock \bibinfo{title}{NIST Privacy Framework: A Tool for Improving Privacy
  through Enterprise Risk Management}.
\newblock \bibinfo{howpublished}{\url{https://www.nist.gov/privacy-framework}}.
\newblock


\bibitem[\protect\citeauthoryear{{National New Generation Artificial
  Intelligence Governance Expert Committee (Ministry of Science and Technology
  convening)}}{{National New Generation Artificial Intelligence Governance
  Expert Committee (Ministry of Science and Technology convening)}}{2019}]%
        {china-responsible-ai}
\bibfield{author}{\bibinfo{person}{{National New Generation Artificial
  Intelligence Governance Expert Committee (Ministry of Science and Technology
  convening)}}.} \bibinfo{year}{2019}\natexlab{}.
\newblock \bibinfo{title}{{Governance Principles for a New Generation of
  Artificial Intelligence: Develop Responsible Artificial Intelligence}}.
\newblock
\newblock


\bibitem[\protect\citeauthoryear{{National Security Commission on Artificial
  Intelligence}}{{National Security Commission on Artificial
  Intelligence}}{2020}]%
        {nscai-q2-2020}
\bibfield{author}{\bibinfo{person}{{National Security Commission on Artificial
  Intelligence}}.} \bibinfo{year}{2020}\natexlab{}.
\newblock \bibinfo{title}{Second Quarter Recommendations}.
\newblock \bibinfo{howpublished}{\url{https://www.nscai.gov}}.
\newblock


\bibitem[\protect\citeauthoryear{Nikitin, Kokoris-Kogias, Jovanovic, Gailly,
  Gasser, Khoffi, Cappos, and Ford}{Nikitin et~al\mbox{.}}{2017}]%
        {nikitin2017chainiac}
\bibfield{author}{\bibinfo{person}{Kirill Nikitin},
  \bibinfo{person}{Eleftherios Kokoris-Kogias}, \bibinfo{person}{Philipp
  Jovanovic}, \bibinfo{person}{Nicolas Gailly}, \bibinfo{person}{Linus Gasser},
  \bibinfo{person}{Ismail Khoffi}, \bibinfo{person}{Justin Cappos}, {and}
  \bibinfo{person}{Bryan Ford}.} \bibinfo{year}{2017}\natexlab{}.
\newblock \showarticletitle{$\{$CHAINIAC$\}$: Proactive software-update
  transparency via collectively signed skipchains and verified builds}. In
  \bibinfo{booktitle}{\emph{26th $\{$USENIX$\}$ Security Symposium
  ($\{$USENIX$\}$ Security 17)}}. \bibinfo{pages}{1271--1287}.
\newblock


\bibitem[\protect\citeauthoryear{Nissenbaum}{Nissenbaum}{1996}]%
        {nissenbaum1996accountability}
\bibfield{author}{\bibinfo{person}{Helen Nissenbaum}.}
  \bibinfo{year}{1996}\natexlab{}.
\newblock \showarticletitle{Accountability in a computerized society}.
\newblock \bibinfo{journal}{\emph{Science and engineering ethics}}
  \bibinfo{volume}{2}, \bibinfo{number}{1} (\bibinfo{year}{1996}),
  \bibinfo{pages}{25--42}.
\newblock


\bibitem[\protect\citeauthoryear{Nissenbaum}{Nissenbaum}{2001}]%
        {nissenbaum2001}
\bibfield{author}{\bibinfo{person}{Helen Nissenbaum}.}
  \bibinfo{year}{2001}\natexlab{}.
\newblock \showarticletitle{{How computer systems embody values}}.
\newblock \bibinfo{journal}{\emph{Computer}} \bibinfo{volume}{34},
  \bibinfo{number}{3} (\bibinfo{date}{March} \bibinfo{year}{2001}),
  \bibinfo{pages}{120--119}.
\newblock
\showISSN{0018-9162}


\bibitem[\protect\citeauthoryear{Nissenbaum}{Nissenbaum}{2005}]%
        {nissenbaum2005values}
\bibfield{author}{\bibinfo{person}{Helen Nissenbaum}.}
  \bibinfo{year}{2005}\natexlab{}.
\newblock \showarticletitle{Values in technical design}.
\newblock In \bibinfo{booktitle}{\emph{Encyclopedia of science, technology, and
  ethics}}. \bibinfo{publisher}{Macmillan New York, NY},
  \bibinfo{pages}{66--70}.
\newblock


\bibitem[\protect\citeauthoryear{{Office of the Secretary of Defense}}{{Office
  of the Secretary of Defense}}{2020}]%
        {dod-official-ethics-principles-2020}
\bibfield{author}{\bibinfo{person}{{Office of the Secretary of Defense}}.}
  \bibinfo{year}{2020}\natexlab{}.
\newblock \bibinfo{title}{Artificial Intelligence Ethical Principles for the
  Department of Defense}.
\newblock \bibinfo{howpublished}{{OSD Memorandum}}.
\newblock


\bibitem[\protect\citeauthoryear{{Organization for Economic Cooperation and
  Development}}{{Organization for Economic Cooperation and
  Development}}{2019}]%
        {oecd-ai-guidelines}
\bibfield{author}{\bibinfo{person}{{Organization for Economic Cooperation and
  Development}}.} \bibinfo{year}{2019}\natexlab{}.
\newblock \bibinfo{title}{Recommendation of the Council on Artificial
  Intelligence}.
\newblock \bibinfo{howpublished}{OECD/LEGAL/0449,
  \url{https://legalinstruments.oecd.org/en/instruments/OECD-LEGAL-0449}}.
\newblock


\bibitem[\protect\citeauthoryear{{Partnership on AI}}{{Partnership on
  AI}}{2019}]%
        {about-ml}
\bibfield{author}{\bibinfo{person}{{Partnership on AI}}.}
  \bibinfo{year}{2019}\natexlab{}.
\newblock \bibinfo{title}{Annotation and Benchmarking on Understanding and
  Transparency of Machine Learning Lifecycles (ABOUT ML)}.
\newblock \bibinfo{howpublished}{\url{https://partnershiponai.com/about-ml}}.
\newblock


\bibitem[\protect\citeauthoryear{Passi and Barocas}{Passi and Barocas}{2019}]%
        {passi2019problem}
\bibfield{author}{\bibinfo{person}{Samir Passi} {and} \bibinfo{person}{Solon
  Barocas}.} \bibinfo{year}{2019}\natexlab{}.
\newblock \showarticletitle{Problem formulation and fairness}. In
  \bibinfo{booktitle}{\emph{Conference on Fairness, Accountability, and
  Transparency}}.
\newblock


\bibitem[\protect\citeauthoryear{Passi and Jackson}{Passi and Jackson}{2018}]%
        {passi2018}
\bibfield{author}{\bibinfo{person}{Samir Passi} {and} \bibinfo{person}{Steven~J
  Jackson}.} \bibinfo{year}{2018}\natexlab{}.
\newblock \showarticletitle{{Trust in Data Science: Collaboration, Translation,
  and Accountability in Corporate Data Science Projects}}.
\newblock \bibinfo{journal}{\emph{Proceedings of the ACM on Human-Computer
  Interaction}} \bibinfo{volume}{2}, \bibinfo{number}{CSCW}
  (\bibinfo{date}{nov} \bibinfo{year}{2018}), \bibinfo{pages}{1--28}.
\newblock
\showISSN{25730142}


\bibitem[\protect\citeauthoryear{Pearson and Charlesworth}{Pearson and
  Charlesworth}{2009}]%
        {pearson2009accountability}
\bibfield{author}{\bibinfo{person}{Siani Pearson} {and} \bibinfo{person}{Andrew
  Charlesworth}.} \bibinfo{year}{2009}\natexlab{}.
\newblock \showarticletitle{Accountability as a way forward for privacy
  protection in the cloud}.
\newblock In \bibinfo{booktitle}{\emph{Cloud computing}}.
  \bibinfo{publisher}{Springer}, \bibinfo{pages}{131--144}.
\newblock


\bibitem[\protect\citeauthoryear{P{\'e}rez, Rubio, and
  S{\'a}enz-Ad{\'a}n}{P{\'e}rez et~al\mbox{.}}{2018}]%
        {perez2018systematic}
\bibfield{author}{\bibinfo{person}{Beatriz P{\'e}rez}, \bibinfo{person}{Julio
  Rubio}, {and} \bibinfo{person}{Carlos S{\'a}enz-Ad{\'a}n}.}
  \bibinfo{year}{2018}\natexlab{}.
\newblock \showarticletitle{A systematic review of provenance systems}.
\newblock \bibinfo{journal}{\emph{Knowledge and Information Systems}}
  (\bibinfo{year}{2018}), \bibinfo{pages}{1--49}.
\newblock


\bibitem[\protect\citeauthoryear{Petersen, Wohlin, and Baca}{Petersen
  et~al\mbox{.}}{2009}]%
        {petersen2009waterfall}
\bibfield{author}{\bibinfo{person}{Kai Petersen}, \bibinfo{person}{Claes
  Wohlin}, {and} \bibinfo{person}{Dejan Baca}.}
  \bibinfo{year}{2009}\natexlab{}.
\newblock \showarticletitle{The waterfall model in large-scale development}. In
  \bibinfo{booktitle}{\emph{International Conference on Product-Focused
  Software Process Improvement}}. Springer, \bibinfo{pages}{386--400}.
\newblock


\bibitem[\protect\citeauthoryear{Porter}{Porter}{1992}]%
        {porter1992quantification}
\bibfield{author}{\bibinfo{person}{Theodore~M Porter}.}
  \bibinfo{year}{1992}\natexlab{}.
\newblock \showarticletitle{Quantification and the accounting ideal in
  science}.
\newblock \bibinfo{journal}{\emph{Social studies of science}}
  \bibinfo{volume}{22}, \bibinfo{number}{4} (\bibinfo{year}{1992}),
  \bibinfo{pages}{633--651}.
\newblock


\bibitem[\protect\citeauthoryear{Rader, Cotter, and Cho}{Rader
  et~al\mbox{.}}{2018}]%
        {rader2018chi}
\bibfield{author}{\bibinfo{person}{Emilee Rader}, \bibinfo{person}{Kelley
  Cotter}, {and} \bibinfo{person}{Janghee Cho}.}
  \bibinfo{year}{2018}\natexlab{}.
\newblock \showarticletitle{Explanations as Mechanisms for Supporting
  Algorithmic Transparency}.
\newblock \bibinfo{journal}{\emph{Proceedings of the International Conference
  on Human Factors in Computer Systems (CHI)}} (\bibinfo{year}{2018}).
\newblock


\bibitem[\protect\citeauthoryear{Raghavan, Barocas, Kleinberg, and
  Levy}{Raghavan et~al\mbox{.}}{2020}]%
        {raghavan2020mitigating}
\bibfield{author}{\bibinfo{person}{Manish Raghavan}, \bibinfo{person}{Solon
  Barocas}, \bibinfo{person}{Jon Kleinberg}, {and} \bibinfo{person}{Karen
  Levy}.} \bibinfo{year}{2020}\natexlab{}.
\newblock \showarticletitle{Mitigating bias in algorithmic hiring: Evaluating
  claims and practices}. In \bibinfo{booktitle}{\emph{Proceedings of the 2020
  Conference on Fairness, Accountability, and Transparency}}.
  \bibinfo{pages}{469--481}.
\newblock


\bibitem[\protect\citeauthoryear{Raji, Smart, White, Mitchell, Gebru,
  Hutchinson, Smith-Loud, Theron, and Barnes}{Raji et~al\mbox{.}}{2020}]%
        {raji2020closing}
\bibfield{author}{\bibinfo{person}{Inioluwa~Deborah Raji},
  \bibinfo{person}{Andrew Smart}, \bibinfo{person}{Rebecca~N White},
  \bibinfo{person}{Margaret Mitchell}, \bibinfo{person}{Timnit Gebru},
  \bibinfo{person}{Ben Hutchinson}, \bibinfo{person}{Jamila Smith-Loud},
  \bibinfo{person}{Daniel Theron}, {and} \bibinfo{person}{Parker Barnes}.}
  \bibinfo{year}{2020}\natexlab{}.
\newblock \showarticletitle{Closing the AI Accountability Gap: Defining an
  End-to-End Framework for Internal Algorithmic Auditing}.
\newblock \bibinfo{journal}{\emph{ACM Conference on Fairness, Accountability,
  and Transparency}} (\bibinfo{year}{2020}).
\newblock


\bibitem[\protect\citeauthoryear{Rees}{Rees}{2009}]%
        {rees2009hostages}
\bibfield{author}{\bibinfo{person}{Joseph~V Rees}.}
  \bibinfo{year}{2009}\natexlab{}.
\newblock \bibinfo{booktitle}{\emph{Hostages of each other: The transformation
  of nuclear safety since Three Mile Island}}.
\newblock \bibinfo{publisher}{University of Chicago Press}.
\newblock


\bibitem[\protect\citeauthoryear{Reisman, Schultz, Crawford, and
  Whittaker}{Reisman et~al\mbox{.}}{2018}]%
        {ainow2018impact}
\bibfield{author}{\bibinfo{person}{Dillon Reisman}, \bibinfo{person}{Jason
  Schultz}, \bibinfo{person}{Kate Crawford}, {and} \bibinfo{person}{Meredith
  Whittaker}.} \bibinfo{year}{2018}\natexlab{}.
\newblock \bibinfo{title}{Algorithmic Impact Assessements: A Practical
  Framework for Public Agency Accountability}.
\newblock \bibinfo{howpublished}{AI Now Institute Report
  \url{https://ainowinstitute.org/aiareport2018.pdf}}.
\newblock


\bibitem[\protect\citeauthoryear{Reyes, Wijesekera, Reardon, On, Razaghpanah,
  Vallina-Rodriguez, and Egelman}{Reyes et~al\mbox{.}}{2018}]%
        {reyes2018won}
\bibfield{author}{\bibinfo{person}{Irwin Reyes}, \bibinfo{person}{Primal
  Wijesekera}, \bibinfo{person}{Joel Reardon}, \bibinfo{person}{Amit
  Elazari~Bar On}, \bibinfo{person}{Abbas Razaghpanah}, \bibinfo{person}{Narseo
  Vallina-Rodriguez}, {and} \bibinfo{person}{Serge Egelman}.}
  \bibinfo{year}{2018}\natexlab{}.
\newblock \showarticletitle{``Won’t somebody think of the children?''
  examining COPPA compliance at scale}.
\newblock \bibinfo{journal}{\emph{Proceedings on Privacy Enhancing
  Technologies}} \bibinfo{volume}{2018}, \bibinfo{number}{3}
  (\bibinfo{year}{2018}), \bibinfo{pages}{63--83}.
\newblock


\bibitem[\protect\citeauthoryear{Rice}{Rice}{1953}]%
        {rice1953classes}
\bibfield{author}{\bibinfo{person}{Henry~Gordon Rice}.}
  \bibinfo{year}{1953}\natexlab{}.
\newblock \showarticletitle{Classes of recursively enumerable sets and their
  decision problems}.
\newblock \bibinfo{journal}{\emph{Trans. Amer. Math. Soc.}}
  (\bibinfo{year}{1953}), \bibinfo{pages}{358--366}.
\newblock


\bibitem[\protect\citeauthoryear{Rosner and Ames}{Rosner and Ames}{2014}]%
        {rosner2014designing}
\bibfield{author}{\bibinfo{person}{Daniela~K. Rosner} {and}
  \bibinfo{person}{Morgan Ames}.} \bibinfo{year}{2014}\natexlab{}.
\newblock \showarticletitle{{Designing for repair?: infrastructures and
  materialities of breakdown}}. In \bibinfo{booktitle}{\emph{Proceedings of the
  17th ACM conference on Computer supported cooperative work {\&} social
  computing - CSCW '14}}. \bibinfo{publisher}{ACM Press},
  \bibinfo{address}{Baltimore}, \bibinfo{pages}{319--331}.
\newblock
\showISBNx{9781450325400}


\bibitem[\protect\citeauthoryear{Rothenberg}{Rothenberg}{1999}]%
        {rothenberg1999avoiding}
\bibfield{author}{\bibinfo{person}{Jeff Rothenberg}.}
  \bibinfo{year}{1999}\natexlab{}.
\newblock \bibinfo{booktitle}{\emph{Avoiding Technological Quicksand: Finding a
  Viable Technical Foundation for Digital Preservation. A Report to the Council
  on Library and Information Resources.}}
\newblock \bibinfo{publisher}{ERIC}.
\newblock


\bibitem[\protect\citeauthoryear{Rubinstein and Good}{Rubinstein and
  Good}{2013}]%
        {rubinstein2013privacy}
\bibfield{author}{\bibinfo{person}{Ira~S Rubinstein} {and}
  \bibinfo{person}{Nathaniel Good}.} \bibinfo{year}{2013}\natexlab{}.
\newblock \showarticletitle{Privacy by design: A counterfactual analysis of
  Google and Facebook privacy incidents}.
\newblock \bibinfo{journal}{\emph{Berkeley Tech. LJ}}  \bibinfo{volume}{28}
  (\bibinfo{year}{2013}), \bibinfo{pages}{1333}.
\newblock


\bibitem[\protect\citeauthoryear{Salvendy}{Salvendy}{2012}]%
        {salvendy2012handbook}
\bibfield{author}{\bibinfo{person}{Gavriel Salvendy}.}
  \bibinfo{year}{2012}\natexlab{}.
\newblock \bibinfo{booktitle}{\emph{Handbook of human factors and ergonomics}}.
\newblock \bibinfo{publisher}{John Wiley \& Sons}.
\newblock


\bibitem[\protect\citeauthoryear{Sandvig}{Sandvig}{2015}]%
        {sandvig2015seeing}
\bibfield{author}{\bibinfo{person}{Christian Sandvig}.}
  \bibinfo{year}{2015}\natexlab{}.
\newblock \showarticletitle{Seeing the Sort: The Aesthetic and Industrial
  Defense of “The Algorithm”}.
\newblock \bibinfo{journal}{\emph{Journal of the New Media Caucus}}
  (\bibinfo{year}{2015}).
\newblock
\newblock
\shownote{ISSN: 1942-017X, [Online]
  \url{http://median.newmediacaucus.org/art-infrastructures-information/seeing-the-sort-the-aesthetic-and-industrial-defense-of-the-algorithm/}.}


\bibitem[\protect\citeauthoryear{Sandvig, Hamilton, Karahalios, and
  Langbort}{Sandvig et~al\mbox{.}}{2014}]%
        {sandvig2014auditing}
\bibfield{author}{\bibinfo{person}{Christian Sandvig}, \bibinfo{person}{Kevin
  Hamilton}, \bibinfo{person}{Karrie Karahalios}, {and} \bibinfo{person}{Cedric
  Langbort}.} \bibinfo{year}{2014}\natexlab{}.
\newblock \showarticletitle{Auditing algorithms: Research methods for detecting
  discrimination on internet platforms}.
\newblock \bibinfo{journal}{\emph{Data and Discrimination: Converting Critical
  Concerns into Productive Inquiry}} (\bibinfo{year}{2014}).
\newblock


\bibitem[\protect\citeauthoryear{Sasson, Chiesa, Garman, Green, Miers, Tromer,
  and Virza}{Sasson et~al\mbox{.}}{2014}]%
        {sasson2014zerocash}
\bibfield{author}{\bibinfo{person}{Eli~Ben Sasson}, \bibinfo{person}{Alessandro
  Chiesa}, \bibinfo{person}{Christina Garman}, \bibinfo{person}{Matthew Green},
  \bibinfo{person}{Ian Miers}, \bibinfo{person}{Eran Tromer}, {and}
  \bibinfo{person}{Madars Virza}.} \bibinfo{year}{2014}\natexlab{}.
\newblock \showarticletitle{Zerocash: Decentralized anonymous payments from
  bitcoin}. In \bibinfo{booktitle}{\emph{2014 IEEE Symp. Sec. \& Priv.}} IEEE,
  \bibinfo{pages}{459--474}.
\newblock


\bibitem[\protect\citeauthoryear{Selbst}{Selbst}{2017}]%
        {selbst2017disparate}
\bibfield{author}{\bibinfo{person}{Andrew Selbst}.}
  \bibinfo{year}{2017}\natexlab{}.
\newblock \showarticletitle{Disparate Impact in Big Data Policing}.
\newblock \bibinfo{journal}{\emph{Georgia Law Review}} \bibinfo{volume}{52},
  \bibinfo{number}{109} (\bibinfo{year}{2017}).
\newblock


\bibitem[\protect\citeauthoryear{Selbst and Barocas}{Selbst and
  Barocas}{2018}]%
        {selbst2018intuitive}
\bibfield{author}{\bibinfo{person}{Andrew~D Selbst} {and}
  \bibinfo{person}{Solon Barocas}.} \bibinfo{year}{2018}\natexlab{}.
\newblock \showarticletitle{The intuitive appeal of explainable machines}.
\newblock \bibinfo{journal}{\emph{Fordham L. Rev.}}  \bibinfo{volume}{87}
  (\bibinfo{year}{2018}).
\newblock


\bibitem[\protect\citeauthoryear{Selbst, Boyd, Friedler, Venkatasubramanian,
  and Vertesi}{Selbst et~al\mbox{.}}{2019}]%
        {selbst2019fairness}
\bibfield{author}{\bibinfo{person}{Andrew~D Selbst}, \bibinfo{person}{Danah
  Boyd}, \bibinfo{person}{Sorelle~A Friedler}, \bibinfo{person}{Suresh
  Venkatasubramanian}, {and} \bibinfo{person}{Janet Vertesi}.}
  \bibinfo{year}{2019}\natexlab{}.
\newblock \showarticletitle{Fairness and abstraction in sociotechnical
  systems}. In \bibinfo{booktitle}{\emph{Conference on Fairness,
  Accountability, and Transparency}}. ACM, \bibinfo{pages}{59--68}.
\newblock


\bibitem[\protect\citeauthoryear{Selbst and Powles}{Selbst and Powles}{2017}]%
        {selbst2017meaningful}
\bibfield{author}{\bibinfo{person}{Andrew~D Selbst} {and}
  \bibinfo{person}{Julia Powles}.} \bibinfo{year}{2017}\natexlab{}.
\newblock \showarticletitle{Meaningful information and the right to
  explanation}.
\newblock \bibinfo{journal}{\emph{International Data Privacy Law}}
  \bibinfo{volume}{7}, \bibinfo{number}{4} (\bibinfo{year}{2017}),
  \bibinfo{pages}{233--242}.
\newblock


\bibitem[\protect\citeauthoryear{Selsam, Liang, and Dill}{Selsam
  et~al\mbox{.}}{2017}]%
        {selsam2017developing}
\bibfield{author}{\bibinfo{person}{Daniel Selsam}, \bibinfo{person}{Percy
  Liang}, {and} \bibinfo{person}{David~L Dill}.}
  \bibinfo{year}{2017}\natexlab{}.
\newblock \showarticletitle{Developing bug-free machine learning systems with
  formal mathematics}.
\newblock \bibinfo{journal}{\emph{arXiv:1706.08605}} (\bibinfo{year}{2017}).
\newblock


\bibitem[\protect\citeauthoryear{Sendak, Elish, Gao, Futoma, Ratliff, Nichols,
  Bedoya, Balu, and O'Brien}{Sendak et~al\mbox{.}}{2020}]%
        {sendak2020human}
\bibfield{author}{\bibinfo{person}{Mark Sendak},
  \bibinfo{person}{Madeleine~Clare Elish}, \bibinfo{person}{Michael Gao},
  \bibinfo{person}{Joseph Futoma}, \bibinfo{person}{William Ratliff},
  \bibinfo{person}{Marshall Nichols}, \bibinfo{person}{Armando Bedoya},
  \bibinfo{person}{Suresh Balu}, {and} \bibinfo{person}{Cara O'Brien}.}
  \bibinfo{year}{2020}\natexlab{}.
\newblock \showarticletitle{``The human body is a black box'' supporting
  clinical decision-making with deep learning}. In
  \bibinfo{booktitle}{\emph{Conference on Fairness, Accountability, and
  Transparency}}. \bibinfo{pages}{99--109}.
\newblock


\bibitem[\protect\citeauthoryear{Shilton}{Shilton}{2018}]%
        {shilton2018}
\bibfield{author}{\bibinfo{person}{Katie Shilton}.}
  \bibinfo{year}{2018}\natexlab{}.
\newblock \showarticletitle{{Values and Ethics in Human-Computer Interaction}}.
\newblock \bibinfo{journal}{\emph{Foundations and Trends in Human-Computer
  Interaction}} \bibinfo{volume}{12}, \bibinfo{number}{2}
  (\bibinfo{year}{2018}), \bibinfo{pages}{107--171}.
\newblock
\showISBNx{9781139168564}
\showISSN{1551-3955}


\bibitem[\protect\citeauthoryear{Shilton, Koepfler, and Fleischmann}{Shilton
  et~al\mbox{.}}{2014}]%
        {shilton2014}
\bibfield{author}{\bibinfo{person}{Katie Shilton}, \bibinfo{person}{Jes~A.
  Koepfler}, {and} \bibinfo{person}{Kenneth~R. Fleischmann}.}
  \bibinfo{year}{2014}\natexlab{}.
\newblock \showarticletitle{{How to see values in social computing: Methods for
  Studying Values Dimensions}}. In \bibinfo{booktitle}{\emph{Proceedings of the
  17th ACM Conference on Computer Supported Cooperative Work {\&} Social
  Computing (CSCW '14)}}. \bibinfo{pages}{426--435}.
\newblock
\showISBNx{9781450325400}


\bibitem[\protect\citeauthoryear{Siegelman and Heckman}{Siegelman and
  Heckman}{1993}]%
        {siegelman1993urban}
\bibfield{author}{\bibinfo{person}{Peter Siegelman} {and} \bibinfo{person}{J
  Heckman}.} \bibinfo{year}{1993}\natexlab{}.
\newblock \showarticletitle{The Urban Institute audit studies: Their methods
  and findings}.
\newblock \bibinfo{journal}{\emph{Clear and Convincing Evidence: Measurement of
  Discrimination in America, Washington}}  \bibinfo{volume}{187}
  (\bibinfo{year}{1993}), \bibinfo{pages}{258}.
\newblock


\bibitem[\protect\citeauthoryear{Star and Ruhleder}{Star and Ruhleder}{1994}]%
        {star1994steps}
\bibfield{author}{\bibinfo{person}{Susan~Leigh Star} {and}
  \bibinfo{person}{Karen Ruhleder}.} \bibinfo{year}{1994}\natexlab{}.
\newblock \showarticletitle{Steps towards an ecology of infrastructure: complex
  problems in design and access for large-scale collaborative systems}. In
  \bibinfo{booktitle}{\emph{Proceedings of the 1994 ACM conference on Computer
  supported cooperative work}}. ACM, \bibinfo{pages}{253--264}.
\newblock


\bibitem[\protect\citeauthoryear{Star and Ruhleder}{Star and Ruhleder}{1996}]%
        {star1996steps}
\bibfield{author}{\bibinfo{person}{Susan~Leigh Star} {and}
  \bibinfo{person}{Karen Ruhleder}.} \bibinfo{year}{1996}\natexlab{}.
\newblock \showarticletitle{Steps toward an ecology of infrastructure: Design
  and access for large information spaces}.
\newblock \bibinfo{journal}{\emph{Information systems research}}
  \bibinfo{volume}{7}, \bibinfo{number}{1} (\bibinfo{year}{1996}),
  \bibinfo{pages}{111--134}.
\newblock


\bibitem[\protect\citeauthoryear{Steinhardt}{Steinhardt}{2016}]%
        {steinhardt2016breaking}
\bibfield{author}{\bibinfo{person}{Stephanie~B Steinhardt}.}
  \bibinfo{year}{2016}\natexlab{}.
\newblock \showarticletitle{{Breaking Down While Building Up: Design and
  Decline in Emerging Infrastructures}}. In
  \bibinfo{booktitle}{\emph{Proceedings of the 2016 CHI Conference on Human
  Factors in Computing Systems - CHI '16}}. \bibinfo{publisher}{ACM Press},
  \bibinfo{address}{New York, New York, USA}, \bibinfo{pages}{2198--2208}.
\newblock
\showISBNx{9781450333627}


\bibitem[\protect\citeauthoryear{Stodden, Leisch, and Peng}{Stodden
  et~al\mbox{.}}{2014}]%
        {stodden2014implementing}
\bibfield{author}{\bibinfo{person}{Victoria Stodden},
  \bibinfo{person}{Friedrich Leisch}, {and} \bibinfo{person}{Roger~D Peng}.}
  \bibinfo{year}{2014}\natexlab{}.
\newblock \bibinfo{booktitle}{\emph{Implementing reproducible research}}.
\newblock \bibinfo{publisher}{CRC Press}.
\newblock


\bibitem[\protect\citeauthoryear{Stodden and Miguez}{Stodden and
  Miguez}{2014}]%
        {stodden2014best}
\bibfield{author}{\bibinfo{person}{Victoria Stodden} {and}
  \bibinfo{person}{Sheila Miguez}.} \bibinfo{year}{2014}\natexlab{}.
\newblock \showarticletitle{Best practices for computational science: Software
  infrastructure and environments for reproducible and extensible research}.
\newblock \bibinfo{journal}{\emph{Journal of Open Research Software}}
  \bibinfo{volume}{2}, \bibinfo{number}{1} (\bibinfo{year}{2014}), 21.
\newblock


\bibitem[\protect\citeauthoryear{Turk, Robert, and Rumpe}{Turk
  et~al\mbox{.}}{2005}]%
        {turk2005assumptions}
\bibfield{author}{\bibinfo{person}{Daniel Turk}, \bibinfo{person}{France
  Robert}, {and} \bibinfo{person}{Bernhard Rumpe}.}
  \bibinfo{year}{2005}\natexlab{}.
\newblock \showarticletitle{Assumptions underlying agile software-development
  processes}.
\newblock \bibinfo{journal}{\emph{Journal of Database Management (JDM)}}
  \bibinfo{volume}{16}, \bibinfo{number}{4} (\bibinfo{year}{2005}),
  \bibinfo{pages}{62--87}.
\newblock


\bibitem[\protect\citeauthoryear{{United States Office of the Director of
  National Intelligence}}{{United States Office of the Director of National
  Intelligence}}{2020}]%
        {ic-ai-ethics}
\bibfield{author}{\bibinfo{person}{{United States Office of the Director of
  National Intelligence}}.} \bibinfo{year}{2020}\natexlab{}.
\newblock \bibinfo{title}{{Artificial Intelligence Ethics Framework for the
  Intelligence Community}}.
\newblock
\newblock


\bibitem[\protect\citeauthoryear{Vaughan}{Vaughan}{1996}]%
        {vaughan1996challenger}
\bibfield{author}{\bibinfo{person}{Diane Vaughan}.}
  \bibinfo{year}{1996}\natexlab{}.
\newblock \bibinfo{booktitle}{\emph{The Challenger launch decision: Risky
  technology, culture, and deviance at NASA}}.
\newblock \bibinfo{publisher}{University of Chicago press}.
\newblock


\bibitem[\protect\citeauthoryear{Veale}{Veale}{2017}]%
        {veale2017logics}
\bibfield{author}{\bibinfo{person}{Michael Veale}.}
  \bibinfo{year}{2017}\natexlab{}.
\newblock \showarticletitle{Logics and practices of transparency and opacity in
  real-world applications of public sector machine learning}.
\newblock \bibinfo{journal}{\emph{{Workshop on Fairness, Accountability, and
  Transparency in Machine Learning (FAT/ML)}}} (\bibinfo{year}{2017}).
\newblock


\bibitem[\protect\citeauthoryear{Vu, Setty, Blumberg, and Walfish}{Vu
  et~al\mbox{.}}{2013}]%
        {vu2013hybrid}
\bibfield{author}{\bibinfo{person}{Victor Vu}, \bibinfo{person}{Srinath Setty},
  \bibinfo{person}{Andrew~J Blumberg}, {and} \bibinfo{person}{Michael
  Walfish}.} \bibinfo{year}{2013}\natexlab{}.
\newblock \showarticletitle{A hybrid architecture for interactive verifiable
  computation}. In \bibinfo{booktitle}{\emph{Proc. IEEE S \& P}}.
\newblock


\bibitem[\protect\citeauthoryear{Wagenknecht, Lee, Lustig, O'Neill, and
  Zade}{Wagenknecht et~al\mbox{.}}{2016}]%
        {wagenknecht2016}
\bibfield{author}{\bibinfo{person}{Susann Wagenknecht}, \bibinfo{person}{Min
  Lee}, \bibinfo{person}{Caitlin Lustig}, \bibinfo{person}{Jacki O'Neill},
  {and} \bibinfo{person}{Himanshu Zade}.} \bibinfo{year}{2016}\natexlab{}.
\newblock \showarticletitle{{Algorithms at Work: Empirical Diversity, Analytic
  Vocabularies, Design Implications}}. In \bibinfo{booktitle}{\emph{Proceedings
  of the 19th ACM Conference on Computer Supported Cooperative Work and Social
  Computing Companion - CSCW '16 Companion}}. \bibinfo{publisher}{ACM Press},
  \bibinfo{address}{New York, New York, USA}, \bibinfo{pages}{536--543}.
\newblock
\showISBNx{9781450339506}


\bibitem[\protect\citeauthoryear{Warden}{Warden}{2018}]%
        {warden2018reproducibility}
\bibfield{author}{\bibinfo{person}{Patrick Warden}.}
  \bibinfo{year}{2018}\natexlab{}.
\newblock \bibinfo{title}{The Machine Learning Reproducibility Crisis}.
\newblock
  \bibinfo{howpublished}{\url{https://petewarden.com/2018/03/19/the-machine-learning-reproducibility-crisis/}}.
\newblock


\bibitem[\protect\citeauthoryear{Waters, Balfanz, Durfee, and Smetters}{Waters
  et~al\mbox{.}}{2004}]%
        {waters2004building}
\bibfield{author}{\bibinfo{person}{Brent~R Waters}, \bibinfo{person}{Dirk
  Balfanz}, \bibinfo{person}{Glenn Durfee}, {and} \bibinfo{person}{Diana~K
  Smetters}.} \bibinfo{year}{2004}\natexlab{}.
\newblock \showarticletitle{Building an Encrypted and Searchable Audit Log}. In
  \bibinfo{booktitle}{\emph{NDSS}}, Vol.~\bibinfo{volume}{4}.
  \bibinfo{pages}{5--6}.
\newblock


\bibitem[\protect\citeauthoryear{Weitzner, Abelson, Berners-Lee, Feigenbaum,
  Hendler, and Sussman}{Weitzner et~al\mbox{.}}{2007}]%
        {weitzner2007information}
\bibfield{author}{\bibinfo{person}{Daniel~J Weitzner}, \bibinfo{person}{Harold
  Abelson}, \bibinfo{person}{Tim Berners-Lee}, \bibinfo{person}{Joan
  Feigenbaum}, \bibinfo{person}{James Hendler}, {and}
  \bibinfo{person}{Gerald~Jay Sussman}.} \bibinfo{year}{2007}\natexlab{}.
\newblock \bibinfo{booktitle}{\emph{Information Accountability}}.
\newblock \bibinfo{type}{{T}echnical {R}eport} MIT-CSAIL-TR-2007-034.
  \bibinfo{institution}{Massachussets Institute of Technology, Computer Science
  and Artificial Intelligence Laboratory}.
\newblock


\bibitem[\protect\citeauthoryear{Wieringa}{Wieringa}{2020}]%
        {wieringa2020account}
\bibfield{author}{\bibinfo{person}{Maranke Wieringa}.}
  \bibinfo{year}{2020}\natexlab{}.
\newblock \showarticletitle{What to account for when accounting for algorithms:
  a systematic literature review on algorithmic accountability}. In
  \bibinfo{booktitle}{\emph{Proceedings of the 2020 Conference on Fairness,
  Accountability, and Transparency}}. \bibinfo{publisher}{ACM},
  \bibinfo{address}{New York, NY}, \bibinfo{pages}{1--18}.
\newblock


\bibitem[\protect\citeauthoryear{Wolf, Zhu, Bullard, Lee, and Brubaker}{Wolf
  et~al\mbox{.}}{2018}]%
        {wolf2018}
\bibfield{author}{\bibinfo{person}{Christine~T Wolf}, \bibinfo{person}{Haiyi
  Zhu}, \bibinfo{person}{Julia Bullard}, \bibinfo{person}{Min~Kyung Lee}, {and}
  \bibinfo{person}{Jed~R Brubaker}.} \bibinfo{year}{2018}\natexlab{}.
\newblock \showarticletitle{{The Changing Contours of "Participation" in
  Data-driven, Algorithmic Ecosystems: Challenges, Tactics, and an Agenda}}. In
  \bibinfo{booktitle}{\emph{Companion of the 2018 ACM Conference on Computer
  Supported Cooperative Work and Social Computing - CSCW '18}}.
  \bibinfo{publisher}{ACM Press}, \bibinfo{address}{New York, New York, USA},
  \bibinfo{pages}{377--384}.
\newblock
\showISBNx{9781450360180}


\bibitem[\protect\citeauthoryear{Wong and Mulligan}{Wong and Mulligan}{2019}]%
        {wong2019bringing}
\bibfield{author}{\bibinfo{person}{Richmond~Y Wong} {and}
  \bibinfo{person}{Deirdre~K Mulligan}.} \bibinfo{year}{2019}\natexlab{}.
\newblock \showarticletitle{Bringing Design to the Privacy Table: Broadening
  “Design” in “Privacy by Design” Through the Lens of HCI}. In
  \bibinfo{booktitle}{\emph{2019 CHI Conference on Human Factors in Computing
  Systems}}. \bibinfo{pages}{1--17}.
\newblock


\bibitem[\protect\citeauthoryear{Young, Magassa, and Friedman}{Young
  et~al\mbox{.}}{2019a}]%
        {young2019toward}
\bibfield{author}{\bibinfo{person}{Meg Young}, \bibinfo{person}{Lassana
  Magassa}, {and} \bibinfo{person}{Batya Friedman}.}
  \bibinfo{year}{2019}\natexlab{a}.
\newblock \showarticletitle{Toward inclusive tech policy design: a method for
  underrepresented voices to strengthen tech policy documents}.
\newblock \bibinfo{journal}{\emph{Ethics and Information Technology}}
  \bibinfo{volume}{21}, \bibinfo{number}{2} (\bibinfo{year}{2019}),
  \bibinfo{pages}{89--103}.
\newblock


\bibitem[\protect\citeauthoryear{Young, Rodriguez, Keller, Sun, Sa,
  Whittington, and Howe}{Young et~al\mbox{.}}{2019b}]%
        {young2019beyond}
\bibfield{author}{\bibinfo{person}{Meg Young}, \bibinfo{person}{Luke
  Rodriguez}, \bibinfo{person}{Emily Keller}, \bibinfo{person}{Feiyang Sun},
  \bibinfo{person}{Boyang Sa}, \bibinfo{person}{Jan Whittington}, {and}
  \bibinfo{person}{Bill Howe}.} \bibinfo{year}{2019}\natexlab{b}.
\newblock \showarticletitle{Beyond open vs. closed: Balancing individual
  privacy and public accountability in data sharing}. In
  \bibinfo{booktitle}{\emph{Proceedings of the Conference on Fairness,
  Accountability, and Transparency}}. \bibinfo{pages}{191--200}.
\newblock


\bibitem[\protect\citeauthoryear{Zhu, Yu, Halfaker, and Terveen}{Zhu
  et~al\mbox{.}}{2018}]%
        {zhu2018}
\bibfield{author}{\bibinfo{person}{Haiyi Zhu}, \bibinfo{person}{Bowen Yu},
  \bibinfo{person}{Aaron Halfaker}, {and} \bibinfo{person}{Loren Terveen}.}
  \bibinfo{year}{2018}\natexlab{}.
\newblock \showarticletitle{{Value-Sensitive Algorithm Design: Method, Case
  Study, and Lessons}}.
\newblock \bibinfo{journal}{\emph{Proceedings of the ACM on Human-Computer
  Interaction}} \bibinfo{volume}{2}, \bibinfo{number}{CSCW}
  (\bibinfo{date}{Nov.} \bibinfo{year}{2018}), \bibinfo{pages}{1--23}.
\newblock
\showISSN{25730142}


\bibitem[\protect\citeauthoryear{Ziewitz}{Ziewitz}{2017}]%
        {ziewitz2017not}
\bibfield{author}{\bibinfo{person}{Malte Ziewitz}.}
  \bibinfo{year}{2017}\natexlab{}.
\newblock \showarticletitle{A not quite random walk: Experimenting with the
  ethnomethods of the algorithm}.
\newblock \bibinfo{journal}{\emph{Big Data \& Society}} \bibinfo{volume}{4},
  \bibinfo{number}{2} (\bibinfo{year}{2017}).
\newblock


\end{thebibliography}


\end{document}